\documentclass[fleqn,usenatbib]{mnras}

    \usepackage{graphicx}	
    \usepackage{amsmath}	
    \usepackage{nccmath}
    \usepackage{amssymb}	
    \usepackage{caption}
    \usepackage{subcaption}
    
    \usepackage{smartdiagram}
    \usesmartdiagramlibrary{additions}
    \usepackage{multirow}
    \usepackage{multicol}

    \title[Fitting spectra and profiles of $\gamma$-ray pulsars]
    {
    Synchro-curvature description of $\gamma$-ray light curves and spectra of pulsars: concurrent fitting
    }
    
    \author[\'I\~niguez-Pascual, Torres \& Vigan\`o]{Daniel \'I\~niguez-Pascual$^{1,2}$\thanks{E-mail: iniguez@ice.csic.es}, Diego F. Torres$^{1,2,3}$\thanks{E-mail: dtorres@ice.csic.es},Daniele Vigan\`o$^{1,2,4}$\thanks{E-mail: vigano@ice.csic.es}\\
    $^{1}$Institute of Space Sciences (ICE, CSIC), Campus UAB, Carrer de Can Magrans s/n, 08193 Barcelona, Spain\\
    $^{2}$Institut d’Estudis Espacials de Catalunya (IEEC), 08034 Barcelona, Spain\\
    $^{3}$Institució Catalana de Recerca i Estudis Avançats (ICREA), E-08010 Barcelona, Spain \\
    $^{4}$Institute of Applied Computing \& Community Code (IAC3), University of the Balearic Islands, Palma, 07122, Spain
    }
    
\begin{document}
    \label{firstpage}
    \pagerange{\pageref{firstpage}--\pageref{lastpage}}
    \maketitle
    
    \begin{abstract}
    We present a concurrent fitting of spectra and light curves of the whole population of detected gamma-ray pulsars.
    Using a synchro-curvature model we compare our theoretical output with the observational data published in the Third Fermi Pulsar Catalog, which has significantly increased the number of known gamma-ray pulsars. 
    Our model properly fits all the spectra and reproduces well a considerable fraction of light curves.
    Light curve fitting is carried out with two different techniques, whose strong points and caveats are discussed. We use a weighted reduced $\chi^2$ of light curves in time domain, and the Euclidean distance of the Fourier transform of the light curves, i.e. transforming the light curves to the frequency domain.
    The performance of both methods is found to be qualitatively similar, but individual best-fit solutions may differ.
    We also show that, in our model based on few effective parameters, the light curve fitting is basically insensitive to the timing and spectral parameters of the pulsar.
    Finally, we look for correlations between model and physical parameters, and recover trends found in previous studies but without any significant correlation involving geometrical parameters.
    \end{abstract}

    \begin{keywords}  
    pulsars: general -- gamma-rays: stars -- X-rays: stars -- acceleration of particles -- radiation mechanisms: non-thermal
    \end{keywords}

    \section{Introduction}
    \label{introduction}

    The population of detected gamma-ray pulsars has been greatly enlarged by the recent release of the Third Fermi Pulsar Catalog \citep{3fpc}, hereafter referred to as 3PC, containing data for 294 objects\footnote{\url{https://fermi.gsfc.nasa.gov/ssc/data/access/lat/3rd_PSR_catalog/}}.
    A recognizable feature of the 3PC is the quality of the gamma-ray light curves presented.
    Comparison of these observational light curves with synthetic ones generated by theoretical models can provide relevant information about the structure of pulsars magnetospheres, for instance about the production sites of high-energy radiation.
    This task has been addressed widely in the last 20 years, either considering gap models \citep{Watters09, Romani10, Venter09, Pierbattista15}, force-free electrodynamics \citep{Bai10a, Kalapotharakos14,Cao19,Benli21,Petri21} or Particle-in-Cell (PIC) magnetospheric simulations \citep{Philippov18, Kalapotharakos18}. Each of these approaches focus on different aspects of pulsar physics, and their versatility, computational cost, and physics included in the models differ from one another, see \cite{review_pulsar_magnetospheres_and_radiation} for a review.

    Very recently, \cite{Cerutti24} has used PIC simulations to fit the light curves of the 3PC, with promising results.
    We shall compare our results with this paper in more detail below.
    Only few studies exist producing together gamma-ray light curves and spectra of pulsars \citep{Cerutti16, Petri19, Kalapotharakos23} and even less comparing those with observational data \citep{Chang19,Yang24}.

    In this paper we present a concurrent fitting of spectra and light curves for the whole population of objects quoted in the 3PC.
    We follow the spectral fitting from our previous works \citep{Vigan_2015b, torres19_systematic_fitting, iniguezpascual22_sc_emitting_regions_2022}, and here we mostly focus on the light curve fitting.
    The latter is performed with two different techniques that we scrutinize in detail, i.e. using a weighted, reduced $\chi^2$ to assess the goodness of the light curve fits in time domain or using the Euclidean distance in frequency domain by comparing the Fourier transform of both the observational and synthetic light curves.
    Section \ref{sc_model} presents a brief summary of the spectral and geometrical models that generate the high-energy  spectra and light curves.
    The comparison of theoretical spectra with observational ones is shown in Section \ref{fitting_spectra}.
    The light curve fitting procedures are discussed in Section \ref{fitting_lightcurves}, and the results shown in Section \ref{results_lightcurve_fitting}.
    We also consider how well a single pulsar can describe the whole light curve variety, despite being generated in systems with different timing and spectral properties.
    Finally, in Section \ref{conclusions} we draw the main conclusions.

    \section{Synchro-curvature model}
    \label{sc_model}
    
    The synchro-curvature model we use in this work has been extensively studied in a series of papers before
    \citep{Vigan_2015,Vigan_2015b,diego_solo,torres19_systematic_fitting,vigano19_light_curves, iniguezpascual22_sc_emitting_regions_2022,iniguezpascual22_period_inference,iniguezpascual24}. In this set of papers, among other things, we enlarged the energy coverage from gamma to X-rays,  included better descriptions of the emission regions, studied how well the spectra can be used to infere the pulsar periods.
    Here we are not presenting additional theoretical improvement of the spectral model and thus, we will just give a very brief summary, referring the reader to these works for more details.
    For an alternative approach see \cite{Kelner15}, who propose a Hamiltonian-based formulation to provide for the particle motion and their synchro-curvature radiation for a given magnetic field configuration, and a more detailed study of the radiation inside and outside the accelerating region.
    
    In our model, we follow the dynamics of a bunch of charged particles moving in the magnetosphere of a pulsar, outside the light cylinder, whose motion is described by the equation of motion of charged particles, which balances the radiative losses and the electromagnetic acceleration given by a parallel electric field of strength $E_{||}$.
    The magnetic field strength along the particles trajectory is parametrized as $B = B_* (x/R_*)^{-b}$ where $B_*= 6.4\times10^{19}\sqrt{P\dot{P}}$G is the surface magnetic field at the pole assuming a dipolar field, where $P$ is the period, $\dot{P}$ the period derivative, $R_*$ is the neutron star radius and $b$ is the magnetic gradient; and the curvature radius of the drift trajectory (i.e. gyration-averaged, see e.g. \cite{iniguezpascual24, Kelner15}) as $r_c = (x/R_{lc})^{\eta}$, where $R_{lc}=\dfrac{c P}{2\pi}$ is the light cylinder radius and $\eta$ is fixed to $0.5$ given its negligible impact on the results found in previous works.
    Having the strength of the electric field component parallel to the magnetic field $E_{||}$ and $b$ as free model parameters, we numerically solve the equation of motion and use the synchro-curvature formulae to obtain the emission of a single particle at each position $\lambda$ \citep{cheng_zhang, compact_formulae}, $dP_{sc}(\lambda)/dE$, which is then convolved with a relative weight given to the particle distribution, $dN/d\lambda$, ruled by a lenghtscale $x_0$, which is the third free parameter of our spectral model. 
    These three parameters, plus a normalization factor $N_0$ that is fixed by observational comparison, fully describe the total radiation emitted by a pulsar with a particular $P$ and $\Dot{P}$, referred to as $dP_{tot}/dE$ as in Eq. (7) of \cite{iniguezpascual22_sc_emitting_regions_2022}.

    Together with the dynamics and emission of the emitting particles, we also compute the geometry of the trajectories they follow.
    In \cite{vigano19_light_curves} and \cite{iniguezpascual24} we presented how to obtain the emission directions of the particles on the emission region, whose shape is given by Eq. (2) of \cite{iniguezpascual24}. 
    This expression depends on the inclination angle, $\psi_\Omega$, the angle between the rotational and magnetic axes.
    Notice that by emitting region we mean the whole set of particle trajectories considered to be generating the radiation.
    For a fixed set of spectral parameters, collecting these emission directions in a sphere centered at the neutron star we build energy-dependent synchro-curvature emission maps, or simply, skymaps, which we refer to as $M_E(\theta_{obs}, \phi_{\Omega},E)$ as in Equation 5 of \cite{iniguezpascual24}.
    This represents the photon flux received by a given observer at a given phase and energy, per unit energy, per unit solid angle, emitted by particles injected all along the region.
    In these skymaps, the x-axis corresponds to the rotational phase of the neutron star $\phi_{\Omega}$ and the y-axis is the viewing angle $\theta_{obs}$, the angle between the plane perpendicular to the rotational axis of the neutron star and the line-of-sight of an observer detecting the pulsar. 
    By symmetry on the region definition, calculations are symmetric around the equatorial observer.
    For a given pulsar, skymaps are mostly determined by $\psi_\Omega$.
    In this work we consider the emission integrated over the Fermi energy range, 100 MeV -- 300 GeV, for the sake of a direct comparison with data.
    The emission maps within this range don't vary much with energy, and we leave the comparison between different sub-ranges for a future work.
    The synthetic, discrete light curve, for a given viewing angle $\theta_{obs}$ is then defined as: $I (\phi_{\Omega}) = \int_{E_{min}}^{E_{max}} M_E(\theta_{obs}, \phi_{\Omega},E) dE$.
    
    For a given pulsar with period $P$ and period derivative $\dot P$, we find the parameters $E_{||}, b, x_0$ which better fit the phase-averaged spectrum (including X-ray data, if available) as in our previous studies, and then look for the values of $\psi_\Omega$ and $\theta_{obs}$ which best fit the Fermi light curve.

    \section{Fitting observational spectra}
    \label{fitting_spectra}
    
    \subsection{Sample selection}
    \label{spectral_sample_selection}
    
    Out of the 294 gamma-ray pulsars presented on the 3PC, there is a fraction that is currently not eligible for fitting.
    This happens either because the pulsars lack a measured $\Dot{P}$, or because their spectra are described with too few data points. 
    With regards to the latter, we only consider fitable pulsars those having 5 consecutive bins with a measured flux (not upper limits) in their spectra. 
    In this way, the fitting sample consists of 129 pulsars, 39 of them having also detected X-ray pulsations, the same sample studied in \cite{CotiZelati20}.

    Even though spectra of many previously unknown gamma-ray pulsar have been released in the 3PC, a dedicated spectral analysis to each pulsar on the catalog has not been performed yet. 
    The spectra given in the catalog relate to those published in the  Incremental Fermi-LAT Fourth Source Catalog \citep{4th_fermilat_catalog_dr3}. 
    For this reason, in some pulsars the spectral data on the 3PC may be less detailed than that publicly available before its release, coming either from \cite{2fpc} (hereafter 2PC), where a dedicated spectral analysis for each pulsar was done, or in dedicated studies of particular pulsars.

    Therefore, we do not take the whole sample of spectra directly from the 3PC, but instead do a mix between those and the spectral data we used in \cite{iniguezpascual22_sc_emitting_regions_2022}, explicitly stating the origin of the data in a case-by-case basis.
    For the new pulsars we obviously take the 3PC data. 
    For those with already-available data, we decide between the spectra from the 3PC or the older ones (meaning 2PC or dedicated studies) based on their quality and following the criterion of having at least 5 consecutive bins with a measured flux (i.e. not upper limits).
    In pulsars for which it is not immediately obvious which set to pick, we choose the data providing the largest number of consecutive data points, always checking that the overall shape between the sets is similar and/or that their difference lies within the errors. 
    For these pulsars we consider systematic errors on the best-fit parameters (being the difference between the best-fit parameters obtained with the 3PC observational spectra and the alternative spectra) in addition to the statistical ones. 
    In a few cases, the spectra of the two sets of data differ widely.
    We do not consider these pulsars in our systematic fitting, but show them in Appendix \ref{app:data_comparison}.
    They are interesting candidates for further study and the light curves are mostly unaffected (see below) by spectral changes.

    \subsection{Results}\label{results_spectral_fitting}
    
    The spectral fitting procedure we have followed in this work is the same as that in \cite{iniguezpascual22_sc_emitting_regions_2022}, and we refer the reader there for more details.
    For each pulsar we let our free spectral parameters to vary and search for the best set of parameters by minimizing the reduced $\chi^2$.
    For pulsars with available X-ray data the three parameters ($E_{||}$, $b$, $x_0$) are let free, while for those pulsars having only gamma-ray data, the magnetic gradient $b$ is fixed to 2.5, since this parameter has a a low impact in the gamma-ray band and mainly affects the X-ray regime of the spectra \citep{diego_solo}.
    The value is chosen based on average best-fit values obtained from previous studies.
    
\begin{figure*}
        \centering
        \includegraphics[width=0.25\textwidth]{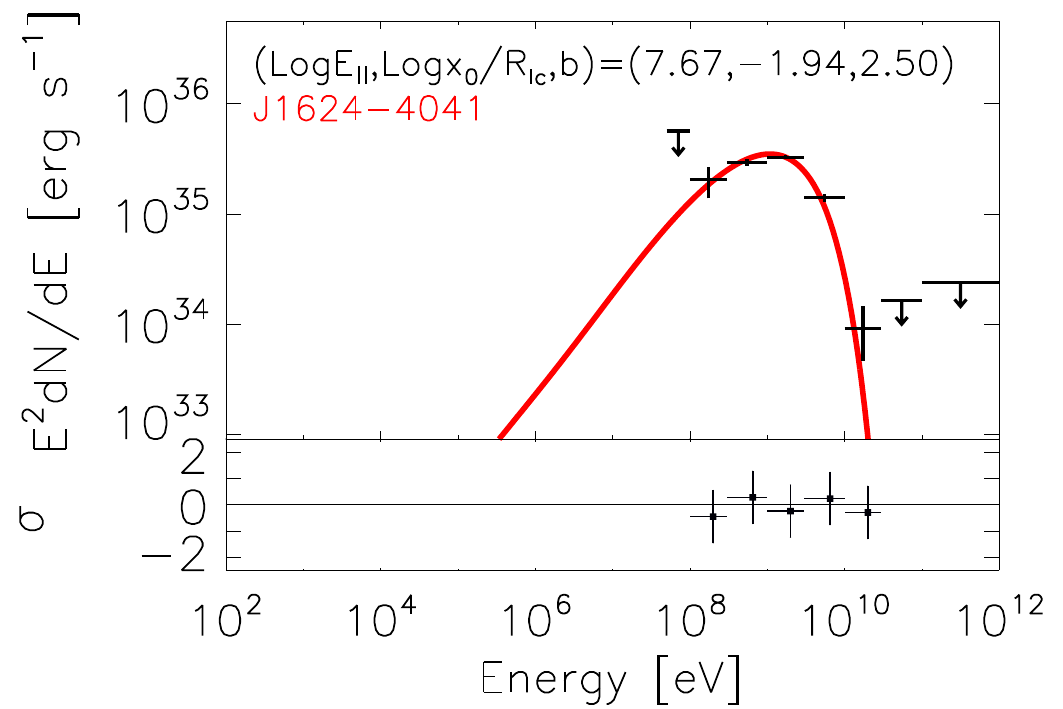}%
        \includegraphics[width=0.25\textwidth]{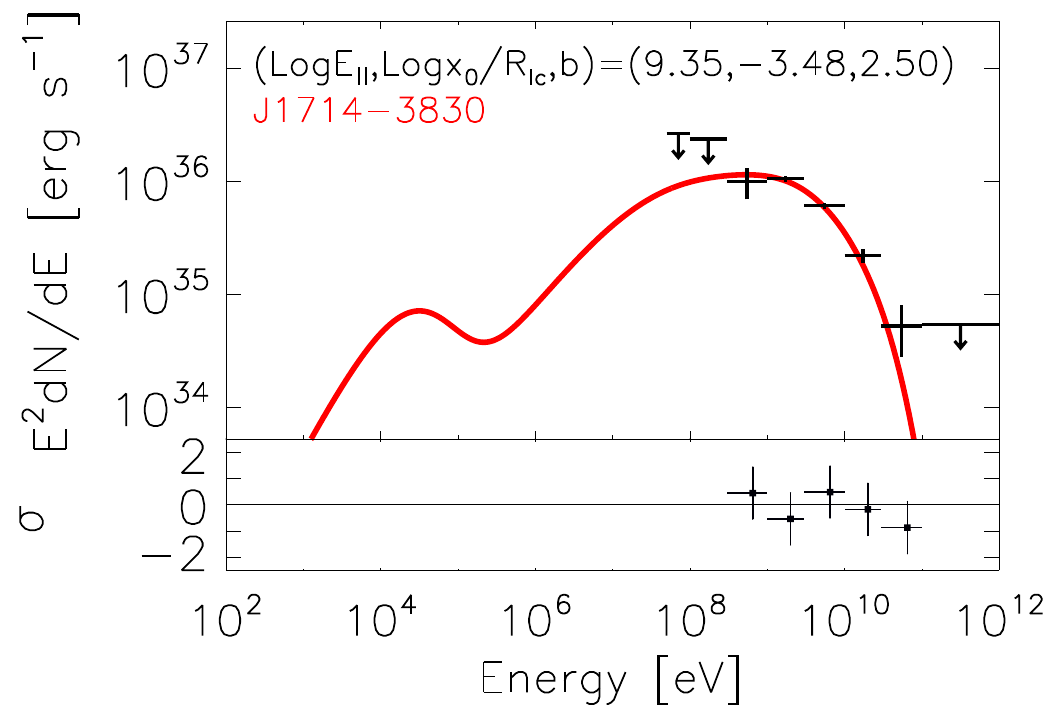}%
        \includegraphics[width=0.25\textwidth]{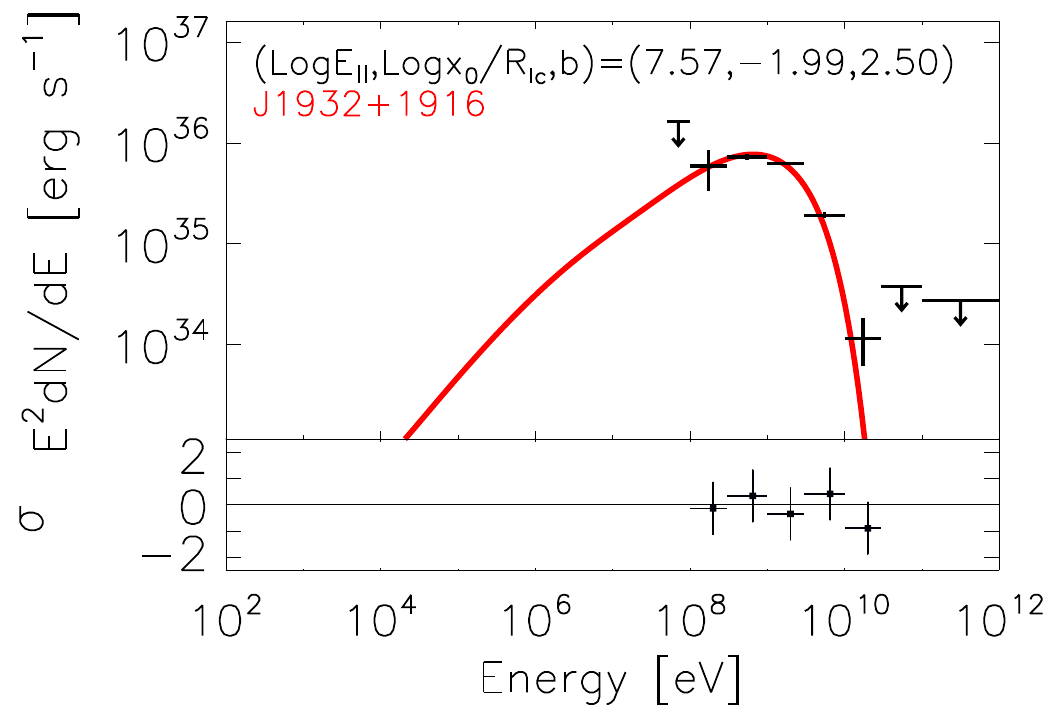}%
        \includegraphics[width=0.25\textwidth]{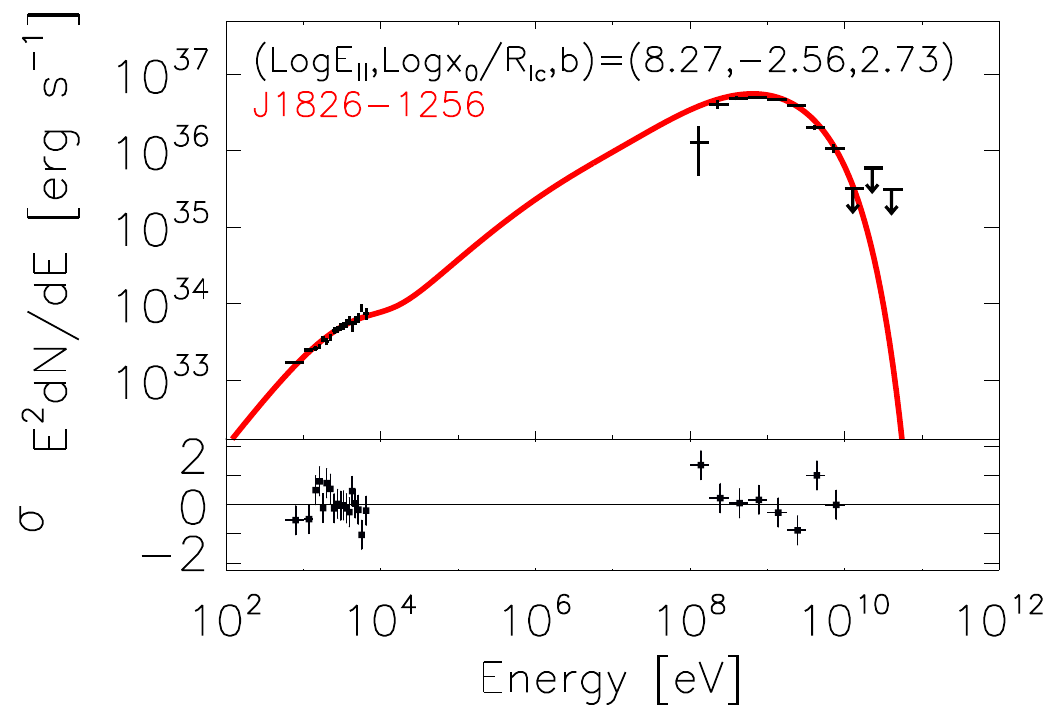}\\
        \includegraphics[width=0.25\textwidth]{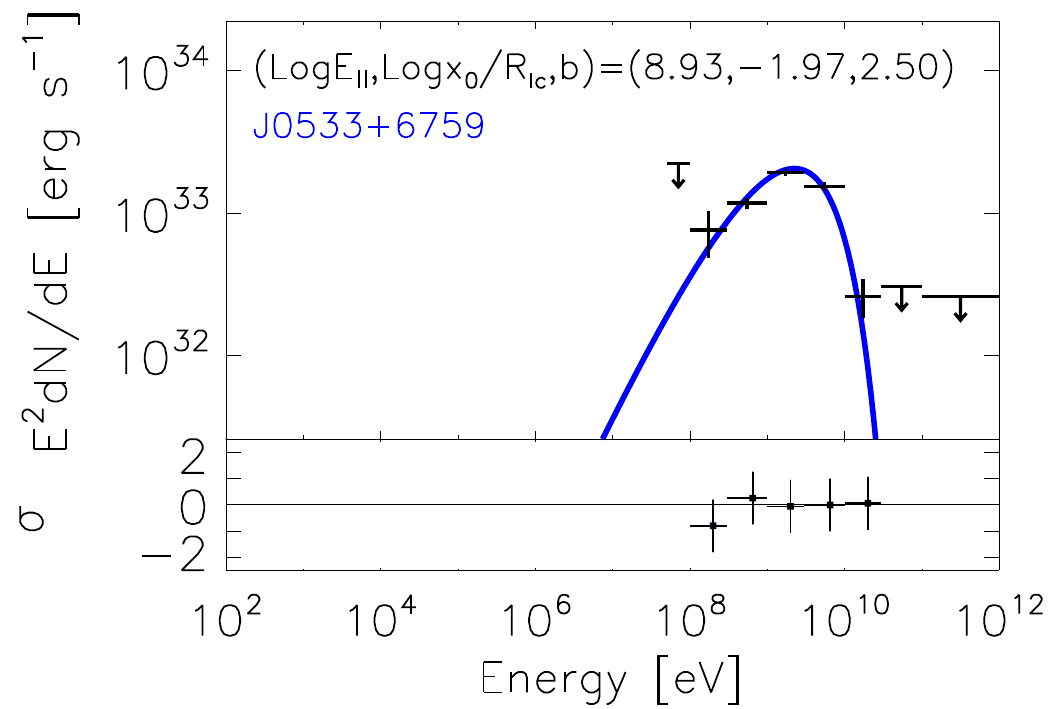}%
        \includegraphics[width=0.25\textwidth]{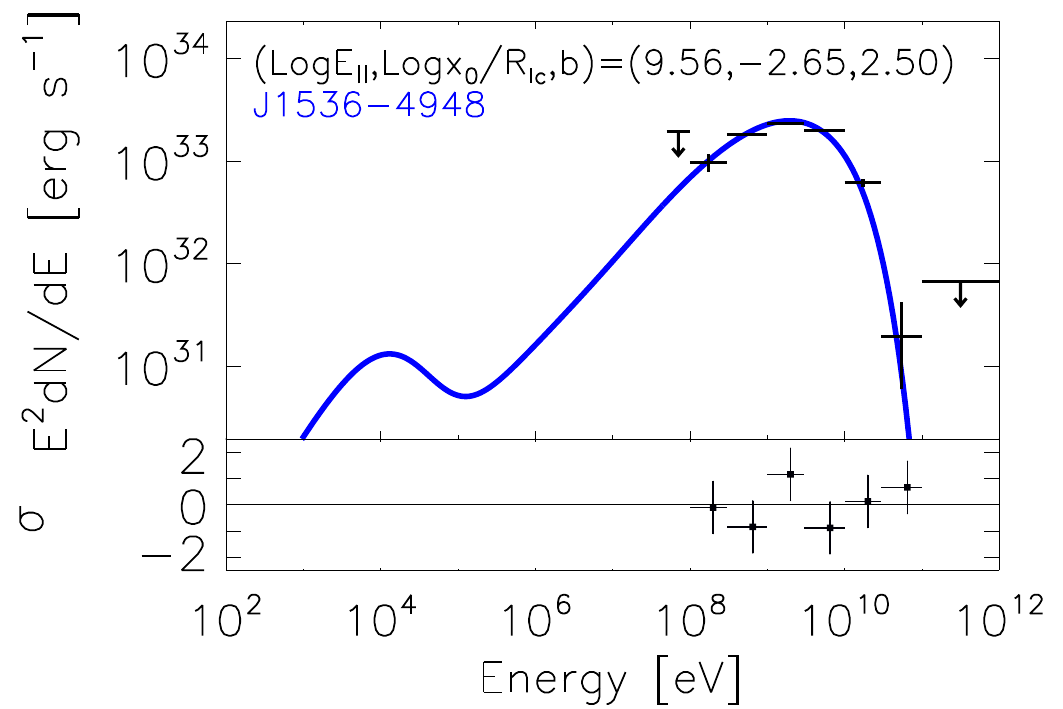}%
        \includegraphics[width=0.25\textwidth]{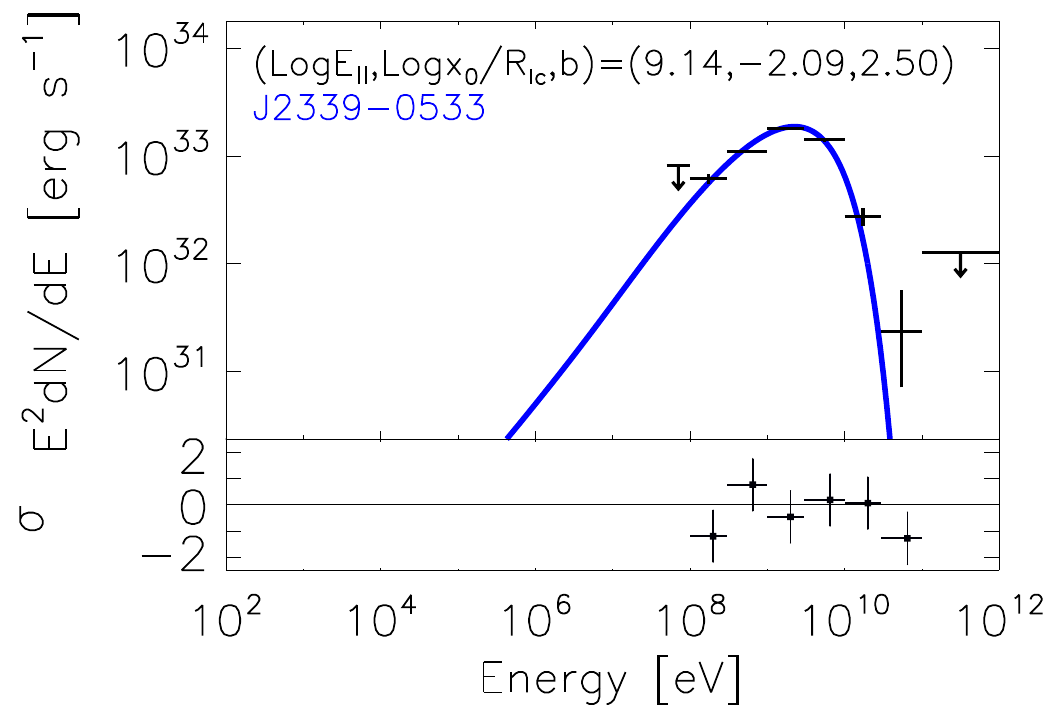}%
        \includegraphics[width=0.25\textwidth]{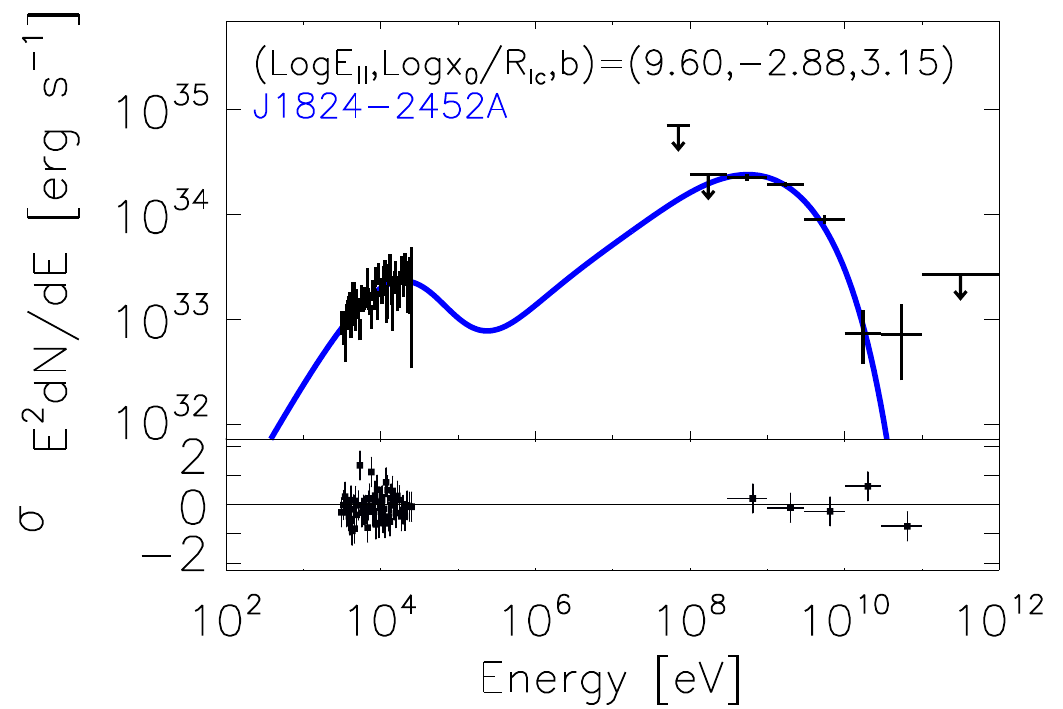}\\
        \caption{Examples of best-fitting spectra of some of the new pulsars reported in gamma-rays in the 3PC. Red (blue) lines are used for young (millisecond) pulsars. }
    \label{fig:spectral_fits}
    \end{figure*}

    Fig. \ref{fig:spectral_fits} shows examples of our spectral fits for some new pulsars presented in the 3PC.
    The model is able to reproduce the spectra of the whole population of gamma-ray pulsars. 
    Thanks to its flexibility, it can resemble the wide variety of gamma-ray spectra, from those with very narrow peaks to those with broader peaks, looking almost flat at energies of  $\sim 1$ GeV.
    Regarding pulsars with X-ray data, results are very similar to \cite{iniguezpascual22_sc_emitting_regions_2022}: the gamma-ray range is reproduced in all the cases, and the X-ray band is well matched in the majority of them with a single set of parameters.

    Values of the best-fit parameters are similar to those obtained in previous works. 
    Small values of lengthscale $x_0$ are obtained, and reinforce the idea that a relevant synchrotron contribution is required to explain the zoo of observed gamma-ray spectra.
    Again, young and millisecond pulsars are equally well fitted.
    The only difference between these two groups is the preference of millisecond pulsars for larger values of the parallel electric field $E_{||}$ (and lower values of the lengthscale $x_0$), as we have already observed in previous works \citep{Vigan_2015b, torres19_systematic_fitting}. 
    For each pulsar, the best-fit values of the spectral parameters ($E_{||}$, $b$, $x_0$) and the normalization $N_0$ are used in the geometrical model to generate light curves, whose fitting is presented in the next section.
    
    \subsection{Searching for correlations between spectral parameters}\label{correlations_spectral_parameters}

    \cite{Vigan_2015b} showed the existence of some trends between spectral and physical parameters of pulsars.
    Now that the number of observed gamma-ray pulsars has increased, it is a good exercise to check whether the found correlations remain or not, and if new ones appear.
    To appoint something as a significant correlation we follow the same criterion as in \cite{Vigan_2015b} and require a Pearson r value larger than 0.85.
    We only consider spectral parameters that have been correctly determined, i.e. those of pulsars whose spectra data fulfills the quality criterion stated in subsection \ref{spectral_sample_selection}.

    The most relevant correlation found is indeed between the spectral parameters $E_{||}$ and $x_0$, with parameters of the best-linear fit very similar to those of the previous study (see Section 5 of \cite{Vigan_2015b} for a discussion on the physical implications of this correlation).
    The correlation between $E_{||}$ and the magnetic field strength at the light cylinder $B_{lc}$ and that of $N_0$ and the theoretical gamma-ray luminosity $L_{theo}$ are now still statistically significant but do not reach our Pearson r threshold, probably due to the enlargement of the population.
    In both cases the parameters of the best linear-fit are very similar to the previous ones.
    The other correlations presented in that paper involved parameters of the phenomenological function PLEC1 \citep{Abdo12_PLEC1} from the 2PC, which has been updated to the PLEC4 \citep{4th_fermilat_catalog_dr3} on the 3PC. 
    No correlations of the PLEC4 model parameters with our (spectral or geometrical) model parameters are found.

    \section{Fitting observational light curves}
    \label{fitting_lightcurves}

    \subsection{Important considerations prior to fitting}
    
    Fitting synthetic light curves to observational ones is not a straightforward task and several considerations have to be taken into account before addressing it. 
    
    \emph{Fermi} gamma-ray light curves in the 3PC differ in the number of phase bins with which they are described.
    For each particular pulsar we then adapt the construction of the theoretical emission maps depending on its number of phase bins. 
    For pulsars with 25, 50 or 100 observational phase bins, we fix the number of synthetic azimuthal bins to 25, 50 or 100, respectively. 
    For pulsars with 200, 400 or 800 observational bins, we set the number of synthetic bins to 100, in order to keep the computational time affordable, since having more bins in the map requires to add more trajectories to the simulation to have an acceptable map resolution. 
    We thus rebin the observational light curves with 200, 400 and 800 bins to have 100 bins in total.
    Notice that by doing this we are losing some information, but what we lose are small-scale features of the observational light curve that our model at its current stage is still not able to reproduce.

    In addition, observational light curves have a certain level of background resulting from detected photons that do not come from the pulsar itself. 
    In the 3PC, a background level is estimated for each pulsar, being computed from the photon weights and indicated in the pulse profile plots by a black dashed line. 
    Our synthetic light curves do not have such background, 
    because the theoretical emission maps represent all the emission coming from the pulsar, and only from the pulsar.
    Therefore we subtract to each observational light curve its corresponding background level obtained from the 3PC before comparing with synthetic predictions.

    We shall impose a data quality criterion before fitting light curves, and do not consider light curves with only 25 observational bins.
    These have very large errors and their shape is generally not well defined yet.
    We also require that a spectral fit can be performed for the pulsars for which we fit the light curves, but here without imposing any restriction on the quality of their spectral data, since in \cite{iniguezpascual24} we demonstrated the small impact of the spectral parameters on the shape of the light curves.
    The population available for our fitting contains 226 gamma-ray light curves after these quality cuts.

    For each one of these 226 pulsars (with particular values of $P$, $\dot{P}$ and spectral parameters) we generate 30 emission maps, evenly sampling $\psi_{\Omega}$ values from $3^{\circ}$ to $90^{\circ}$ (we avoid the case of an aligned rotator, $\psi_{\Omega} = 0^{\circ}$, because no pulsed emission would be seen), and consider 51 observers, $\theta_{obs}$, evenly distributed between $-90^{\circ}$ and $90^{\circ}$.

    \subsection{Fitting procedures}
    \label{fitting_procedures}
    
    We compare the whole synthetic sample of light curves we have generated for each particular pulsar with the corresponding observational light curve.
    Below, we present and discuss different metrics used here to do such a comparison.
    
    \subsubsection{Fitting procedure in the time domain}
    \label{fitting_procedure_time}
    
    The reduced $\chi^2$ method can be used similarly to what is done in the spectral fitting, here to assess the goodness of the fit of $I^{obs}_{i}$ by $I^{syn}_{i}$, i.e. the intensities (in weighted counts/bin units) of the observational and synthetic light curves, respectively.
    In our synthetic light curves, phase $\phi_{\Omega}=0$ corresponds to the plane containing the rotational and magnetic axes, but this is arbitrary.
    For this reason, synthetic and observational light curves are not necessarily aligned. 
    In order to find the best alignment, we rotate each synthetic light curve and compute the reduced $\chi^2$ value for all rotations (as many as the number of observational bins), keeping the smallest one as the reduced $\chi^2$ of that synthetic light curves set.
    We reverse each synthetic light curve to obtain an additional one available for fitting. 
    Two pulsars with the same geometrical parameters but spinning in opposite directions would a priori create different light curves. However, since an inversion in the direction of rotation of the star is equivalent to a pulsar turned upside down and our region possesses an equatorial symmetry, in our model one light curve would be the reversed version of the other.   
    In addition, we normalize the synthetic light curves to the observational ones by using the standard, analytical linear regression formula: a normalization factor $N_{lc}$ is found by solving $\partial \chi^2 / \partial N_{lc} = 0$ and multiplied to our synthetic light curves. 

\begin{figure*}
        \centering
        \includegraphics[width=0.33\textwidth]{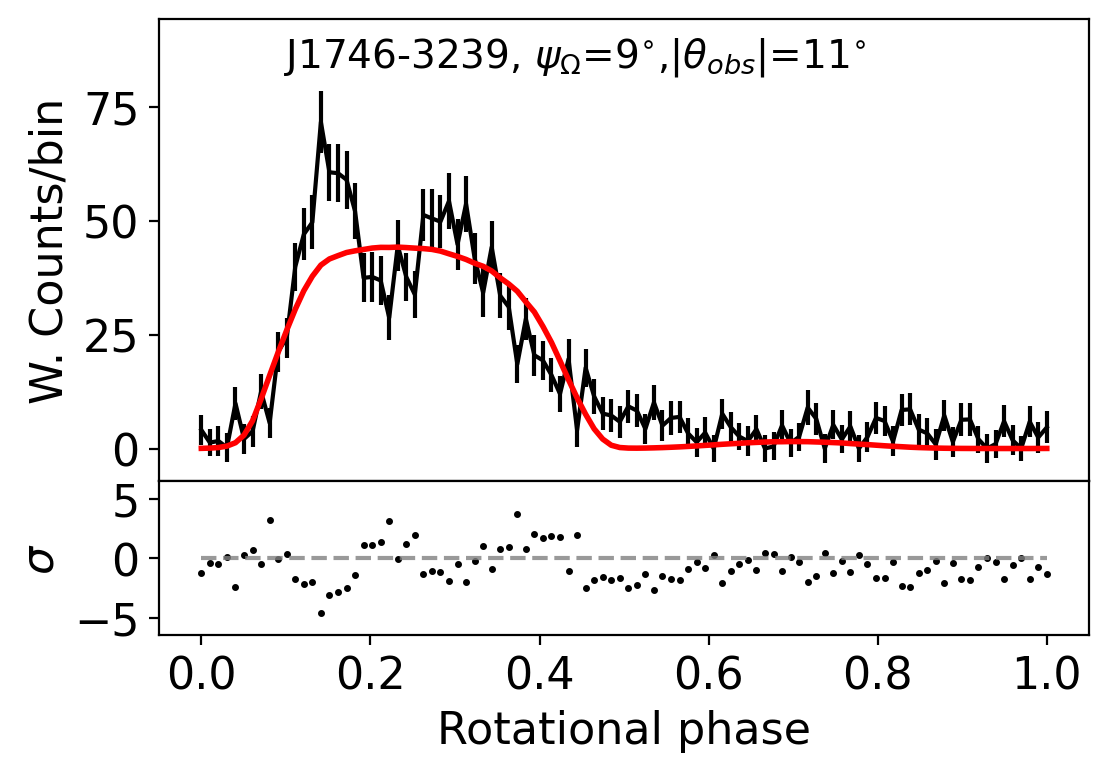}%
        \includegraphics[width=0.33\textwidth]{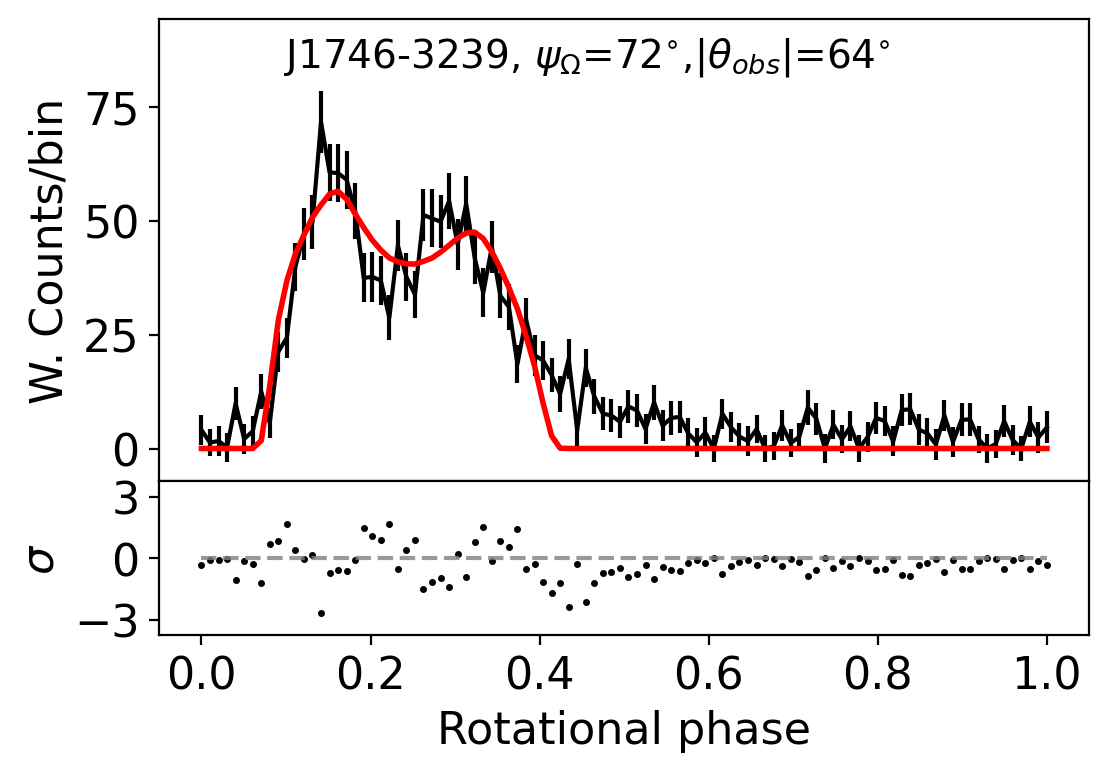}%
        \includegraphics[width=0.33\textwidth]{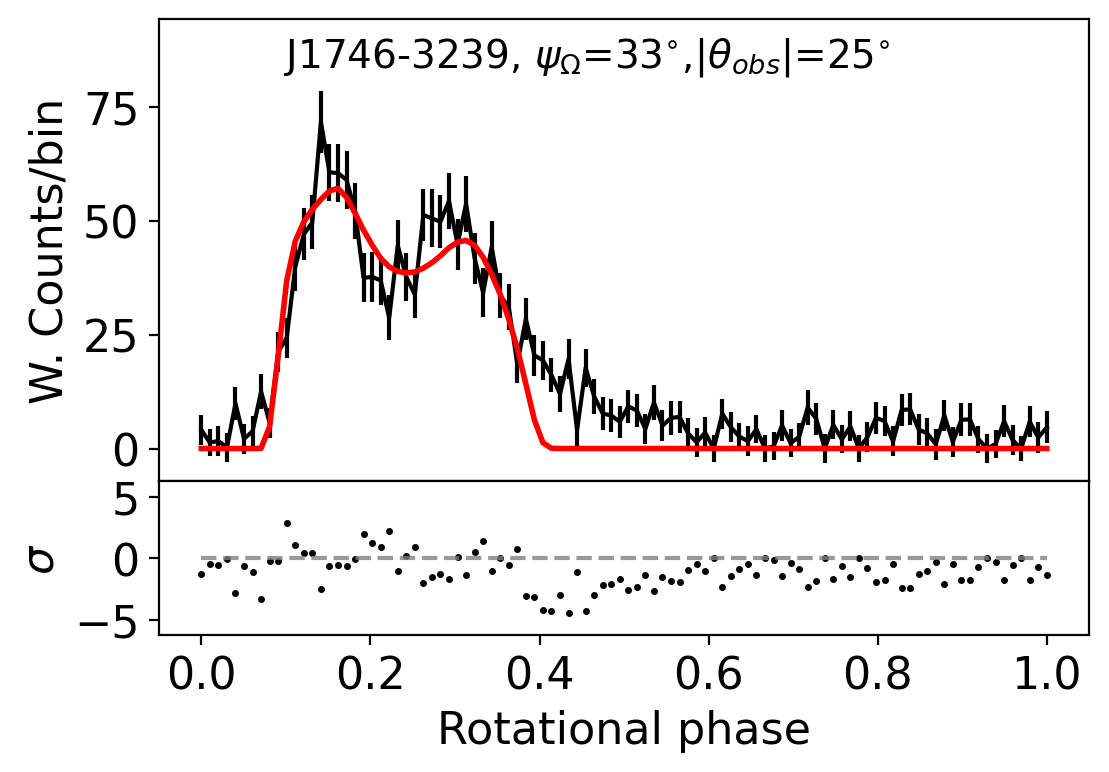}\\
        \includegraphics[width=0.33\textwidth]{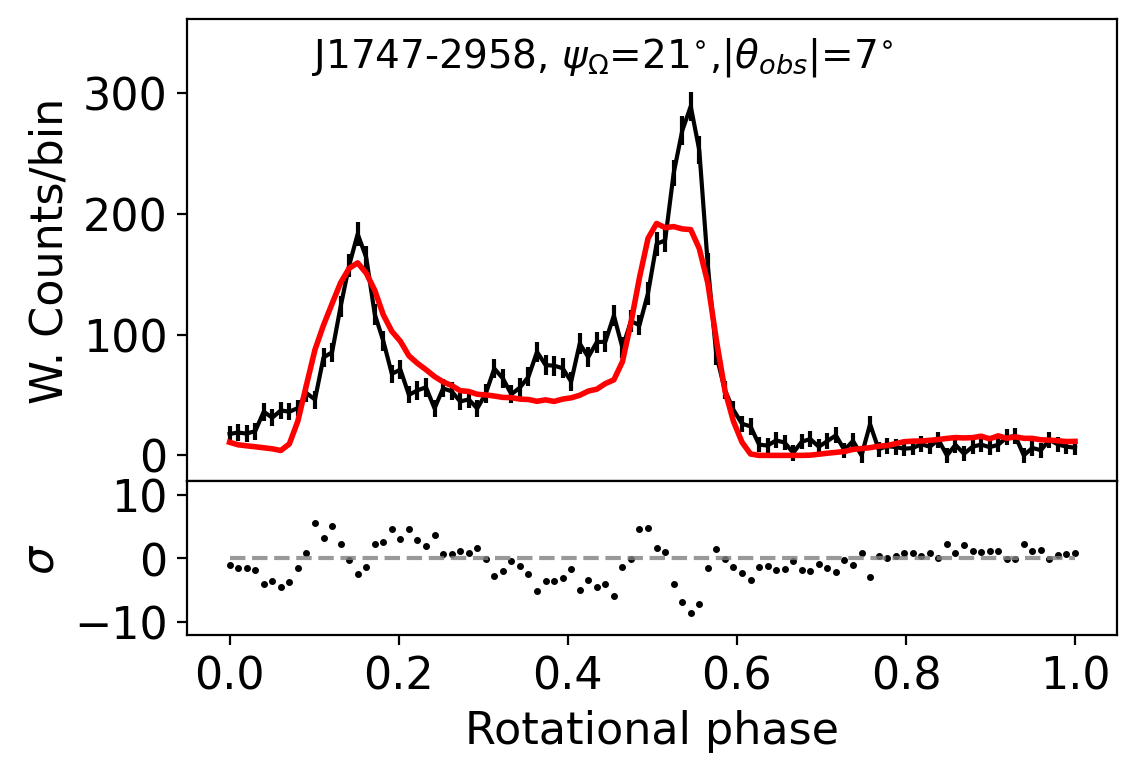}%
        \includegraphics[width=0.33\textwidth]{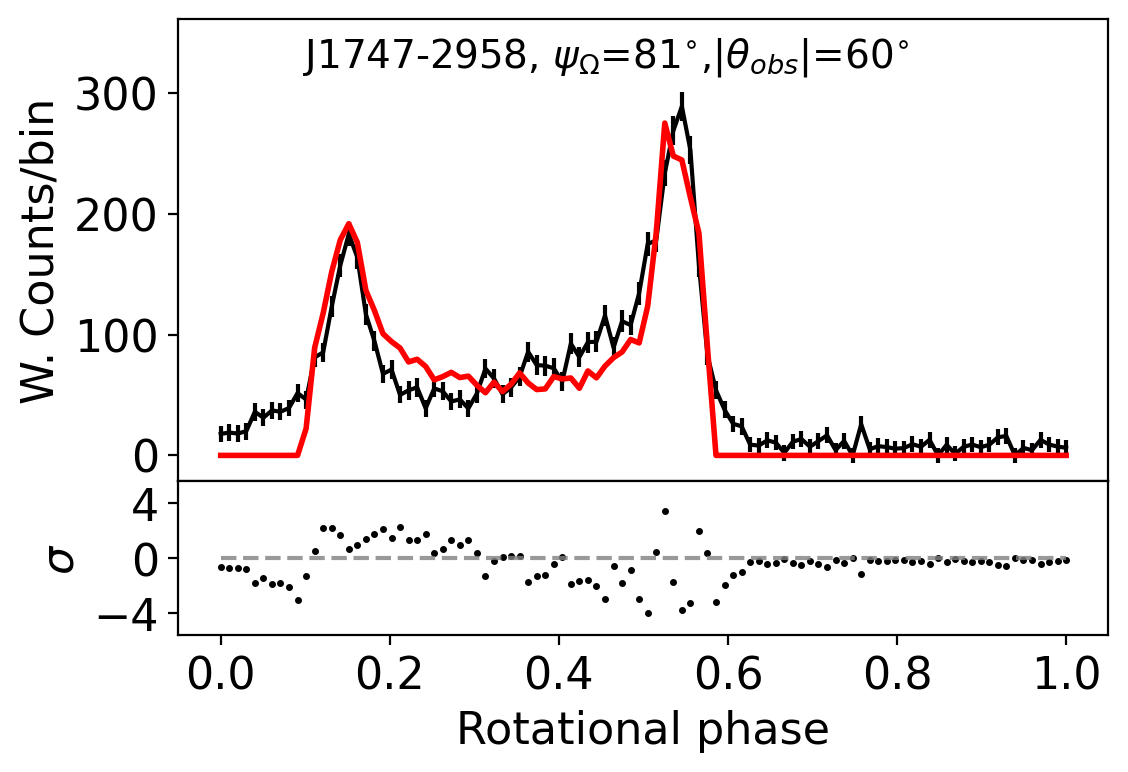}%
        \includegraphics[width=0.33\textwidth]{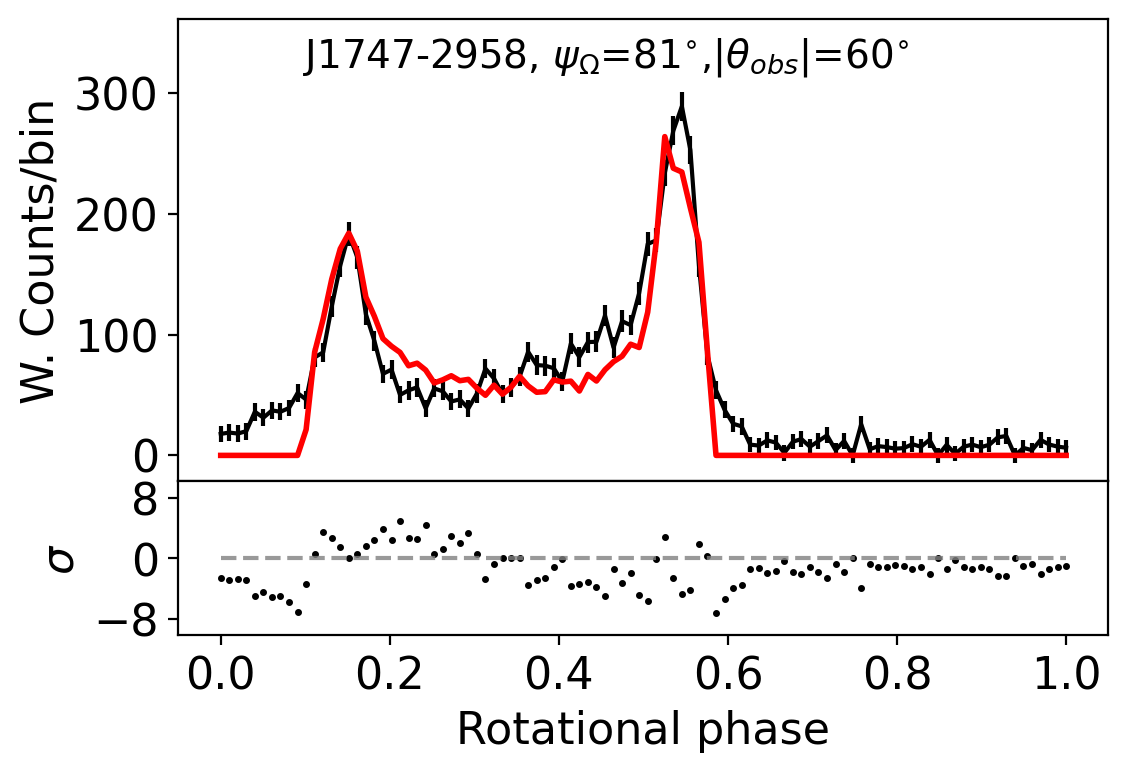}\\
        \caption{Examples of the caveats intrinsically associated with a fitting with $\chi^2$ and possible solutions, for the pulsars J1746 (top row) and J1747-2958 (bottom row). 
        From left to right: best-fitting light curve with a normal $\chi^2$, best-fitting light curve with a weighted $\chi^2$ and best-fitting light curve recovered with the inverse transform from fitting in the frequency domain.
    }
    \label{fig:chi2_caveats}
    \end{figure*}

    However, fitting light curves with $\chi^2$ has some intrinsic caveats (see \cite{garcia25_dtw} for a deeper discussion). 
    For example, a small phase variation of one peak in multi-peaked light curves results in a large $\chi^2$ value, even though the two light curves compared are visually very similar. 
    Fig. \ref{fig:chi2_caveats} shows two examples of cases in which $\chi^2$ selects a best fit for which the peak structure is not well reproduced, and therefore it is not convincing from a human eye perspective.

    In addition, in light curves with narrow peaks, $\chi^2$ can be biased towards synthetic light curves reproducing the observational bins with emission close to the background but failing to reproduce the peaks: since all bins contribute with equal weights to $\chi^2$, a good adjust to many low-flux bins could give an overall small value for $\chi^2$ even when the peaks are not matched at all. 
    In order to reduce this effect, an approach similar to the one proposed in \cite{Churazov96_weighted_chi2} can be taken and consider a weighted reduced $\chi^2$ (we shall keep calling it $\chi^2$ for simplicity), in which each bin’s squared difference is weighted by the normalized flux of the observational light curve, $\bar{I}^{obs}_{i}$, in order to give more weight to the bins with higher flux, as
    \begin{equation}
        \overline{\chi_w^2} = \frac{1}{n - 2} \sum_{i} \frac{(I^{obs}_{i} - I^{syn}_{i})^2 \bar{I}^{obs}_{i}} {(\delta I^{obs}_{i})^2} \hspace{0.3cm} ,
        \label{eq:chi2}
    \end{equation}
    where $n$ is the number of bins of a light curve in time domain, the $2$ value represents the number of free parameters in the fitting, $I^{obs}_{i}$ and $I^{syn}_{i}$ are the intensities of the observational and synthetic light curves, respectively, and $\delta I^{obs}_{i}$ is the observational error.
    In this way, more weight is given to the peaks, and we are promoting to properly reproduce the peaks rather than the low-flux level parts of the light curves.
    Middle panels of Fig. \ref{fig:chi2_caveats} present the best-fitting light curves found with the weighted $\chi^2$ showing that the caveats associated to the normal $\chi^2$ no longer appear.
    Notice that the weighted $\chi^2$ is also not a perfect metric: the peak tails of the chosen best-fit synthetic light curves deviate more from the observational ones than in the regular $\chi^2$ case.
    Hereafter, we will use Eq. \ref{eq:chi2} for the light curve fitting, keeping in mind in any case the unavoidable freedom in choosing a given metric for the comparison, and how the choice is driven by the above mentioned visual, subjective considerations about when a fit is satisfactory or not.

    \subsubsection{Fitting procedure in the frequency domain}
    \label{fitting_procedure_frequency}
    
\begin{figure*}
        \centering
        \includegraphics[width=0.25\textwidth]{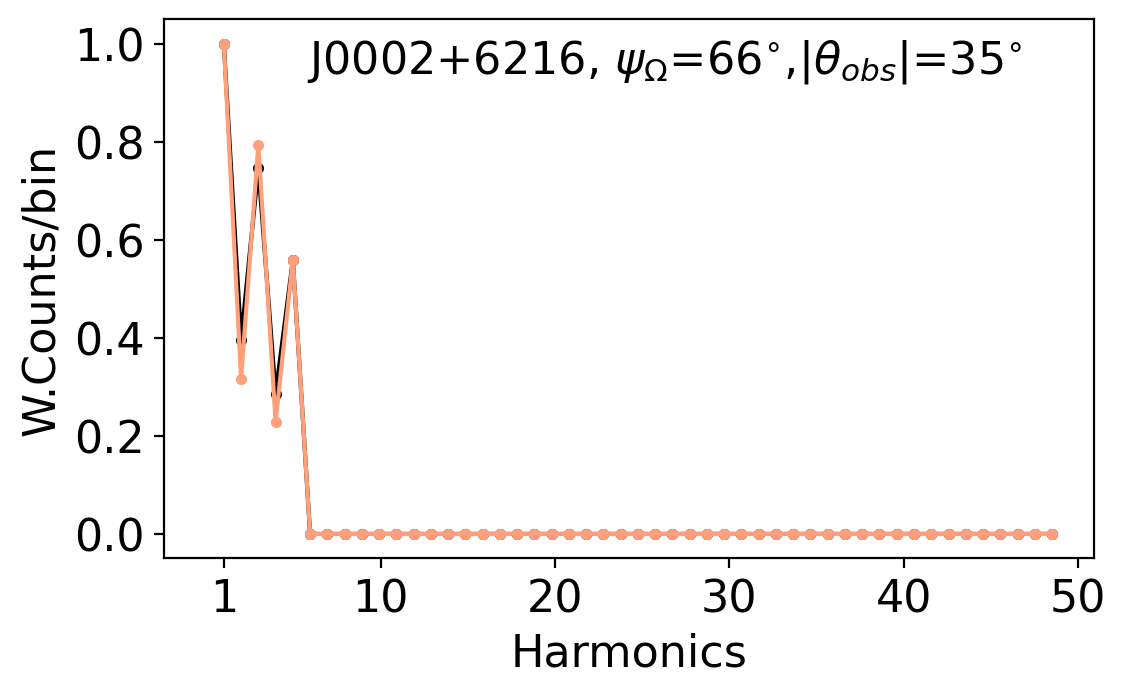}%
        \includegraphics[width=0.25\textwidth]{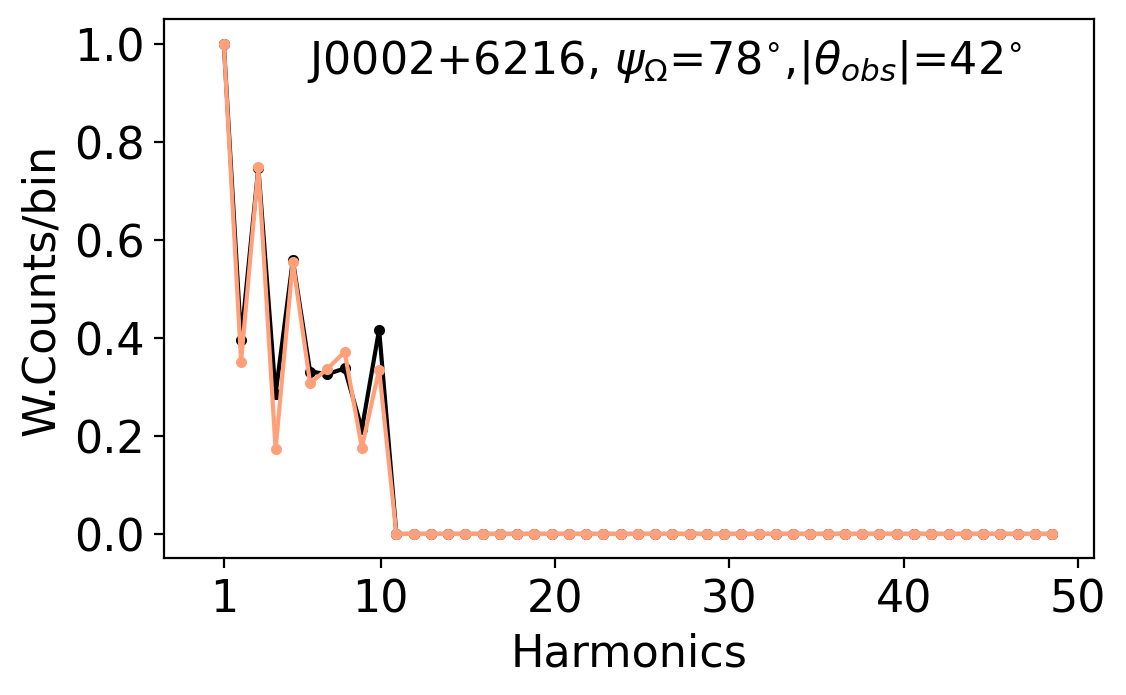}%
        \includegraphics[width=0.25\textwidth]{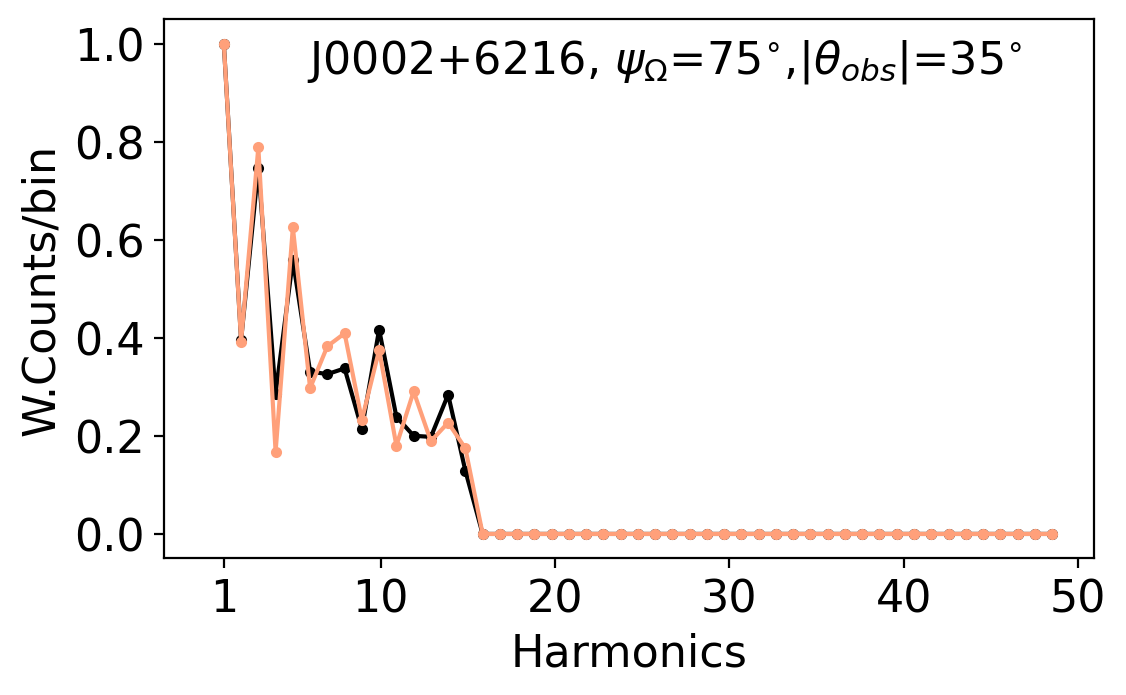}%
        \includegraphics[width=0.25\textwidth]{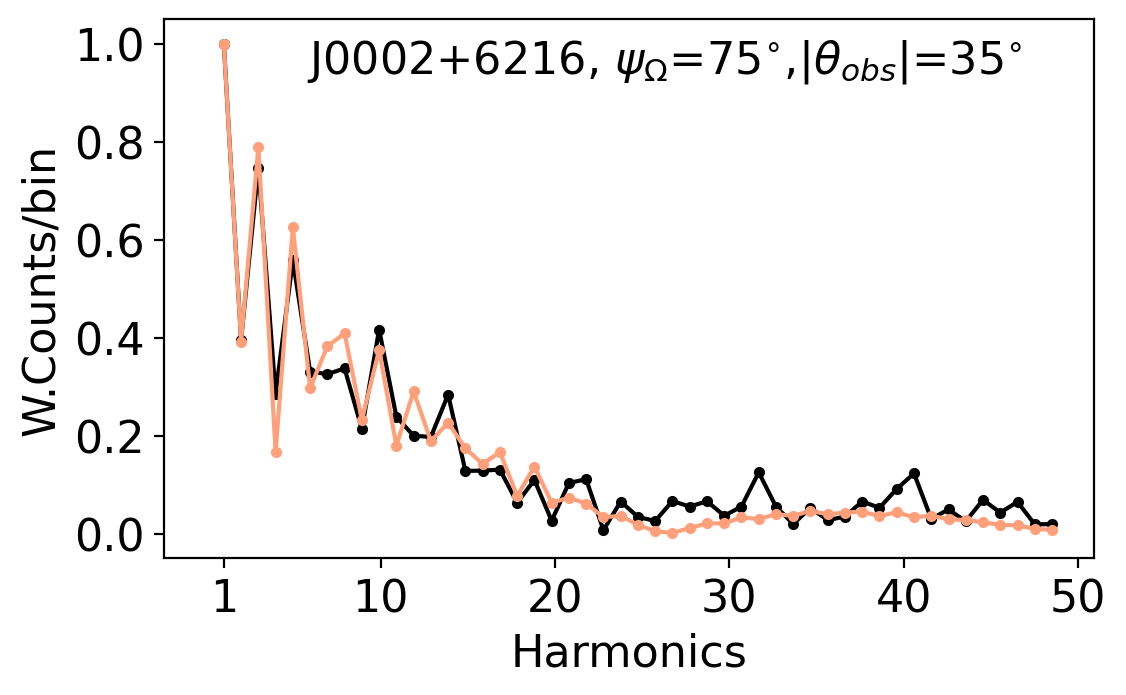}\\
        \includegraphics[width=0.25\textwidth]{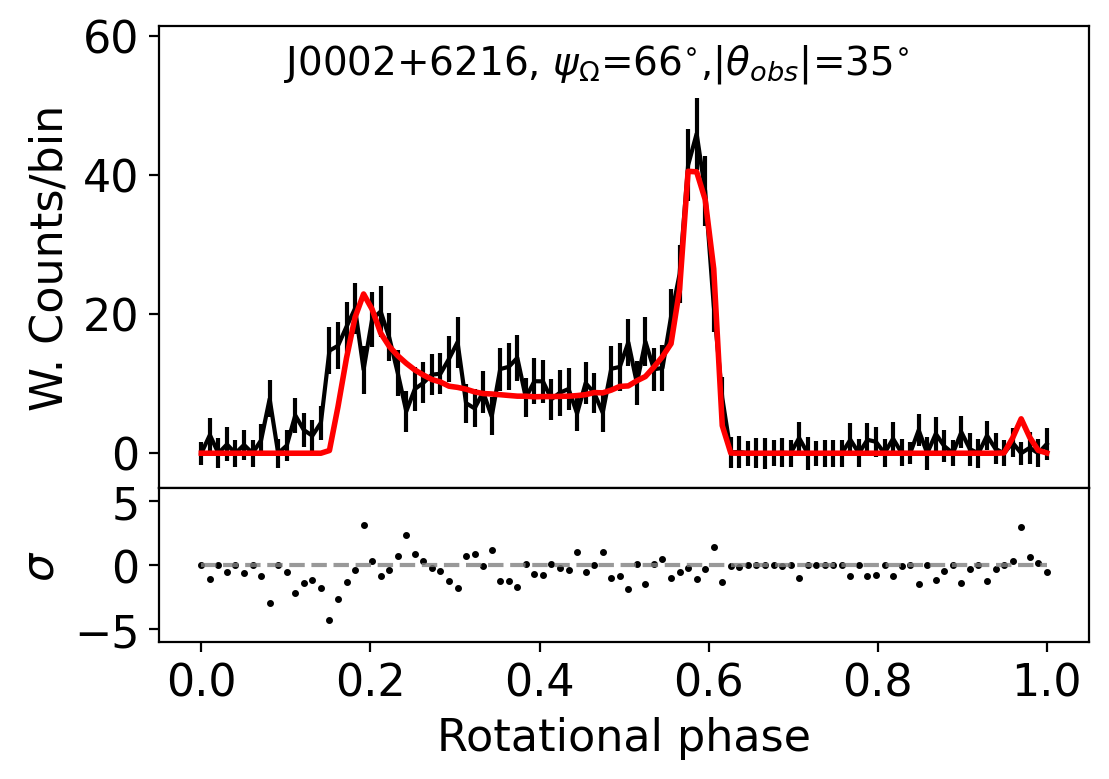}%
        \includegraphics[width=0.25\textwidth]{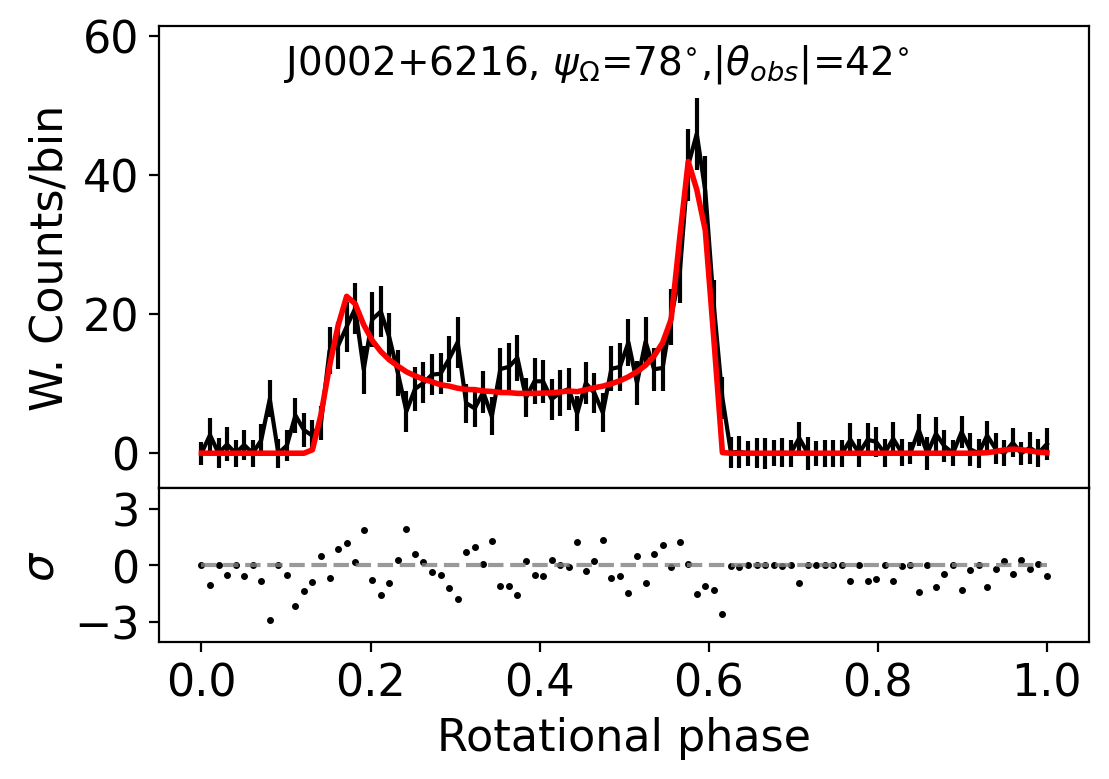}%
        \includegraphics[width=0.25\textwidth]{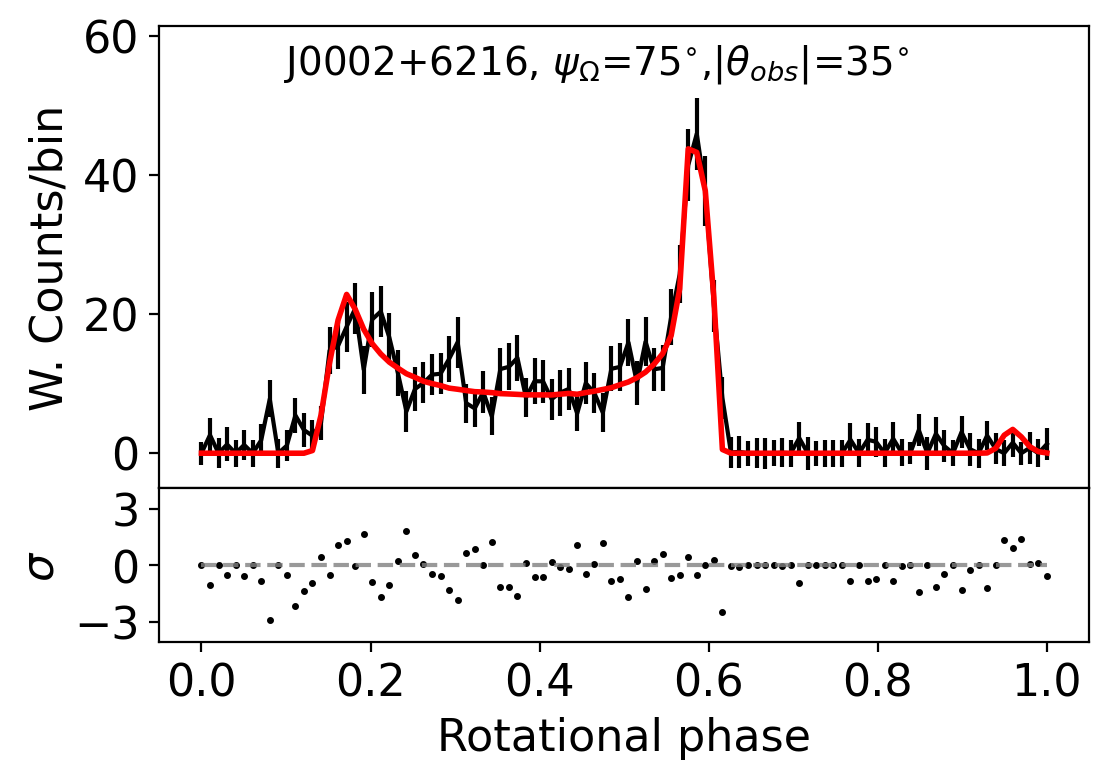}%
        \includegraphics[width=0.25\textwidth]{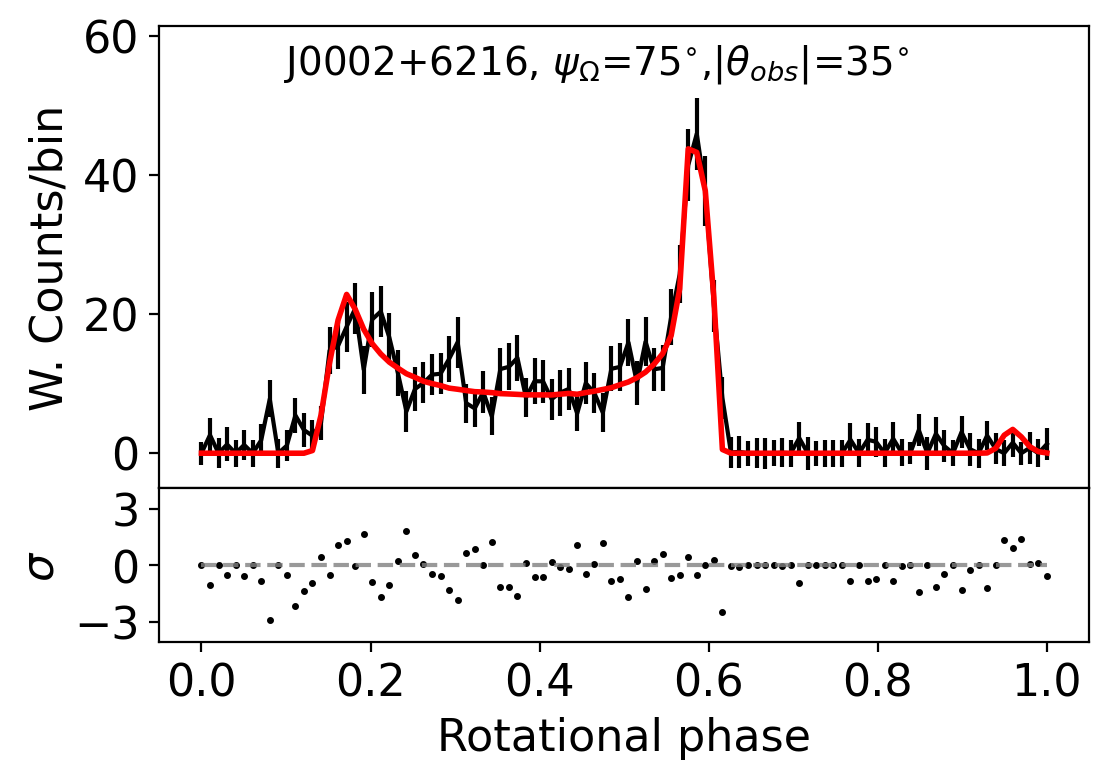}\\
        \includegraphics[width=0.25\textwidth]{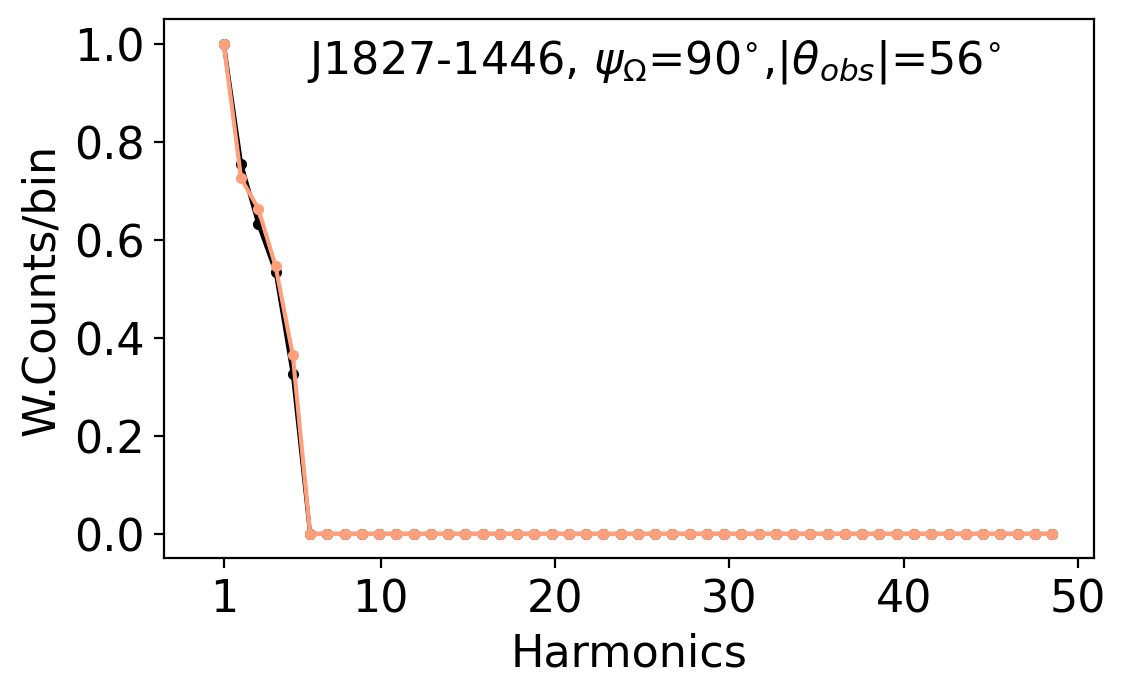}%
        \includegraphics[width=0.25\textwidth]{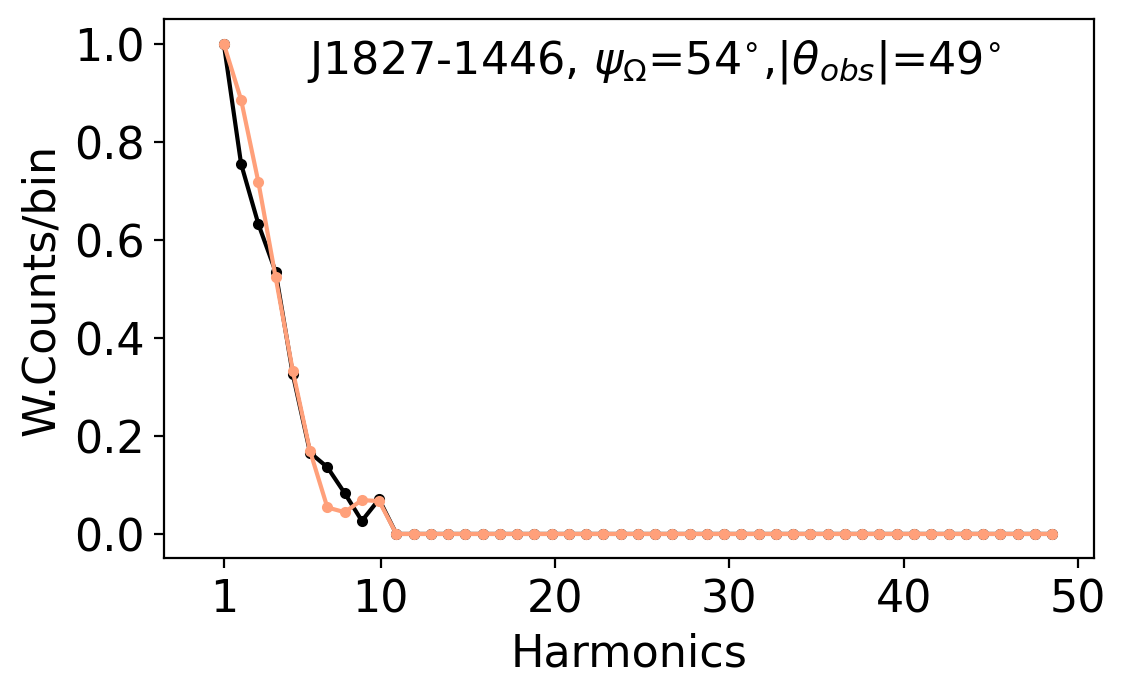}%
        \includegraphics[width=0.25\textwidth]{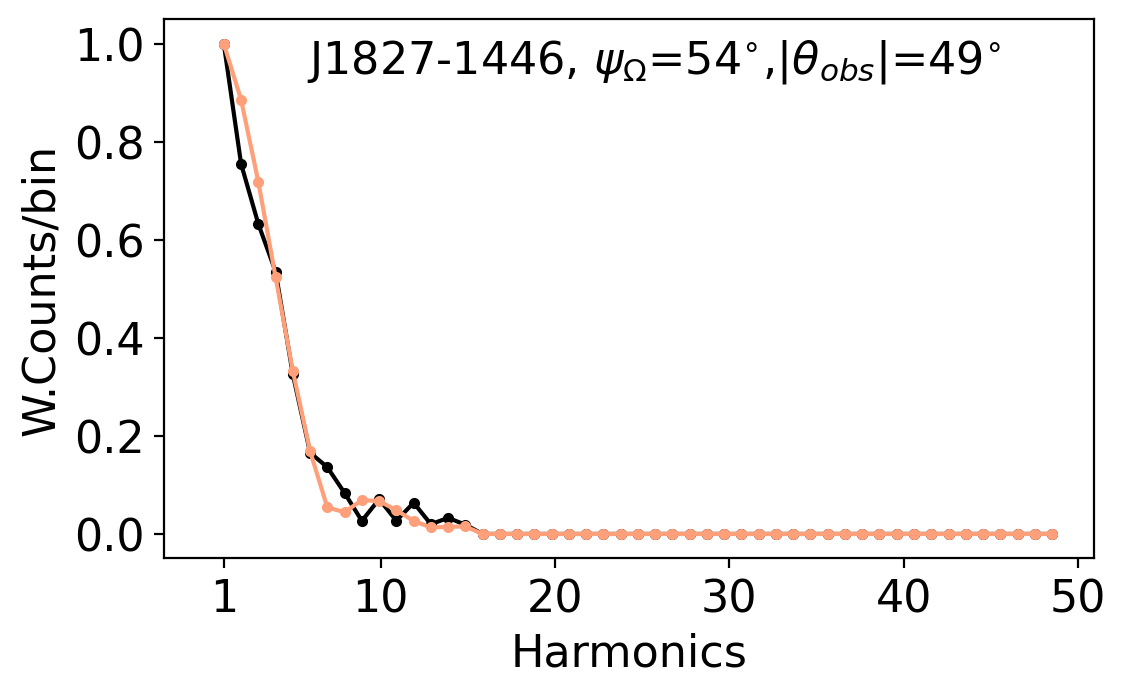}%
        \includegraphics[width=0.25\textwidth]{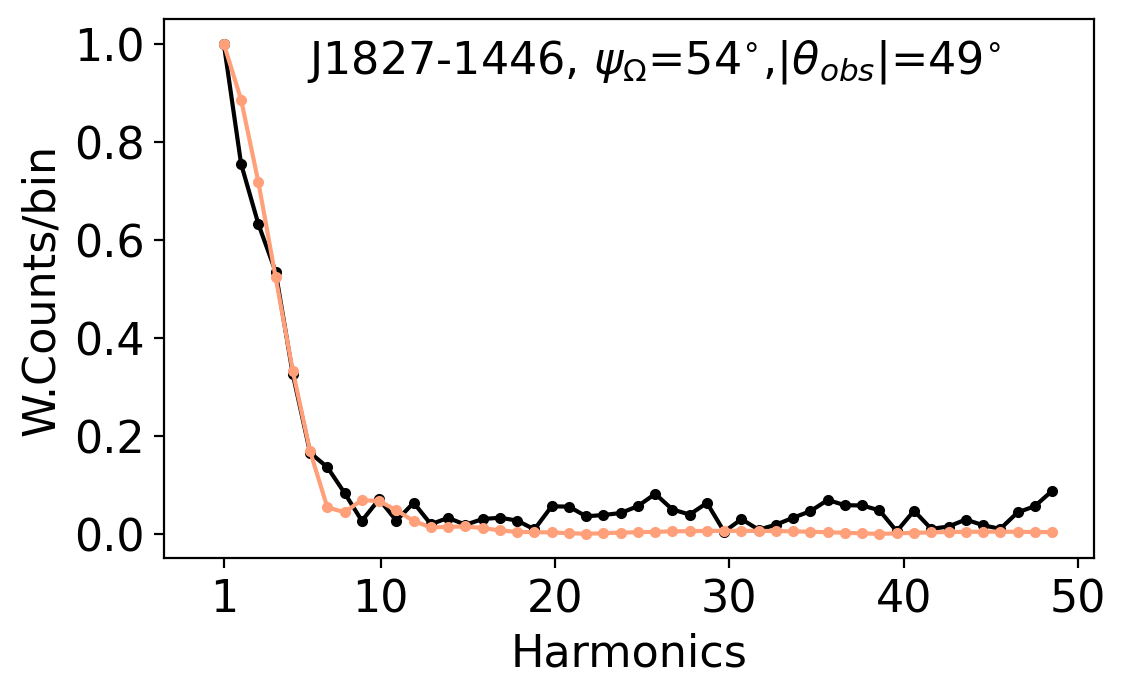}\\
        \includegraphics[width=0.25\textwidth]{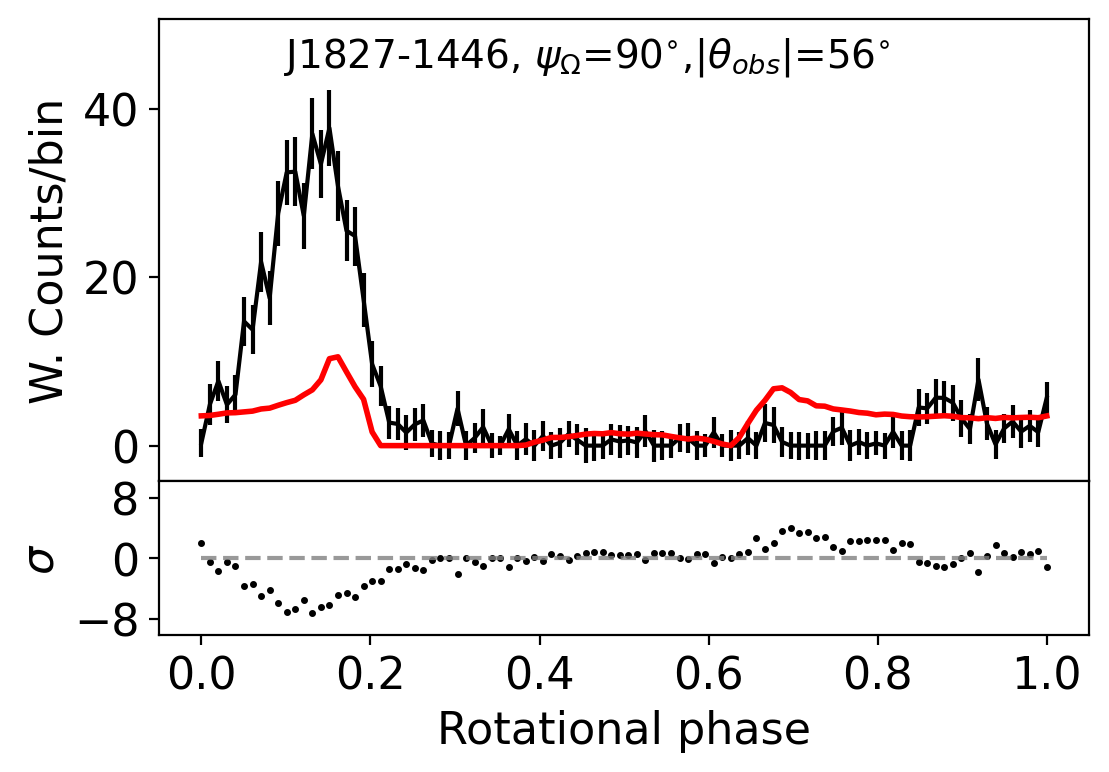}%
        \includegraphics[width=0.25\textwidth]{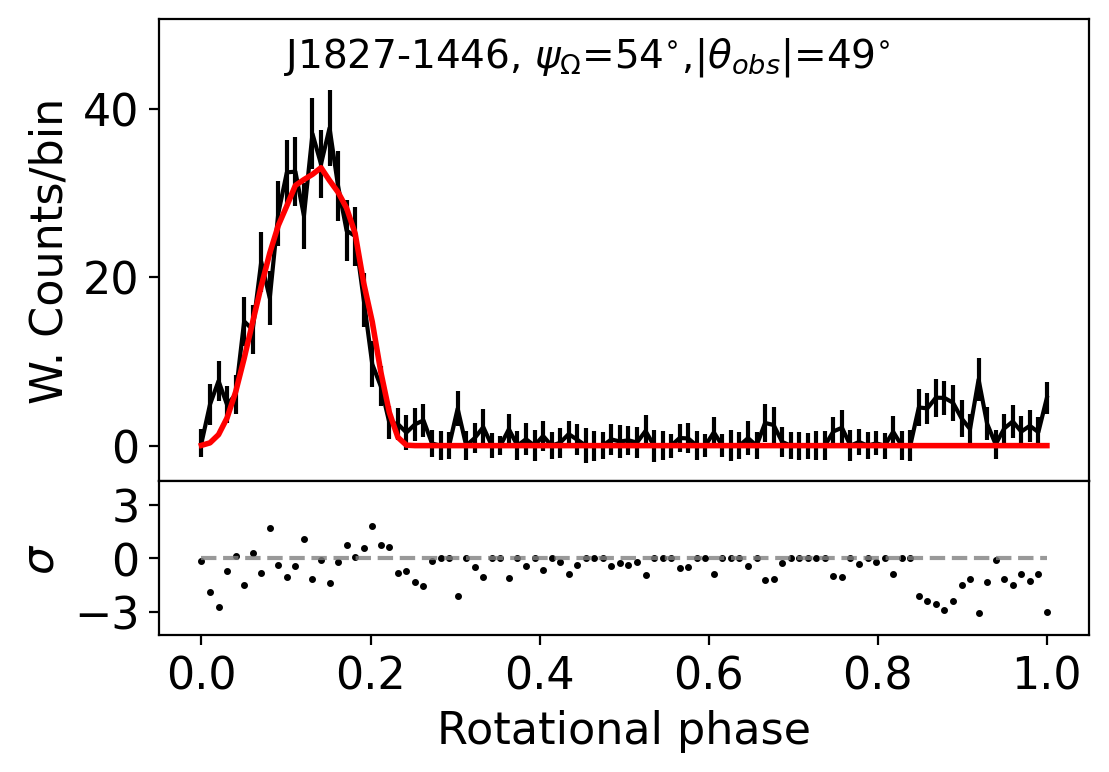}%
        \includegraphics[width=0.25\textwidth]{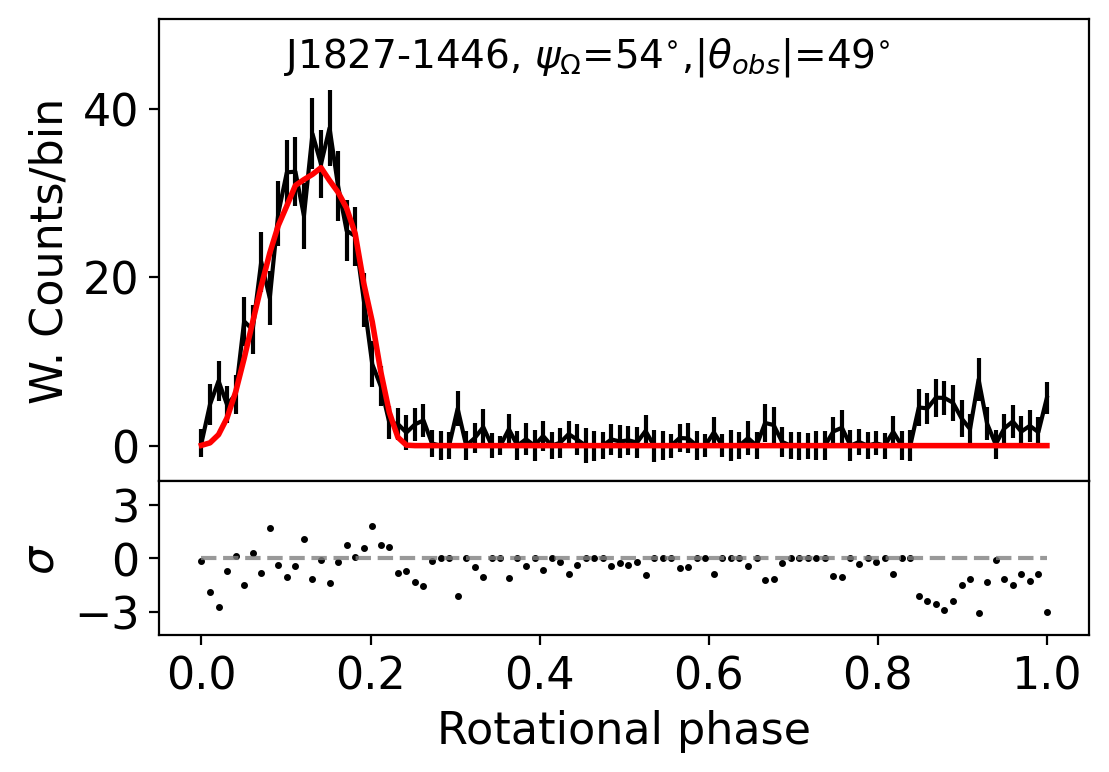}%
        \includegraphics[width=0.25\textwidth]{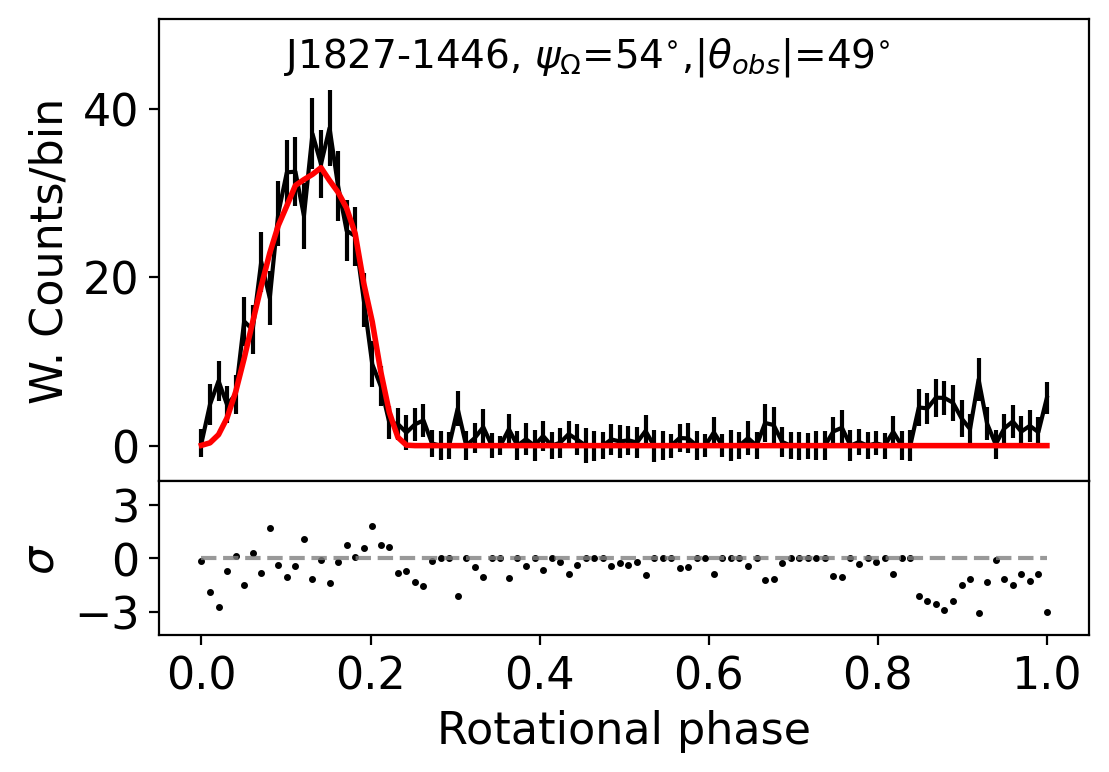}\\
        \caption{Examples of best-fitting light curves in frequency domain for different values of $k_{max}$ for two pulsars, J0002 +6216 on first (Fourier transforms) and second (recovered light curve in time domain) rows and J1827-1446 on third (Fourier transforms) and fourth (recovered light curve in time domain) rows. In all rows, from left to right: $k_{max} = 5$, $k_{max} = 10$, $k_{max} = 15$ and case without filtering.}
    \label{fig:fft_fitting_kmax}
    \end{figure*}

    In parallel, we apply a Fast Fourier Transform (FFT) to all light curves existing in time domain to bring them into frequency domain, $\hat{I}^{obs}_{k} $ and $\hat{I}^{syn}_{k}$, whose amplitude is
    \begin{equation}
        || \hat{I}_k || = K \Bigg| \Bigg| \sum_{j=0}^{n-1} I_{i} e^{-i 2\pi \, k \, j/n} \Bigg| \Bigg| \hspace{0.3cm} ,
        \label{eq:fft}
    \end{equation}
    where $K$ is a normalization factor.
    We normalize the amplitude of the observational Fourier transform to its maximum.
    Each synthetic transform is normalized to match the peak value of the observational one, which is usually at k = 1 or 2, with few cases being at 3 or 4. 
    $k$ is the harmonic, which goes from 1 to $n/2$ (we neglect $k = 0$, the average flux of the light curve).
    Harmonics are the sub-signals forming the complete light curve in time domain, having $k + 1$ nodes. 
    The amplitude of the Fourier transform of a light curve with a single peak roughly symmetric in the rise and decay, will be dominated by odd harmonics, while that of a light curve with a double (roughly symmetric in rise and decay) peak, with a 0.5 phase separation, will be dominated by even harmonics. 
    Different phase separations, the presence of sub-structures and asymmetries in the rise and decay features can redistribute the amplitude to the harmonics with the opposite parity. 
    The width of the peak(s), and the relative peak flux between two peaks, will also determine how the amplitude is distributed over the harmonics.
    In order to assess the impact of the higher harmonics in the fit, we have tried to filter out the Fourier transforms by only considering the harmonics lower than a given maximum harmonic $k_{max} = 5, 10 $ or $15$.
    Even with the extreme case of $k_{max} = 5$, in several cases the shape of the best-fitting light curve barely differs from the complete best-fitting transform.
    Larger values of $k_{max}$ leaves untouched a growing number of best-fit solutions, and for $k_{max}$ the differences are negligible.
    Fig. \ref{fig:fft_fitting_kmax} shows a couple of examples.
    This points out the negligible effect of the higher harmonics in the FFT fitting, which is mostly defined by the lowest harmonics.
    For completeness, we keep all the harmonics in the FFT fitting.

    We compare the amplitudes\footnote{Considering the amplitude instead of the power (which is equal to the squared amplitude) is arbitrary. We have performed tests using the power of the Fourier transforms in the fitting and very similar results are obtained.} of the observational and synthetic Fourier transforms, $||\hat{I}^{obs}_{k}||$ and $||\hat{I}^{syn}_{k}||$, respectively,  with a simple euclidean distance (ED) as comparative metric,
    \begin{equation}
        ED = \sqrt{ \sum_{k} \left(||\hat{I}^{obs}_{k}|| - ||\hat{I}^{syn}_{k}||\right)^2} \hspace{0.3cm} .
        \label{eq:euclidean_distance}
    \end{equation}

    Comparing light curves in frequency domain focuses more on the symmetries and overall structures of the light curves. 
    Moreover, it has computational benefits, since it's phase invariant. 
    The cost of the Fourier transform is less than the cost of ranging all the possible phase alignments and inversion of the curve in the time domain. 
    See Appendix \ref{app:computational_cost} for further computational costs considerations.

    \subsubsection{Fitting using dynamic time warping}

    \cite{garcia25_dtw} has recently introduced the Dynamic Time Warping (DTW) as a method to assess the similarity between light curves, which we could in principle use for our light curve fitting. 
    DTW compares the sequence of data points between two time series without requiring alignment in phase \cite{Berndt1994UsingDTW}. 
    As a result, it is flexible to local stretching or compression in time, allowing it to recognize structural resemblance between light curves regardless of the values of the phases associated with each point.
    Therefore, a good fit can be assigned to e.g. light curves having similar overall structures of peaks and sub-peaks, but very different peak widths, or phase separations between peaks.
    Furthermore, its computational requirements, lasting five orders of magnitude more of time than the fitting in frequency domain and three times more than the fitting in time domain (see Appendix \ref{app:computational_cost}), makes its usage for light curve fitting still not agile enough.
    Therefore, DTW application for light curve fitting, although promising, with the current implementation is not practically feasible.

    \section{Results}\label{results_lightcurve_fitting}
    
    Figure \ref{fig:fitting_results_selected_ones_1} shows examples of best-fitting light curves resulting 
    from both temporal and frequency techniques, for some selected pulsars. 
    All fits for those pulsars with at least 50 observational phase bins in their light curves are shown in the Supplementary Material Online of this paper.
    Table \ref{tab:parameters_pulsars_lightcurves_1} and Fig. \ref{fig:psiomega_and_thetaobs_of_time_vs_frequency_fitting} show the best-fit values of the geometrical parameters for all the pulsars fitted.
    In general, we note that results of the fits obtained with $\chi^2$ in time domain and with ED in frequency domain are qualitatively similar. 
    
    \begin{figure*}
          \centering
           \includegraphics[width=1.0\textwidth]{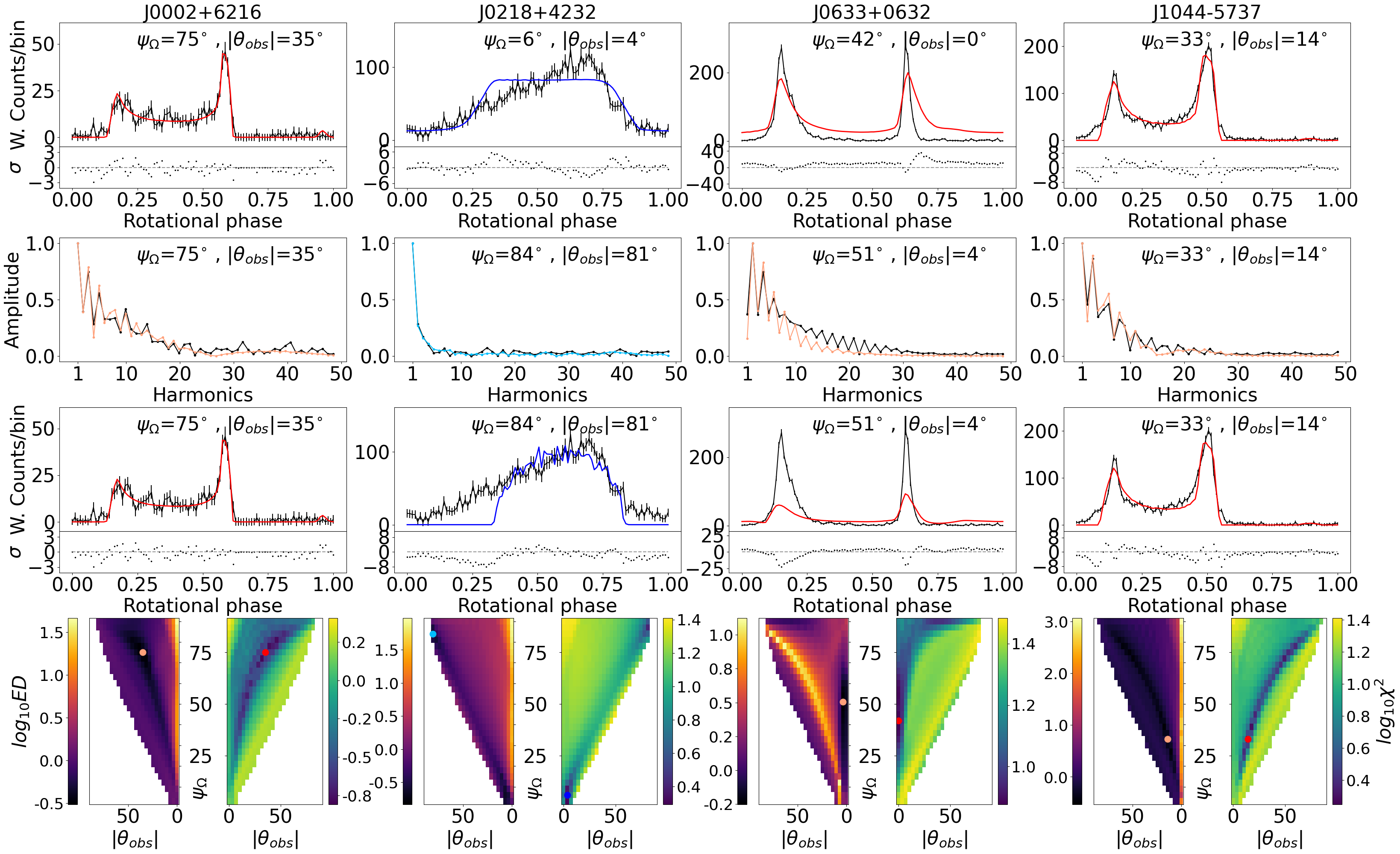}\\\medskip
           \centering
           \includegraphics[width=1.0\textwidth]{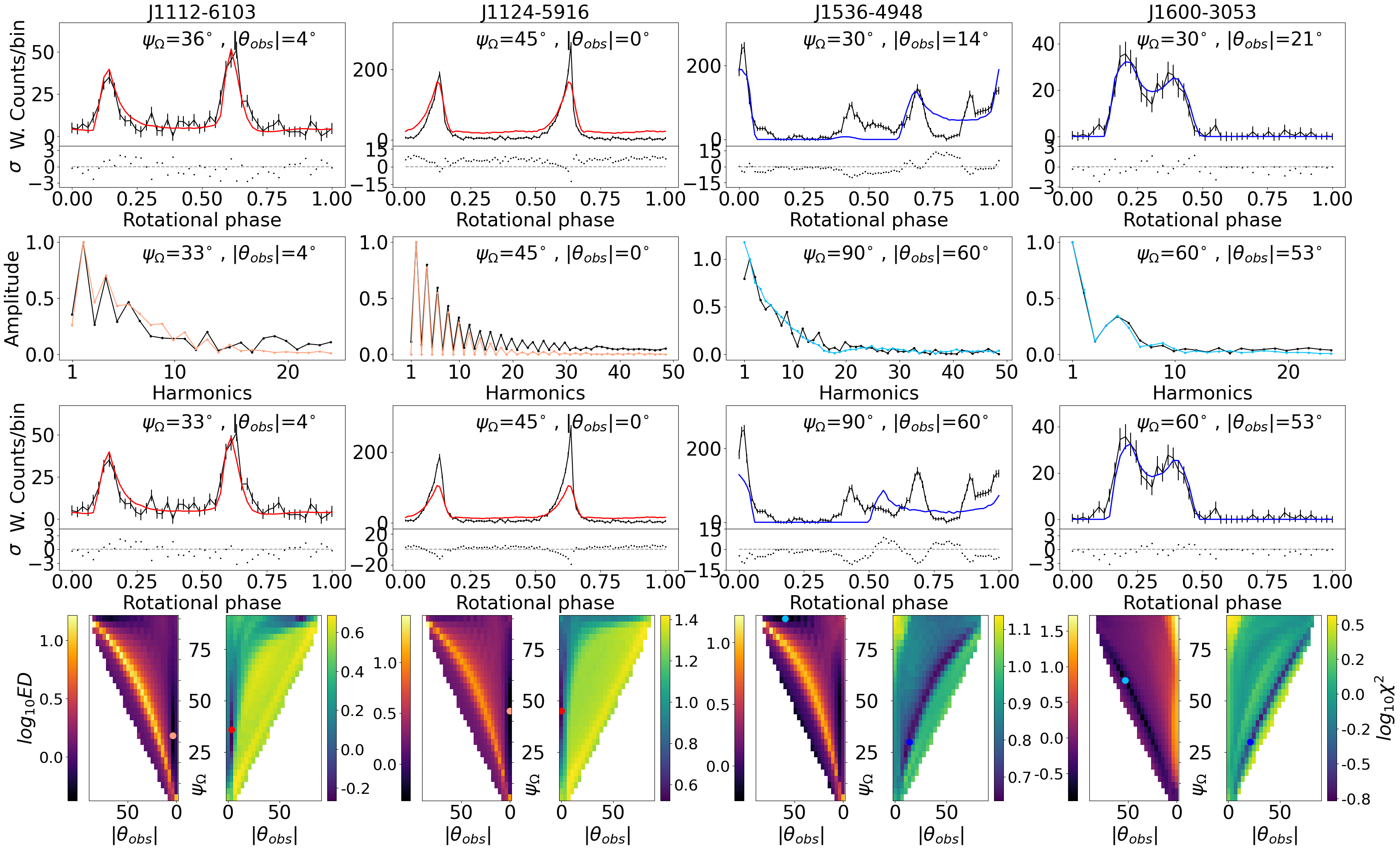}\\
           \caption{Best-fit light curves obtained in time and frequency domain for two sets (four rows each) of four pulsars (one per column).
           The first and fifth rows correspond to fits in the time  domain.
           Second and sixth rows show the best-fit Fourier transform in the frequency domain fits, and third and seventh the light curve recovered with the inverse transform (with the best alignment in phase and spin direction (see Section \ref{fitting_procedure_time}).
           The fourth and eighth rows show the contour plots  of both $\chi^2$ in time domain and ED in frequency domain. 
           Dots correspond to best-fit geometries from the fitting in time domain (dark colors) and from the fitting in frequency domain (light colors). 
           Red (blue) lines correspond to synthetic light curves of young (millisecond) pulsars, and black lines to observational lightcurves.}
           \label{fig:fitting_results_selected_ones_1}
    \end{figure*}

    \addtocounter{figure}{-1}  
    
    \begin{figure*}
          \centering
           \includegraphics[width=1.0\textwidth]{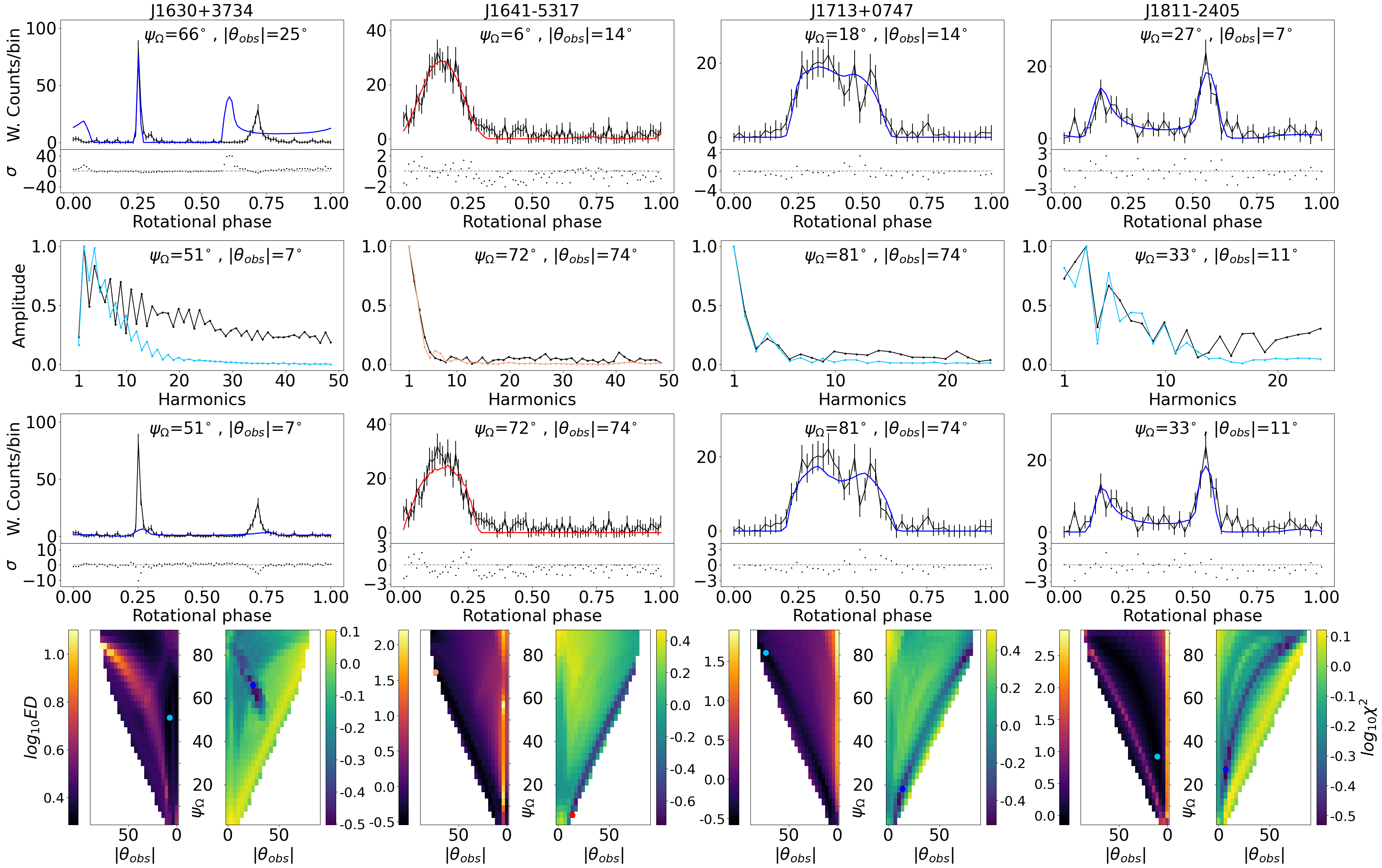}\\\medskip
           \centering
           \includegraphics[width=1.0\textwidth]{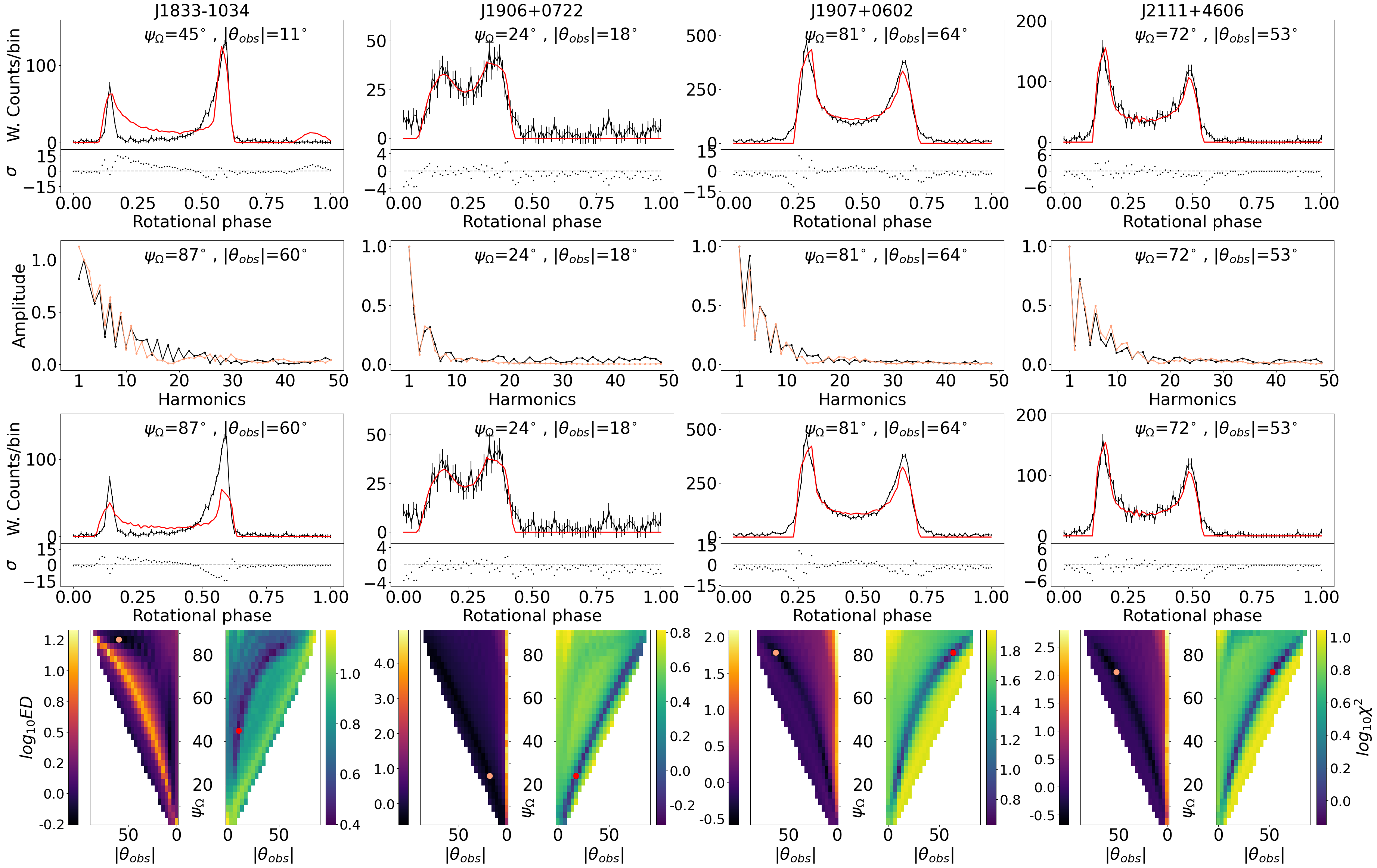}\\
           \caption{- \emph{continued}}
           \label{fig:fitting_results_selected_ones_2}
    \end{figure*}
    
    In many cases the best-fitting synthetic light curves visually resembles well the observational one, with some remarkable fits, such as those of J1044-5737, J1112-6103, J1600-3053, J1641-5317, J1811-2405, J1906+0722, J1907+0602 or J2111+4606. 
    Note the diversity of light curve shapes that we can reproduce, for both young and millisecond pulsars.
    In most of the cases we are able to catch the overall structure of the observational light curves, reproducing number of peaks and their widths, separation of peaks, and even peak flux ratio in some cases.
    In several cases we also reproduce the so-called bridge emission, which, in our model, comes from the outermost parts of the trajectories, which are more curled.
    
    However, our model is not able to capture the small scale features of the light curves of some pulsars, as J0002+6216 or J1713+0747.
    Another aspect in which the model struggles to reproduce is when very asymmetric peaks are present, as in J0218+4232.
    In other cases, like J1124-5916 and other with high-quality data (i.e. small errors), we are not able to reproduce specific flux ratios and/or cusped peaks.
    With only two free geometrical parameters at play, our emission region does not have the complexity that the real magnetosphere surely possess, meaning that we cannot generate light curves as complex as those observed for some pulsars.
    In a few cases,  such as J1536-4948 or J1630+3734, the model is not providing a good fit.

    Looking at the performance of both fitting techniques, we note that both of them find good fits in a number of cases and fail in others. 
    In about a fifth of the sample, both methods give compatible best-fit parameters (within one step in the sweeping grid of values of each parameter), while in the rest of the sample these values noticeably differ.
    This difference is highlighting that the degeneracy in the parameters is large, for both methods. In other words, when one compares the general shape of the contour plots, they are generally similar, but each metric can find a specific value for the best-fit within the best-fitting 'valleys' in the space parameters (fourth and eighth rows of Fig. \ref{fig:fitting_results_selected_ones_1}). Therefore, the individual best-fit values should not be taken as relevant constraints.

\begin{figure*}
        \centering
        \includegraphics[width=0.22\textwidth]{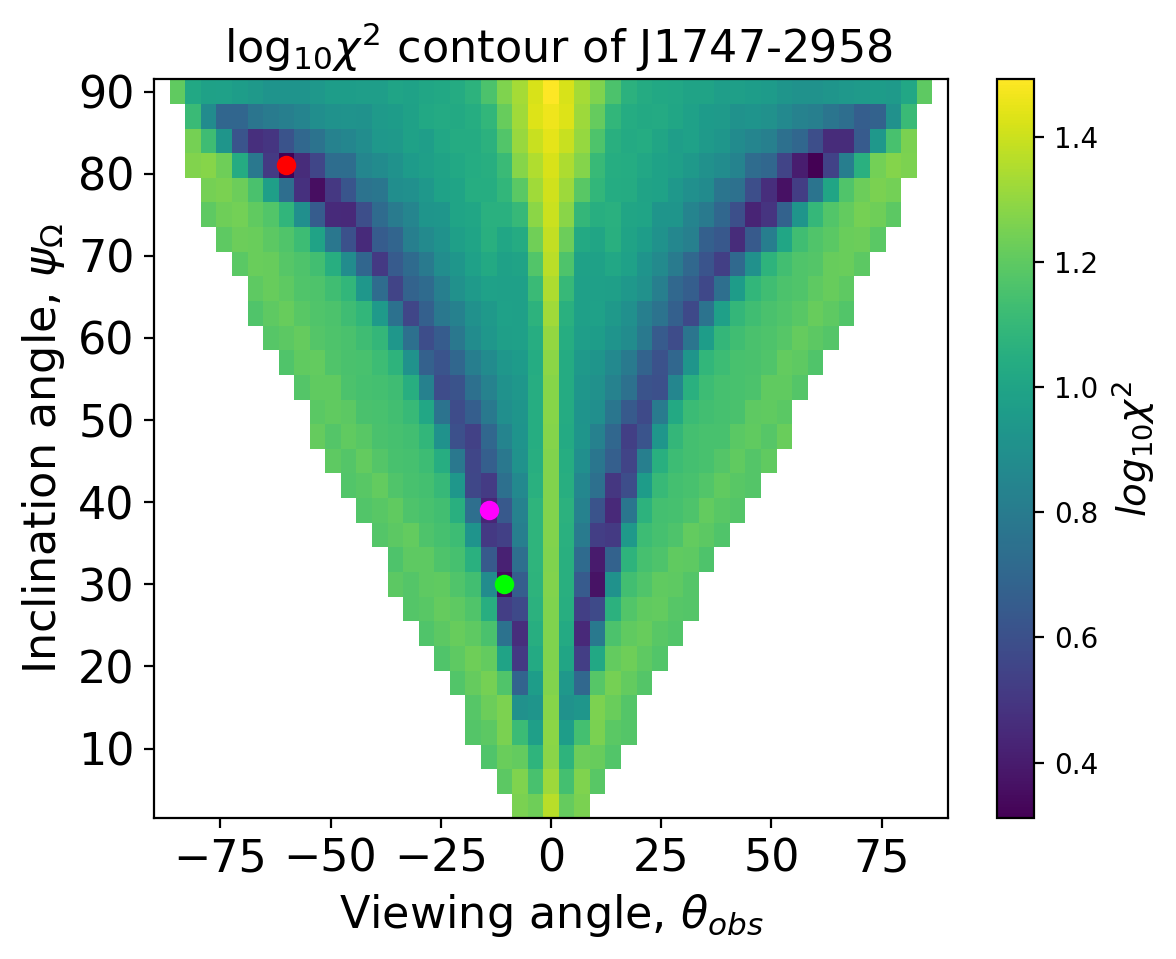}%
        \includegraphics[width=0.25\textwidth]{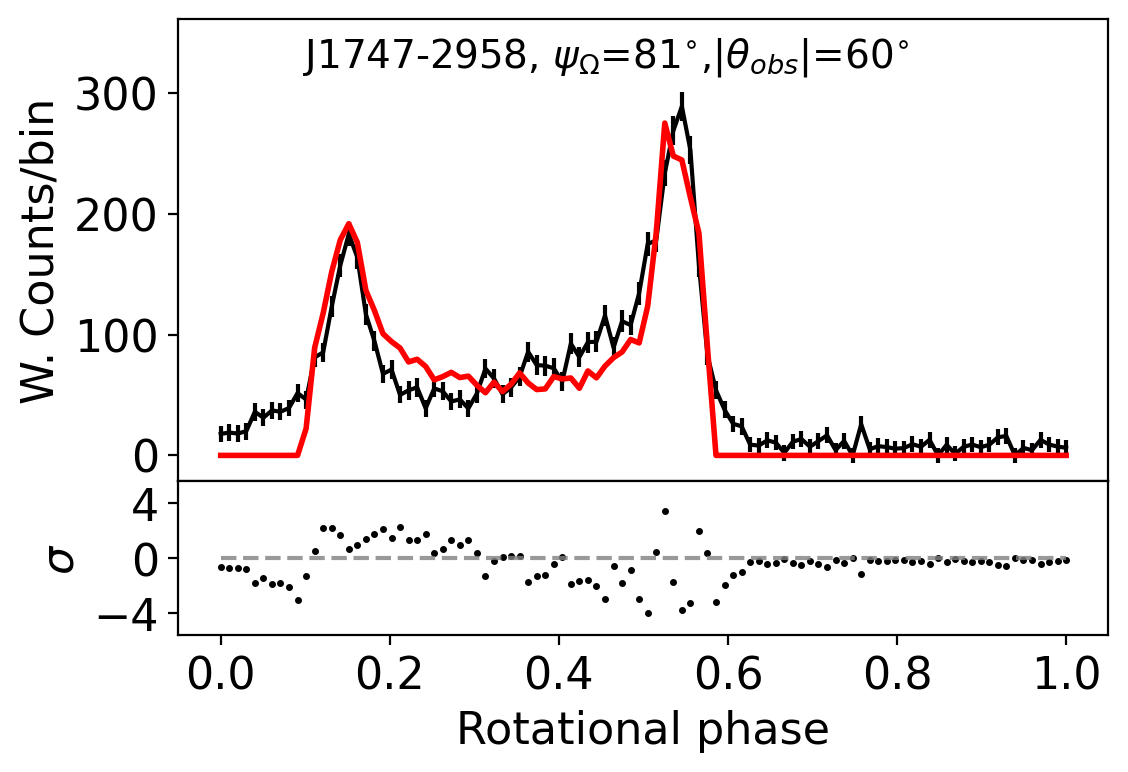}%
        \includegraphics[width=0.25\textwidth]{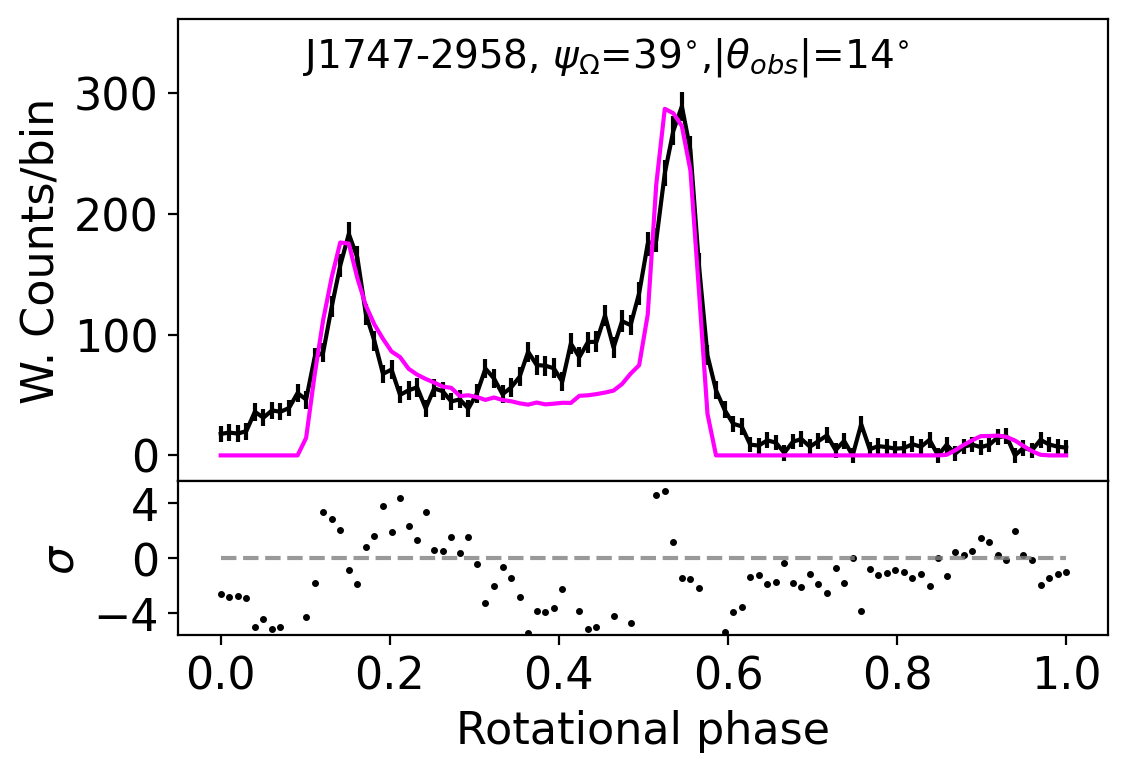}%
        \includegraphics[width=0.25\textwidth]{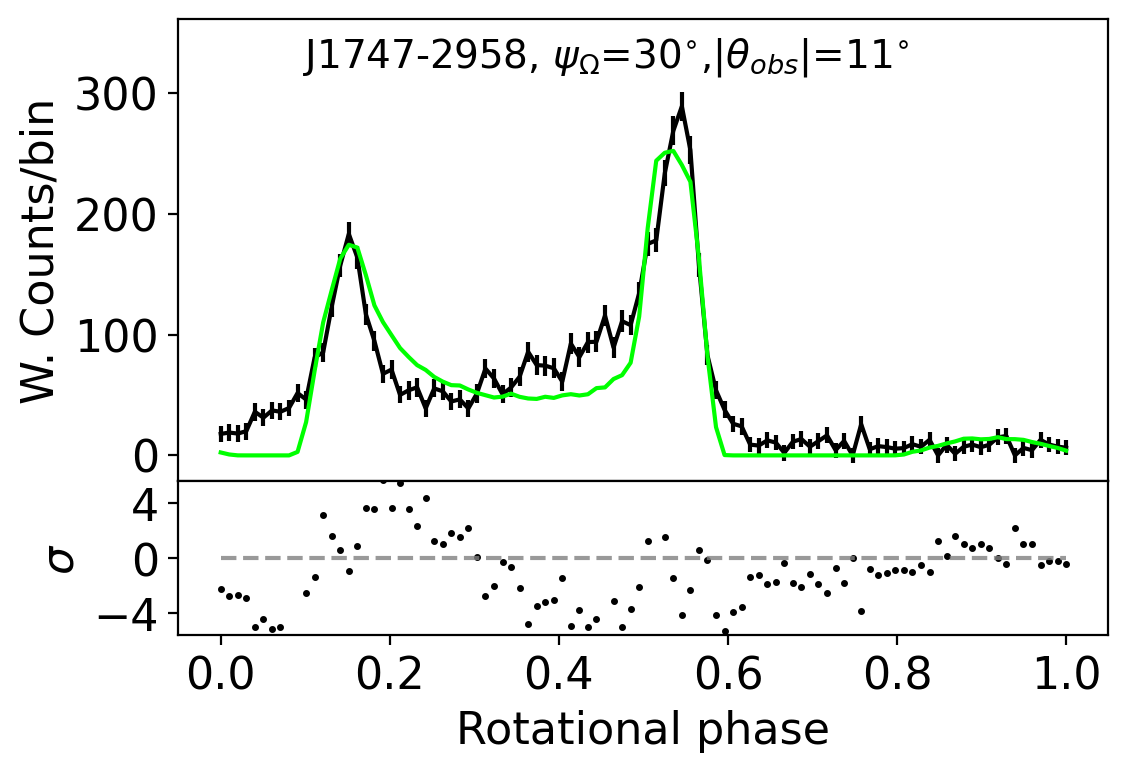}\\
        \caption{Examples of the degeneracy in the contour plots. Left panel shows the $\chi^2$ contour plot of J1747-2958 
        with dots corresponding to the light curves depicted on the other three panels. 
        From left to right: $\chi^2$ best-fitting light curve in red, and other light curves with geometries on the $\chi^2$ valley, in fuchsia and green.
    }
    \label{fig:degeneracy}
    \end{figure*}

    Figure \ref{fig:fitting_results_selected_ones_1} also presents the contour plots of $\chi^2$ and ED for each pulsar, showing the values of these statistics in the space of parameters $\theta_{obs}$--$\psi_{\Omega}$. 
    Due to the equatorial symmetry in the skymaps, we only show $\theta_{obs}$ from $0^{\circ}$ to $90^{\circ}$.
    In many pulsars, the contours from both techniques are not identical but show different regions of low $\chi^2$ and ED. 
    This underscores the distinct performance and intrinsic focus of both methods.
    The most relevant features of these plots are in fact the valleys present in many cases, both in the $\chi^2$ and the ED ones, pointing out a degeneracy in the geometrical parameters. 
    Along the valley of parameter degeneracy, good fitting geometrical values can be found in almost the whole range of $\theta_{obs}$ or $\psi_{\Omega}$, as can be seen in the $\chi^2$ contour plot of J1747-2958 in Fig. \ref{fig:degeneracy}. 
    In this figure we also show some other light curves whose set of parameters lie in the valley, apart from the best-fitting one under
    the $\chi^2$ methodology.
    We see that the shapes of the not-best-fitting light curves (fuchsia and green dots) are rather similar, even though their geometrical parameters are very different.
    For this reason, the uncertainty around the best-fitting values is large, and the latter should not be regarded as strong constraints.
    Models with more asymmetry in the shape of the emitting region would likely break this degeneracy.
    
    We have investigated how the results we found are affected by the grid resolution of the geometrical parameters by performing fits of several pulsars with a thinner grid, considering 90 values of $\psi_{\Omega}$ and 151 values of $\theta_{obs}$, i.e. increasing the resolution a factor 3.
    The results were not significantly changed. The best fit light curves chosen were the same or with best-fit parameters close to those of the best-fit light curves with the default resolution, and the contour plots were basically the same.
    In particular, neither new substructures appeared in the contour, nor it helps breaking the degeneracy.


    \subsection{Fitting all light curves with a single set of maps}\label{fitting_with_one_map}

\begin{figure}
           \centering
           \includegraphics[width=0.4\textwidth]{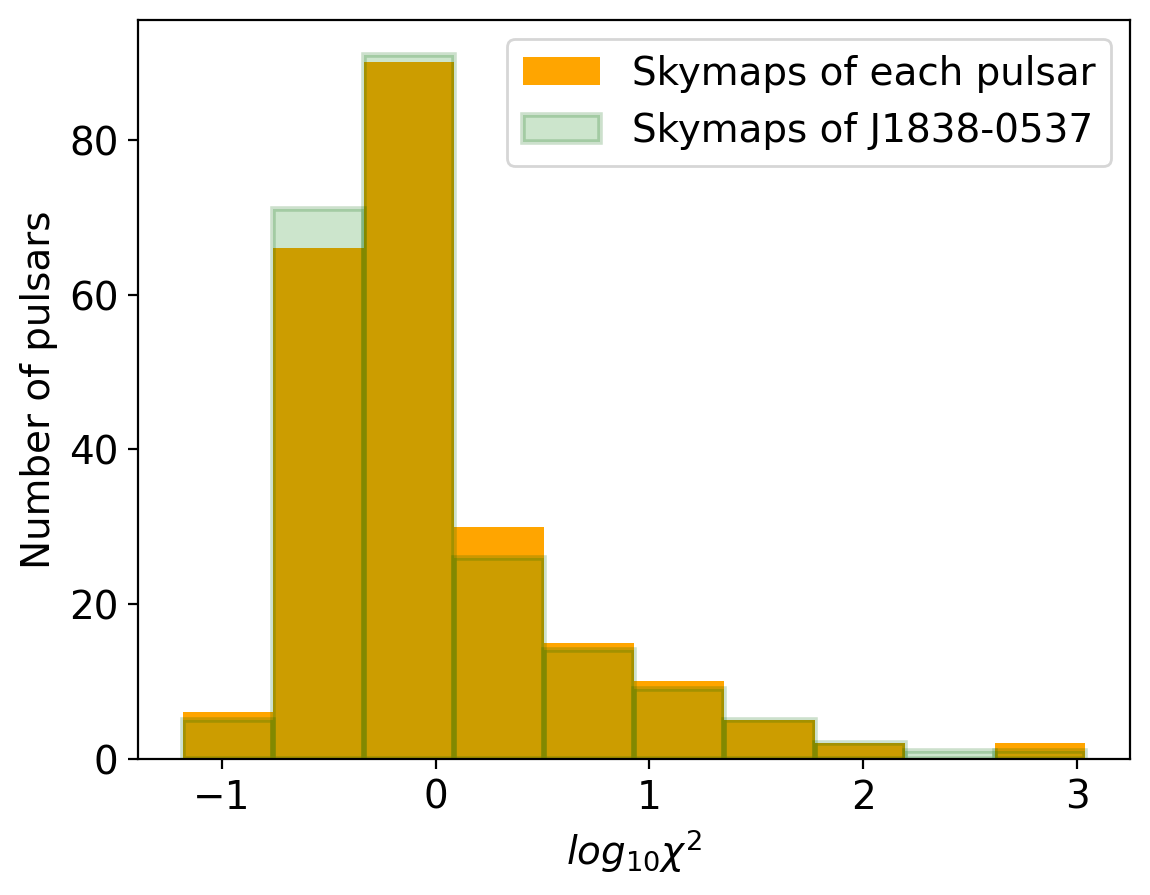}\\
           \centering
           \includegraphics[width=0.4\textwidth]{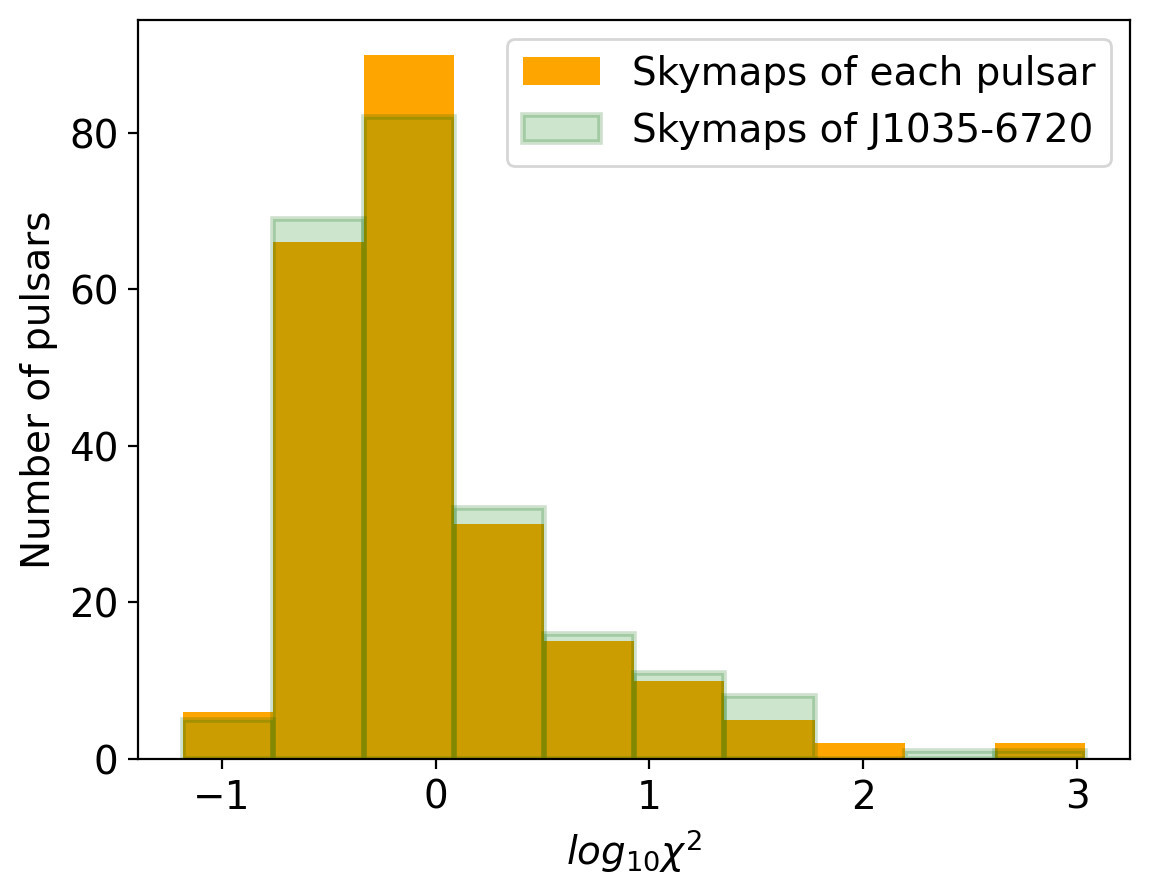}
           \caption{Comparison of the distributions of minimum $\chi^2$ values from the fitting where all observational light curves are compared with each pulsars' skymaps (orange histograms) and with only one set of skymaps from a given, arbitrary pulsar (green histograms). The green histogram in the top (bottom) panel corresponds to the fitting done with the skymaps of J1838-0537 (J1035-6720).}
           \label{fig:chi2_distribution_normal_one_fittings}
    \end{figure}

    In \cite{iniguezpascual24} we stated the idea that, in our model, the spectral and timing properties barely affect the set of light curves, which are controlled basically by the two geometrical parameters. In other words, the emission map (and the set of associated light curve) are insensitive to the spectra that fits a specific pulsar.
    This implies that a set of synthetic light curves produced for a fixed set of best-fit spectral parameters but varying the geometry (inclination, observer line of sight) is also representative of the global set of all observed pulsars, for which the values of such angles are also randomly distributed. 
    We can explore this further asking whether it is possible to fit the  observational light curves of all $\gamma$-ray pulsars with just a set of synthetic emission maps in which timing and spectral parameters are fixed to those of a given, arbitrary pulsar.
    To answer this question, we have fitted the whole sample of light curves generated for a randomly-selected young pulsar, J1838-0537, to the whole set of observational light curves in the same way as explained before.
    We find that the fitting results are qualitatively very similar to those of the fitting in which each observational light curve is compared to the synthetic light curves generated for its corresponding pulsar. 
    The same conclusions we have presented in this section in terms of quality of the fits are also found now, independently of which fitting method is used.
    Considering $\chi^2$ in time domain, the distribution of the minimum $\chi^2$ values found in the normal fitting and in the one using a single set of skymaps are almost identical, as shown in Figure \ref{fig:chi2_distribution_normal_one_fittings}. 
    This exercise has been repeated using the maps from a millisecond pulsar, J1035-6720, and the exact same results are obtained.
    These results, again, back the statement that geometry is the defining aspect for the light curves shapes; and that the global structure of the magnetosphere as well as the emission processes and location of the accelerating region must be roughly the same, regardless of different 
    timing properties, spectra.
    We stress that this might not be the case if also X-ray light curves were to be taken into account, since a strong link between X-$\gamma$ spectral differences and X-$\gamma$ light curve differences is expected, since the synchrotron radiation is spread to a broader angles.
    Our result is in agreement with the recent observational proof by \cite{garcia25_dtw}. These authors quantified  3PC light curve similarity and showed it can be very high for very different pulsars; even some MSPs and young pulsars share detailed light curve morphology.

    \subsection{Searching for correlations between geometrical parameters}\label{correlations_geometrical_parameters}
    
    The geometrical parameters show no correlation with any other parameter of the pulsar or the model, and all pairs have a Pearson coefficient $ r $ smaller than our threshold of 0.85, i.e. there is no significant correlation of the geometry with spectral parameters, timing quantities, or PLEC4 parameters. 
    The former reinforces the idea that there is a decoupling between spectral and geometrical parameters, backing a conclusion obtained in \cite{iniguezpascual24}: spectral and timing considerations are negligible in shaping gamma-ray light curves, being geometrical ones much more relevant.
    This conclusion may, of course, be affected by adding further complexities in our model.

    \section{Conclusions}
    \label{conclusions}
    
    In this paper we have presented the first concurrent fitting of spectra and light curves for the whole population of gamma-ray pulsars. 
    Using an effective radiative model based on synchro-curvature radiation and the computation of the dynamics and emission of charged particles in a pulsar's magnetosphere, we generate synthetic spectra. 
    Intertwined with the spectral calculation, we describe the geometry of the trajectory of these particles and define a particular shape for the region where they move, which allows us to produce light curves. 
    Theoretical spectra and light curves are then compared with observational spectral and light curves data to assess the ability of our models to reproduce them and to find the best-fitting values of the free model parameters, three in the spectral model and two in the geometrical model.

    The spectral model fits well the whole population of gamma-ray spectra, obtaining similar results to previous systematic fittings carried out by the authors.
    For the first time we have fitted the individual gamma-ray light curves. 
    We have made use of two different techniques: using the reduced $\chi^2$ (weighted with the normalized flux of the observational light curve) to evaluate the goodness of the fits of light curves in time domain, and computing the euclidean distance to compare the light curves in frequency domain, i.e. to compare their Fourier transforms. 
    Despite the caveats that both methods may have, the two of them give qualitatively similar results. 
    With both fitting methods we have obtained good fits in many cases.
    The best-fitting values of the geometrical parameters found with the two techniques, which in both cases have a considerable degeneracy and thus cannot be regarded as strong constraints, are compatible in a fraction of the sample, but differ in the rest of the population.
    If one instead looks at the parameter space (fourth and eighth rows of Fig. \ref{fig:fitting_results_selected_ones_1}), the regions with relatively low metric values are similar. 
    The two methods provide therefore similar results in terms of selecting the most likely loci in the geometrical parameter space, but with quantitative differences when one wants to fine tune the values well below such degeneracy-driven uncertainties (which are not easy to be assessed in a statistically sound way). 
    Additional observational constraints, like radio or X-ray light curves, could potentially break the degeneracy and practically make the use of the two techniques more powerful.
    
    For most pulsars we capture the general structures of the observational light curves, such as number of peaks and separation between peaks, or even flux ratio between peaks or width of the peaks, but fail to reproduce the small scale features. 
    In a few pulsars, the observational light curves are not found in our set, either because they simply don't exist or because the fitting techniques miss them.
    Regarding the first possibility, the effective nature of the  geometrical model provides versatility at an acceptable computational cost, but it is still lacking the complexity that a real magnetosphere may have. 
    The model does not generate very complex synthetic light curves, for example with very asymmetric peaks or containing significant substructures (e.g. small peaks near larger ones).
    An improvement of the model, making it more realistic but maintaining the effective approach and the computational cost acceptable, is planned for the future.

    We have found the distributions of best-fit viewing angles for both fitting methods to differ from a uniform distribution, as it should be a priori, showing an overabundance of small viewing angles. 
    We argue that the reason is the way we model the light curves and the assumptions we make, as well as the fact that we are considering the fits from all pulsars together, without taking their goodness into account.

    We have also done the exercise of fitting the whole population of gamma-ray light curves with just a single sample of emission maps, generated for a specific a set of $P$, $\dot{P}$ and spectral parameters. 
    Overall results of this fitting are very similar to those obtained in the fitting in which each observational light curve is fitted with its corresponding sample of skymaps.
    This seems to point towards a unique magnetospheric structure of pulsars' magnetosphere and a unique process generating the radiation, with pulsars' detected emission being solely defined by the inclination angle of the pulsar and the viewing angle of the observer. 
    We also have to admit the possibility that the assumptions in our model, in particular the effective weight to enhance the small parts of the trajectories which give synchrotron-dominated emission, might lead to this effect, by construction. 
    Future improvement of the underlying physics of the model, which will allow to convert effective parameters into more physical ones, will shed more light into this.

    At this point is relevant to comment on the only fitting of 3PC gamma-ray light curves done so far, very recently published in \cite{Cerutti24}, in order to contextualize our work in the present panorama of modeling of pulsar high-energy emission.
    The results presented there and in the present work are qualitatively similar but are based on different assumptions and approaches.
    Both can fit well some light curves and capture the global structure of most of the cases, but fail when fitting complex light curves (with asymmetric peaks or small scale features).
    While we use an effective model, austere, and with low computational cost, 
    \cite{Cerutti24} develop magnetospheric PIC simulations in which the electromagnetic fields and charged particles are self-consistently computed and are computationally very costly. 
    The qualitative similarity of the results obtained with such different (and complementary) models provides support for the current general understanding of the generation of the 
    pulsar high-energy emission.

    Finally, we have studied the possible correlations between pulsar parameters in the same way we did in \cite{Vigan_2015b}.
    The relevant correlation between the $E_{||}$ and $x_0$ is significantly found again, but there is no significant correlation between the PLEC4 model parameters from the 3PC and our model parameters.
    No correlation of the geometrical parameters with any other parameter or physical quantity is found.

    \section*{Acknowledgements}
    
    We thank C. R. García for the useful discussions. 
    This work has been done for/in the context of grants PID2021-124581OB-I00; PID2024-155316NB-I00 and 2021SGR00426. This work was also supported by the Spanish program Unidad de Excelencia María de Maeztu CEX2020-001058-M and by MCIU with funding from European Union NextGeneration EU (PRTR-C17.I1). 
    DIP has been supported by the FPI predoctoral fellowship PRE2021-100290 from the Spanish Ministerio de Ciencia, Innovación y Universidades and his work has been carried out within the framework of the doctoral program in Physics of the Universitat Autònoma de Barcelona. 
    DV is funded by the European Research Council (ERC) under the European Union’s Horizon 2020 research and innovation programme (ERC Starting Grant IMAGINE, No. 948582). 
    %

    \section*{Data Availability}
    
    In the text we cite the sources of the data we use in our study,  no new observational data is herein presented. Any additional theoretical detail required is available from the authors upon reasonable request. 
    
    \bibliographystyle{mnras}
    \bibliography{fitting_spectra_lightcurves_gammaray_pulsars}

    \appendix
    \section{Comparing old and 3PC gamma-ray data"}
    \label{app:data_comparison}

\begin{figure*}
        \centering
        \includegraphics[width=0.25\textwidth]{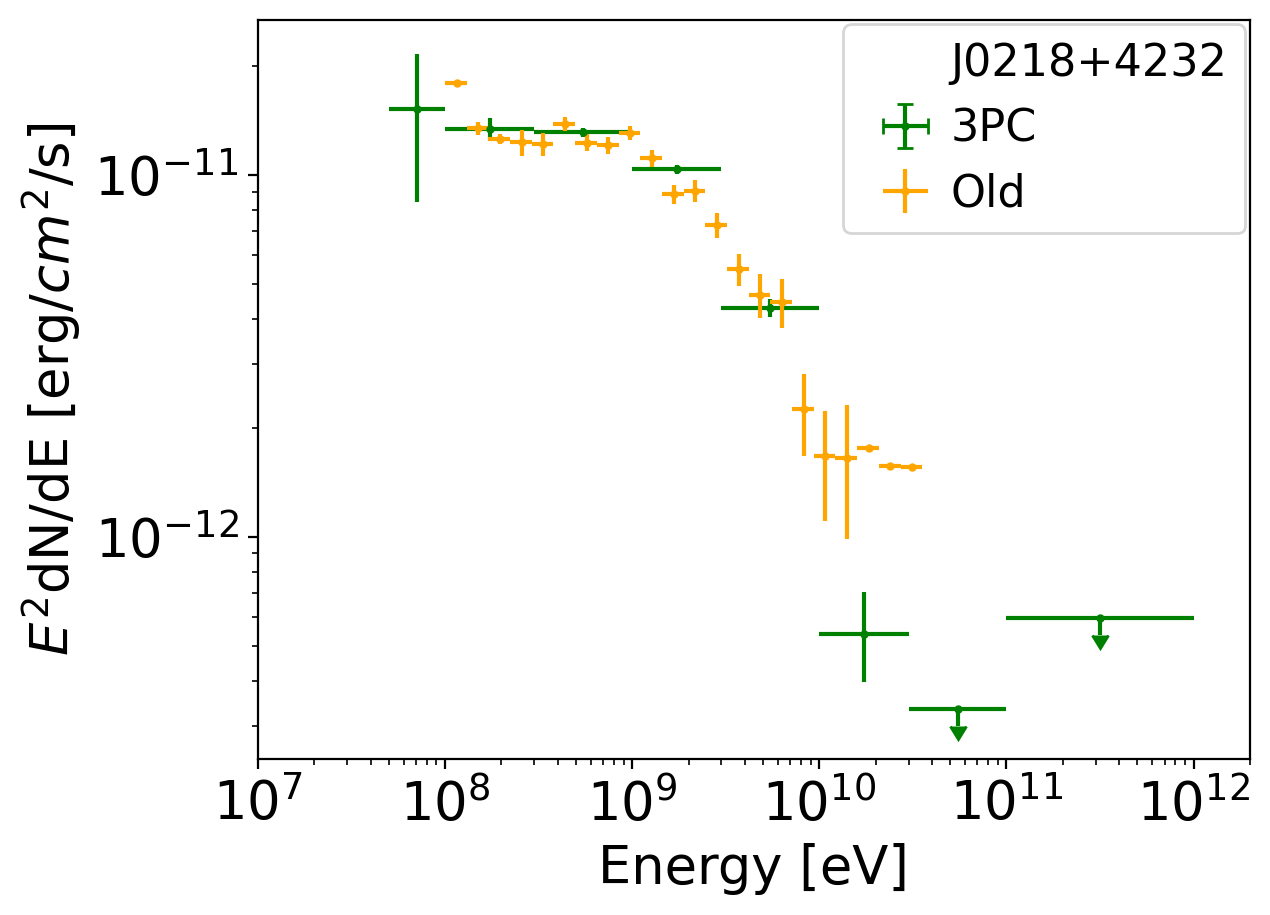}%
        \includegraphics[width=0.25\textwidth]{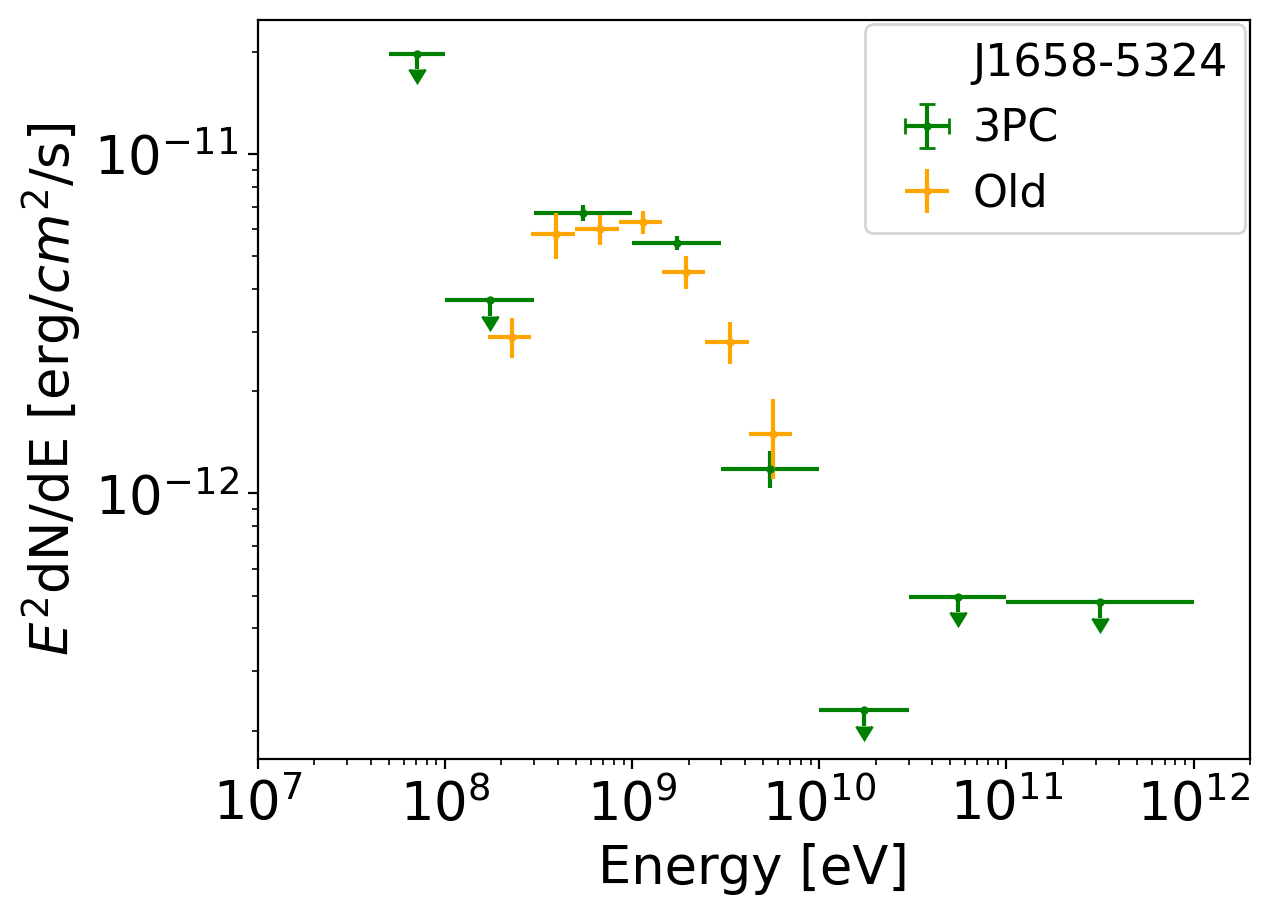}%
        \includegraphics[width=0.25\textwidth]{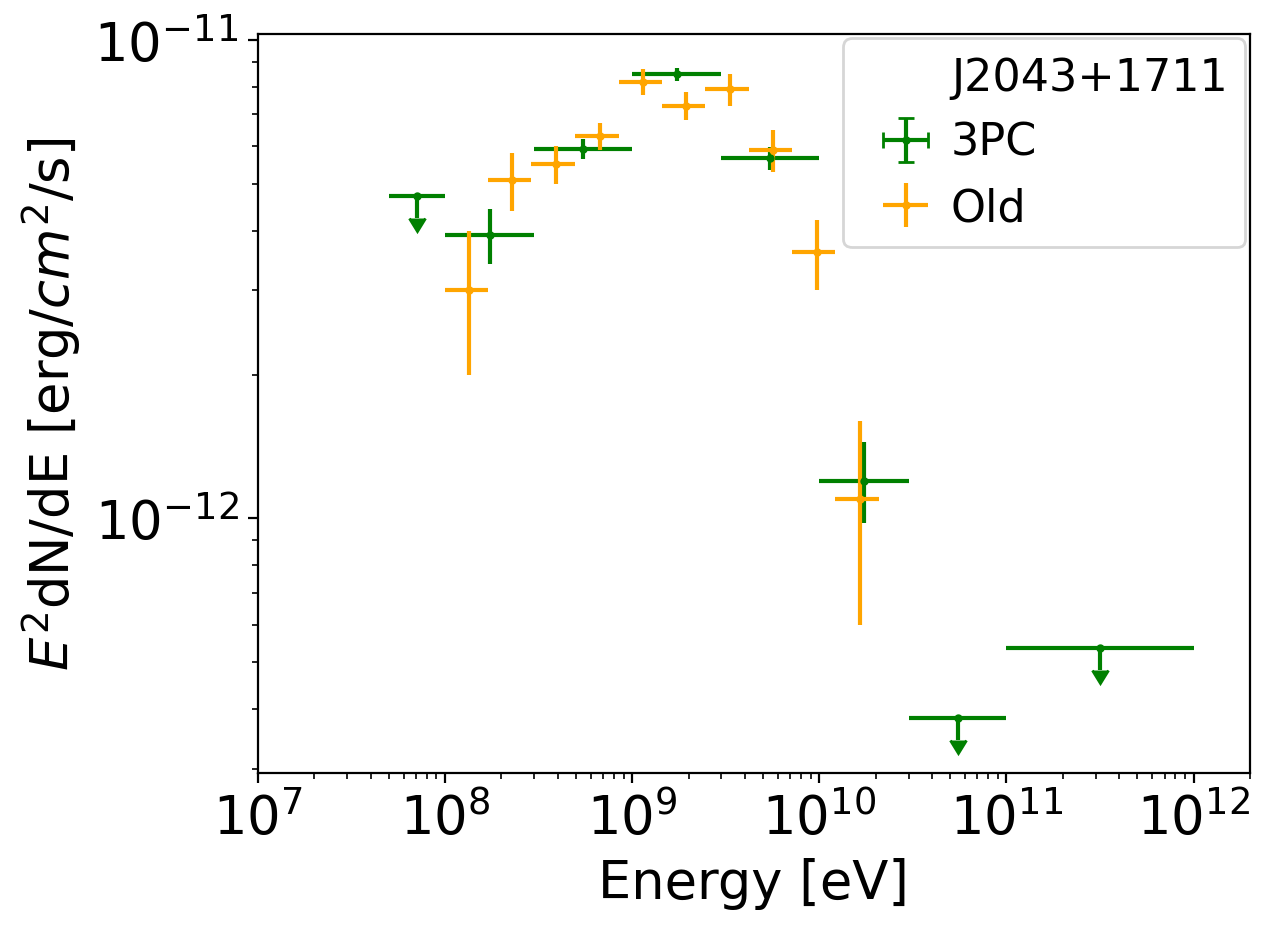}%
        \includegraphics[width=0.25\textwidth]{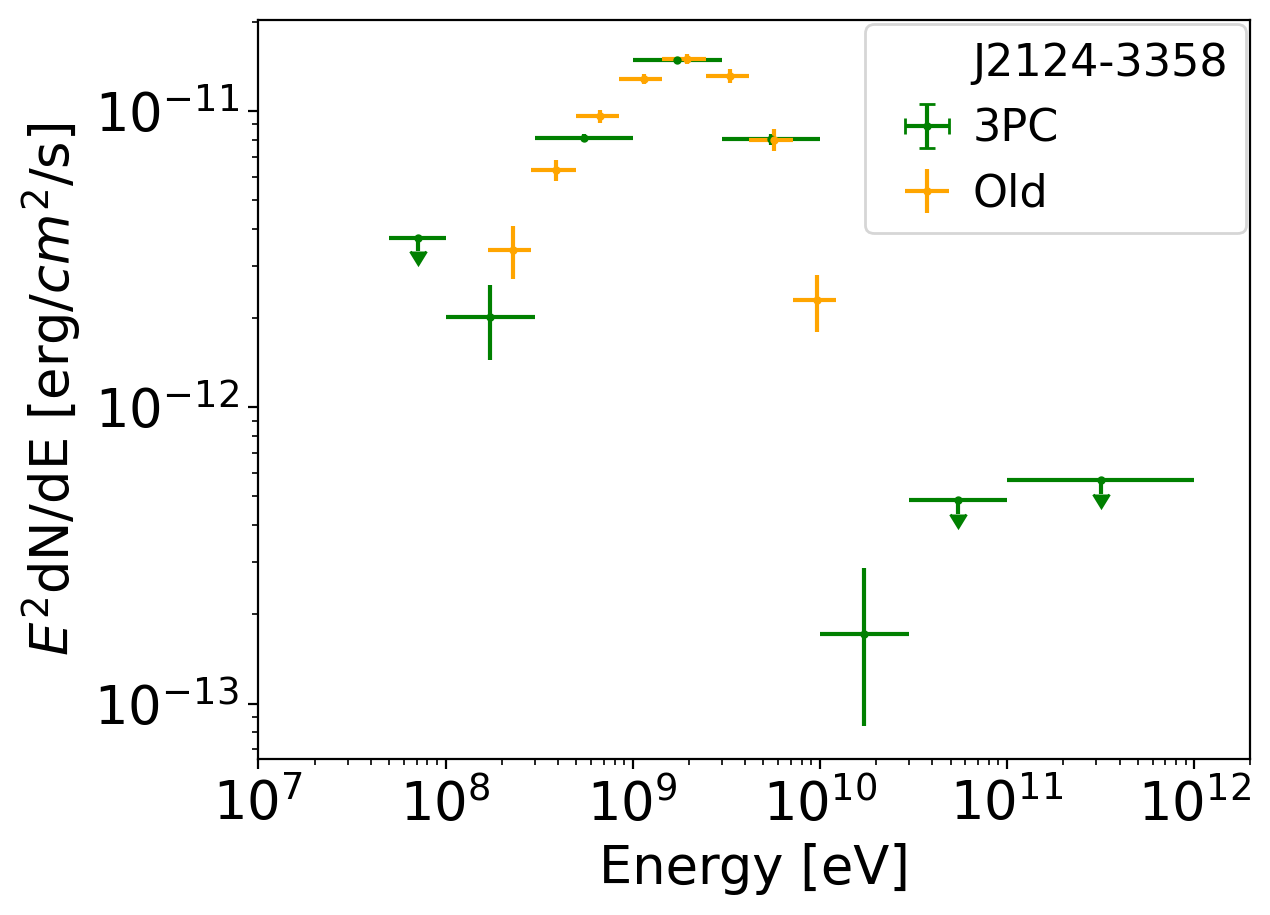}\\
        \includegraphics[width=0.25\textwidth]{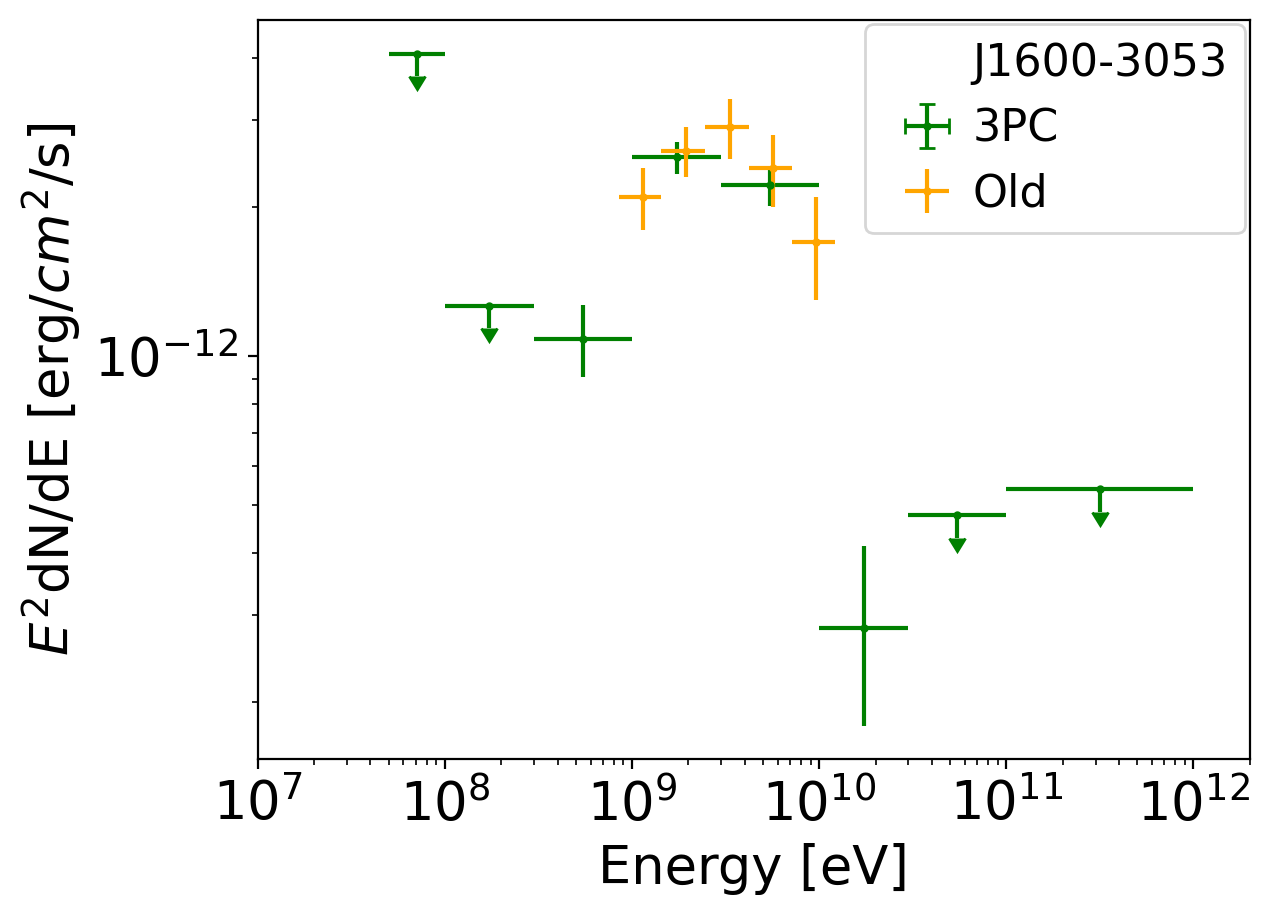}%
        \includegraphics[width=0.25\textwidth]{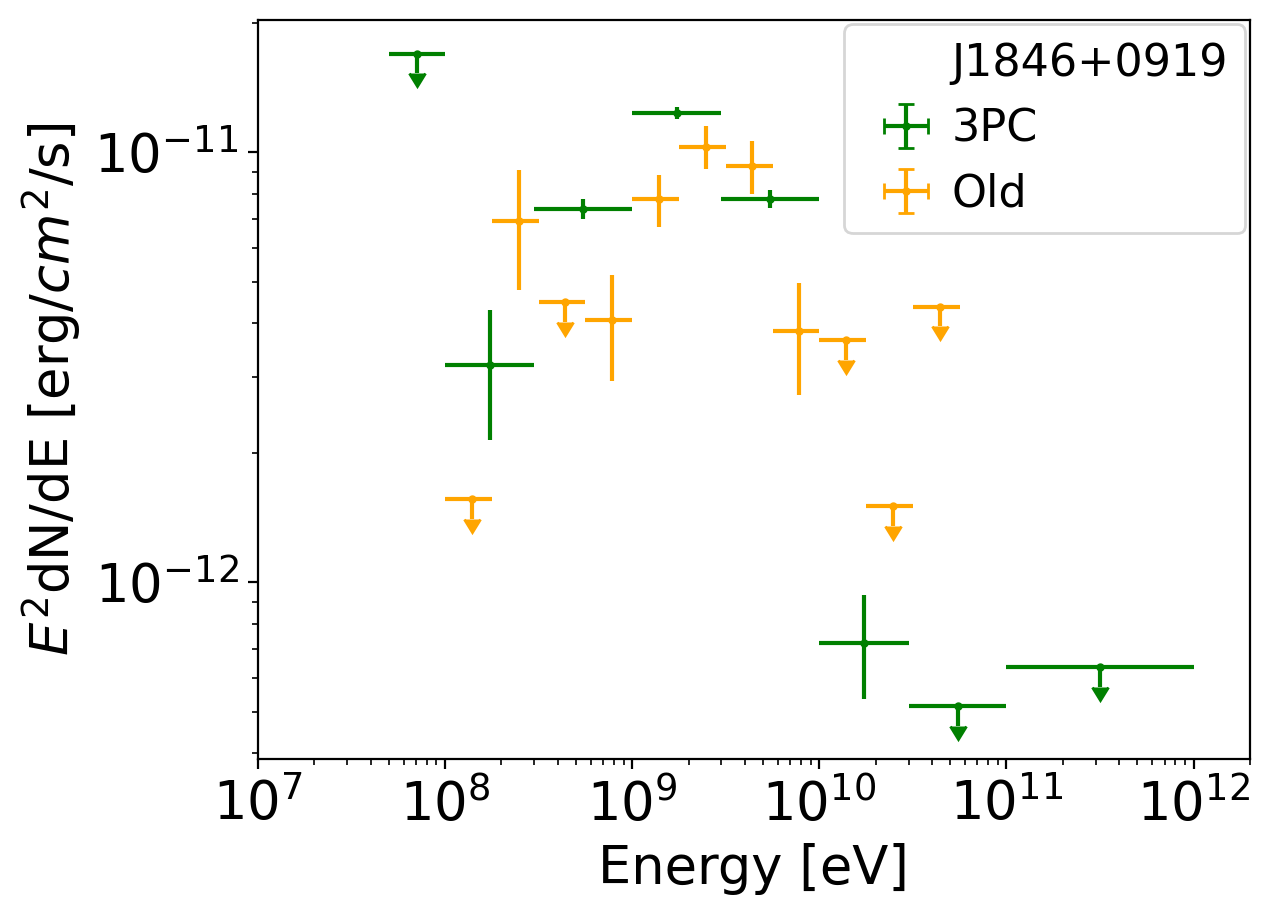}%
        \includegraphics[width=0.25\textwidth]{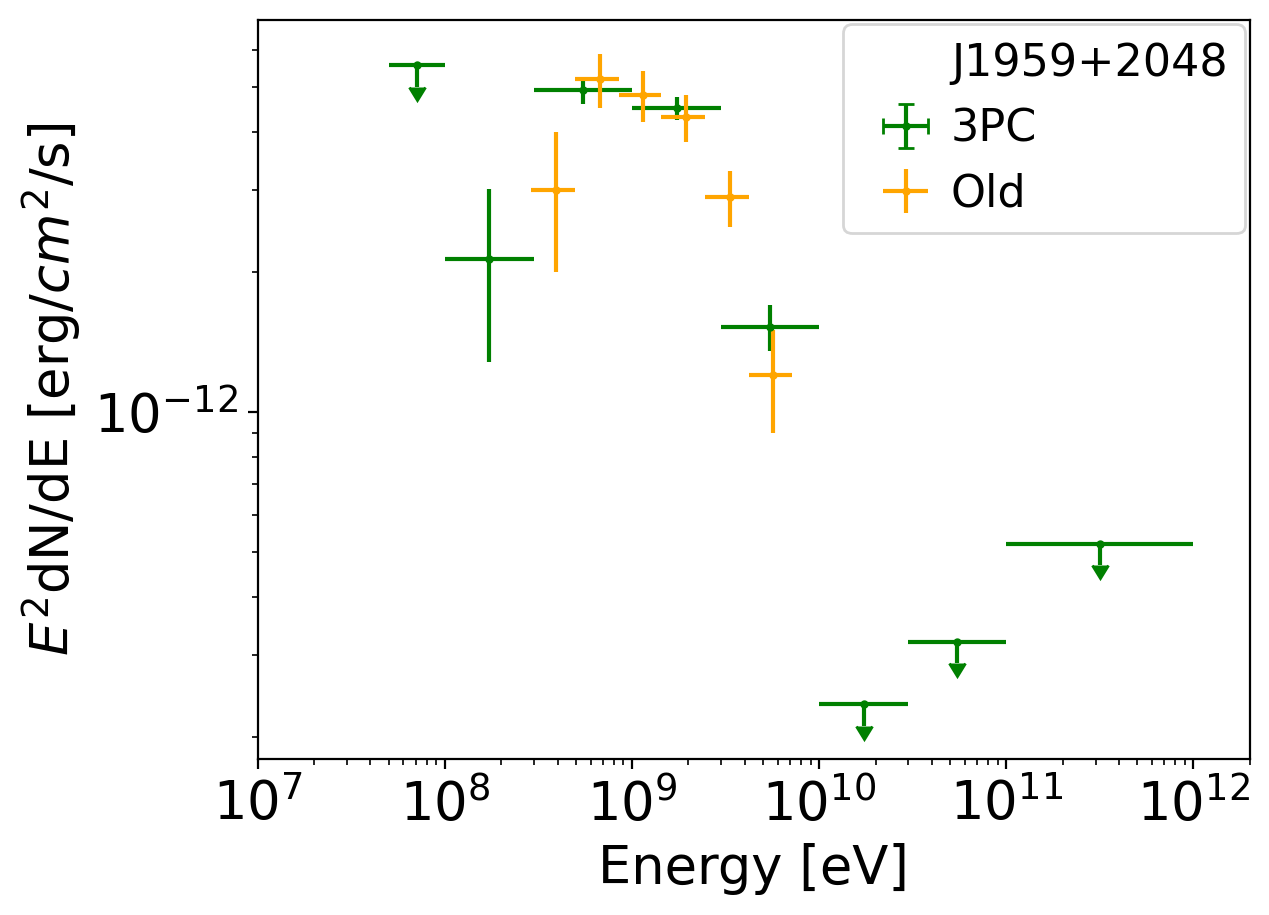}%
        \includegraphics[width=0.25\textwidth]{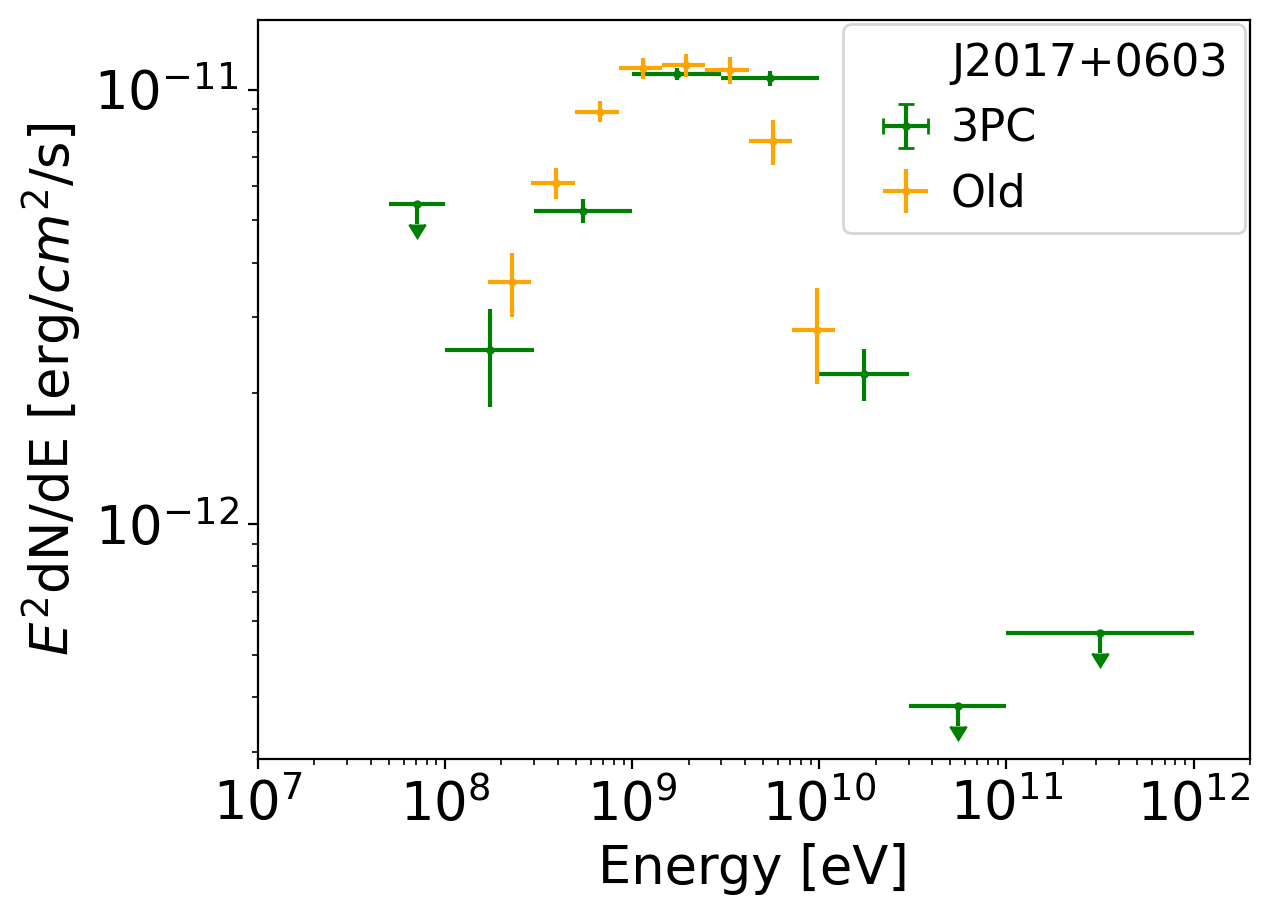}\\
        \includegraphics[width=0.25\textwidth]{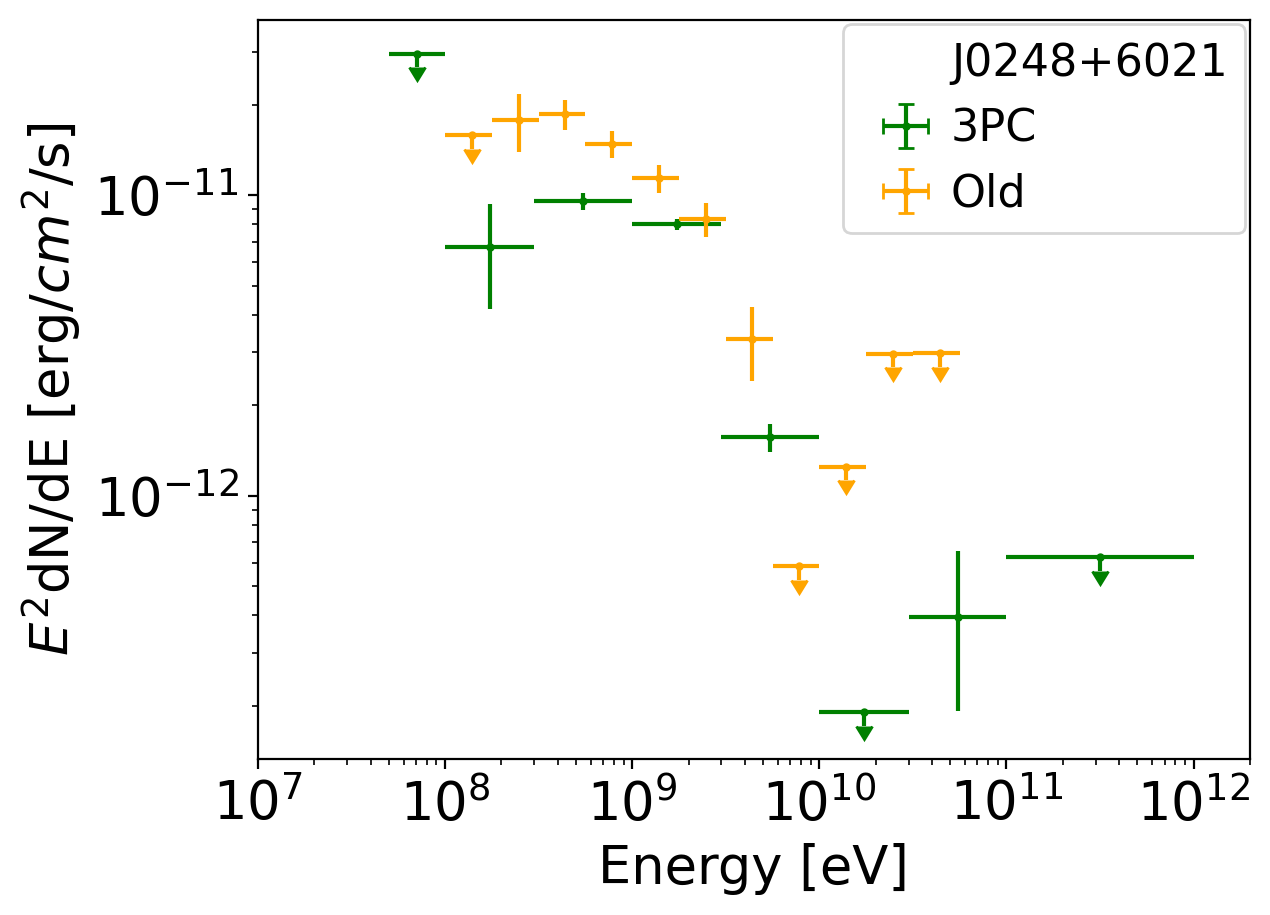}%
        \includegraphics[width=0.25\textwidth]{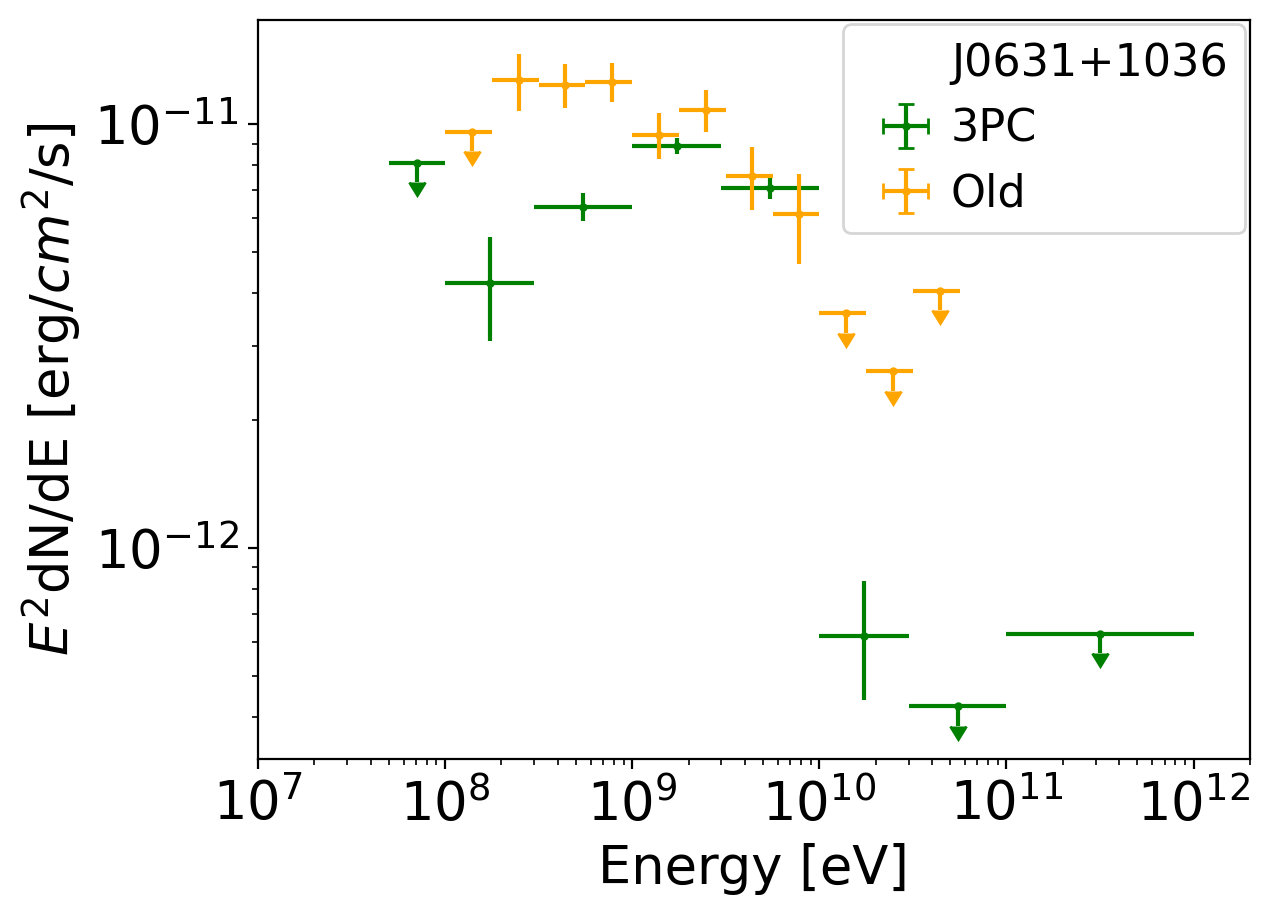}%
        \includegraphics[width=0.25\textwidth]{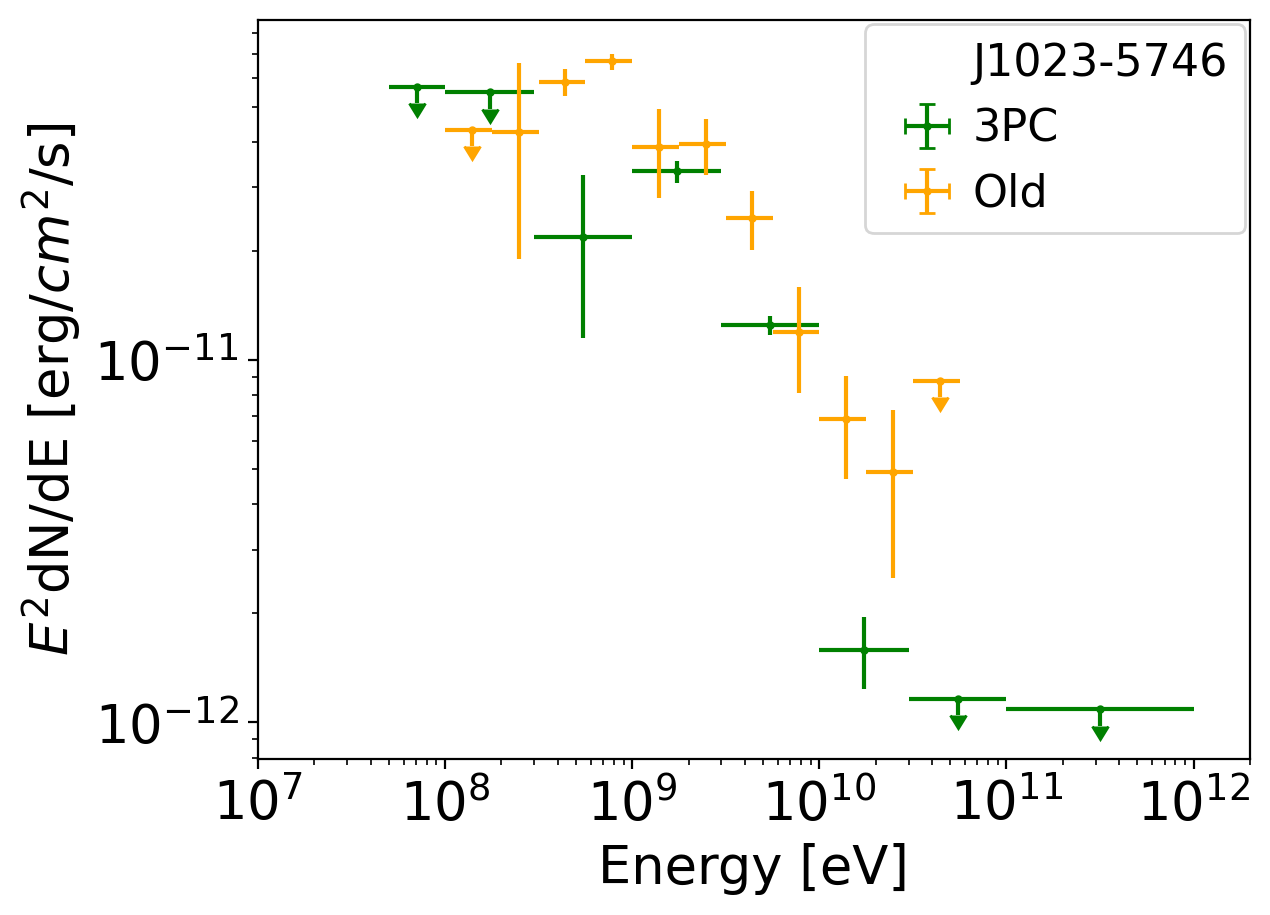}%
        \includegraphics[width=0.25\textwidth]{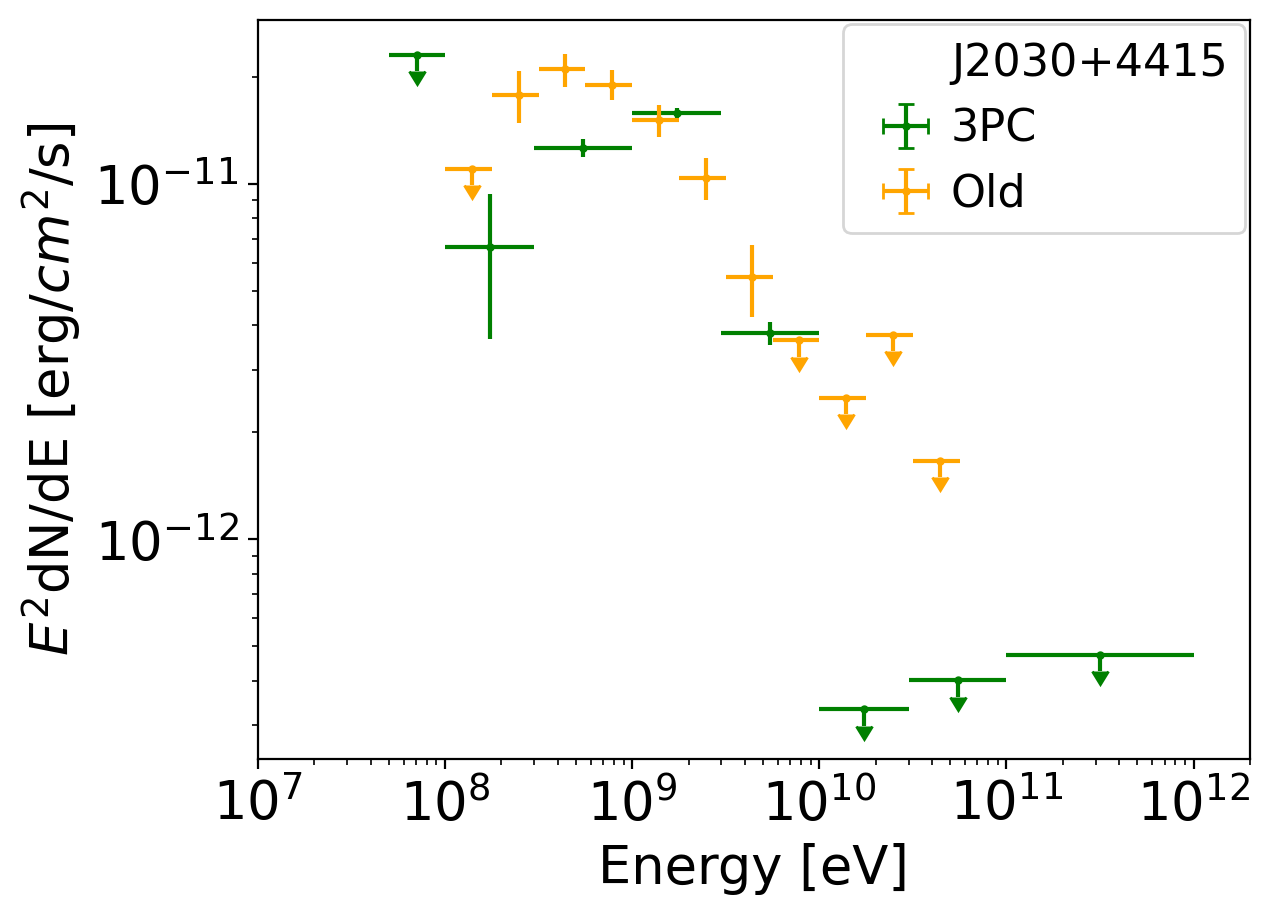}\\
        \caption{Comparison of spectral data from the 2PC or dedicated studies prior to the 3PC (in orange, labeled as \emph{Old} for clarity) and from the 3PC (in green). The top row shows pulsars in which the old data is clearly better defining the spectra.
        In the middle row we show the pulsars in which there is a doubt about which data set is a better description.
        The bottom row shows pulsars for which both data sets are very different.
        }
    \label{fig:comparison_gammaray_data}
    \end{figure*}

    As mentioned in the text, the spectral data presented in the 3PC for some pulsars differs from the data previously available (either in the 2PC or in dedicated studies), in shape but also in quality. In many cases one of the two sets of spectral data is clearly preferable than the other, for example because one of the two has more data points or smaller errors. 
    The top row of Fig \ref{fig:comparison_gammaray_data} shows the compared spectra of some pulsars in which the data previous to the 3PC (labeled for the sake of clarity as \emph{Old}) is preferable to the given in the 3PC.
    In some cases it is not as clear, since both data sets have a very similar shape and quality. 
    In these cases we take the data set with more continuous data points.
    Some examples are shown in middle row of Fig \ref{fig:comparison_gammaray_data}: for PSR J1600-3053, PSR J1959+2048, PSR J2017+0603 we choose the older data set and for PSR J1846+0919 the 3PC one.
    However, in some cases it is not possible to decide which set to choose, because shapes are very different (e.g. the spectra peaks at very different energies or have very different steepness at low energies). This occurs in 4 pulsars: J0248+6021, J0631+1036, J1023-5746 and J2030+4415. The bottom row of Fig. \ref{fig:comparison_gammaray_data} shows these spectra.
    
    \section{Computational cost of the light curve fitting}
    \label{app:computational_cost}

    The three light curve fitting techniques we have introduced in subsection \ref{fitting_procedures} differ substantially in their computational requirements. For the three of them we use the same machine, the Ladon High-Performance Computing Cluster at the Institute of Space Sciences, ICE-CSIC.
    The fastest method is the euclidean distance in frequency domain, requiring $\sim 0.37$ minutes to fit the whole population of 226 pulsars.
    The weighted reduced $\chi^2$ needs around two orders of magnitude more of time to complete the fitting, with $\sim 25.42$ minutes of computational time.
    Finally, DTW application to all pulsars (see its caveats as a fitting technique in the main text) would last $\sim 1.17\times10^4$ minutes, almost three orders of magnitude more than $\chi^2$ in time domain and five more than ED in frequency domain.
    This huge computational requirement, amounting more than eight days, even though is affordable, limits a lot the feasibility to carry on different tests, as we have done in this work.

    \section{Light curve fitting results}
    \label{app:data_light_curve_fits}
    
\begin{figure*}
        \centering
        \includegraphics[width=0.5\textwidth]{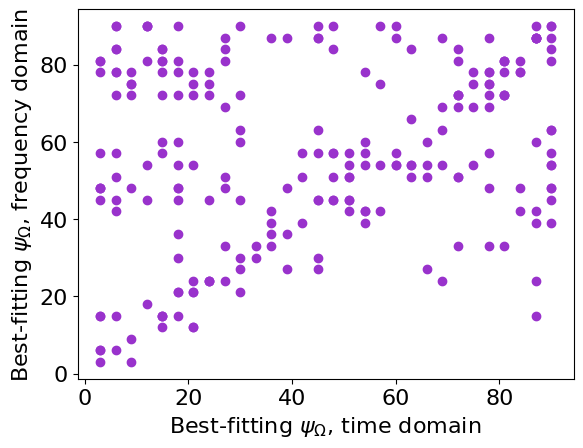}%
        \includegraphics[width=0.5\textwidth]{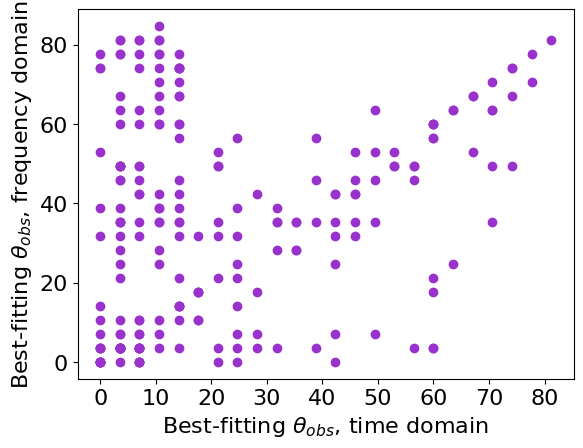}\\
    \caption{Comparison of the best-fitting values of $\psi_{\Omega}$ (left) and $\theta_{obs}$ (right) obtained in the fitting in time vs frequency domains.}
    \label{fig:psiomega_and_thetaobs_of_time_vs_frequency_fitting}
    \end{figure*}

    In the Supplementary Material Online of this paper are shown the plots of the best-fitting light curves in time and frequency domain, together with the contour plots of the statistics used in the fits, $\chi^2$ and euclidean distance, respectively.
    Table \ref{tab:parameters_pulsars_lightcurves_1} presents the best-fit geometrical values for all pulsars fitted, obtained with the two methods explained in the text.
    These are also compared in Fig. \ref{fig:psiomega_and_thetaobs_of_time_vs_frequency_fitting}.

    \begin{table*}
        \centering
        \caption{Best-fitting geometrical parameters (for fits in both time and frequency domain) of the pulsars on the 3PC.
        Notice that, due to the degeneracy between the geometrical parameters and the difficulty of defining an error of the parameters, pointed out in the text and shown in Fig. \ref{fig:degeneracy}, these values are not strong constraints.
        }
        \begin{tabular}{crccc|crccc}
            \hline\hline
            \multirow{2}{*}{Pulsar}  & \multicolumn{2}{c}{Time domain}  & \multicolumn{2}{c}{Frequency domain} & \multirow{2}{*}{Pulsar}  & \multicolumn{2}{c}{Time domain}  & \multicolumn{2}{c}{Frequency domain} 
            \\
             &$\psi_{\Omega} [^{\circ}]$ & $|\theta_{obs}| [^{\circ}]$ & $\psi_{\Omega} [^{\circ}]$ & $|\theta_{obs}| [^{\circ}]$ &  & $\psi_{\Omega} [^{\circ}]$ & $|\theta_{obs}| [^{\circ}]$ & $\psi_{\Omega} [^{\circ}]$ & $|\theta_{obs}| [^{\circ}]$   \\\hline
             J0002+6216  &  $81$  &  $46$  &  $75$  &  $35$  &  J1112\textminus6103 &  $30$  &  $4$  &  $33$  &  $4$   \\
            J0007+7303  &  $21$  &  $14$  &  $21$  &  $14$  &  J1119\textminus6127 &  $9$  &  $11$  &  $30$  &  $25$   \\
            J0023+0923  &  $24$  &  $11$  &  $57$  &  $35$  &  J1124\textminus3653 &  $66$  &  $35$  &  $87$  &  $53$   \\
            J0030+0451  &  $81$  &  $49$  &  $81$  &  $18$  &  J1124\textminus5916 &  $48$  &  $0$  &  $45$  &  $0$   \\
            J0034\textminus0534  &  $15$  &  $7$  &  $69$  &  $53$  &  J1125\textminus5825 &  $81$  &  $49$  &  $66$  &  $32$   \\
            J0101\textminus6422  &  $18$  &  $4$  &  $87$  &  $49$  &  J1135\textminus6055 &  $9$  &  $7$  &  $78$  &  $74$   \\
            J0102+4839  &  $24$  &  $4$  &  $27$  &  $4$  &  J1139\textminus6247 &  $27$  &  $21$  &  $27$  &  $21$   \\
            J0106+4855  &  $33$  &  $35$  &  $57$  &  $4$  &  J1142+0119 &  $84$  &  $74$  &  $72$  &  $60$   \\
            J0205+6449  &  $51$  &  $4$  &  $48$  &  $4$  &  J1151\textminus6108 &  $81$  &  $53$  &  $87$  &  $39$   \\
            J0218+4232  &  $6$  &  $4$  &  $84$  &  $81$  &  J1203\textminus6242 &  $36$  &  $4$  &  $39$  &  $4$   \\
            J0248+4230  &  $18$  &  $4$  &  $60$  &  $28$  &  J1207\textminus5050 &  $3$  &  $4$  &  $57$  &  $0$   \\
            J0251+2606  &  $39$  &  $4$  &  $45$  &  $4$  &  J1208\textminus6238 &  $42$  &  $4$  &  $42$  &  $4$   \\
            J0307+7443  &  $9$  &  $7$  &  $48$  &  $46$  &  J1221\textminus0633 &  $72$  &  $71$  &  $45$  &  $42$   \\
            J0312\textminus0921  &  $66$  &  $49$  &  $54$  &  $32$  &  J1227\textminus4853 &  $6$  &  $4$  &  $42$  &  $32$   \\
            J0318+0253  &  $45$  &  $42$  &  $45$  &  $35$  &  J1231\textminus1411 &  $81$  &  $53$  &  $72$  &  $21$   \\
            J0340+4130  &  $27$  &  $14$  &  $57$  &  $28$  &  J1231\textminus5113 &  $36$  &  $0$  &  $30$  &  $0$   \\
            J0357+3205  &  $81$  &  $71$  &  $75$  &  $74$  &  J1231\textminus6511 &  $3$  &  $11$  &  $78$  &  $67$   \\
            J0359+5414  &  $9$  &  $14$  &  $75$  &  $78$  &  J1301+0833 &  $87$  &  $78$  &  $87$  &  $71$   \\
            J0418+6635  &  $18$  &  $7$  &  $21$  &  $7$  &  J1302\textminus3258 &  $90$  &  $71$  &  $81$  &  $71$   \\
            J0437\textminus4715  &  $15$  &  $18$  &  $15$  &  $18$  &  J1311\textminus3430 &  $18$  &  $7$  &  $69$  &  $42$   \\
            J0514\textminus4408  &  $18$  &  $18$  &  $87$  &  $49$  &  J1312+0051 &  $21$  &  $11$  &  $45$  &  $46$   \\
            J0533+6759  &  $57$  &  $0$  &  $57$  &  $0$  &  J1335\textminus5656 &  $45$  &  $35$  &  $54$  &  $42$   \\
            J0554+3107  &  $90$  &  $71$  &  $78$  &  $64$  &  J1350\textminus6225 &  $42$  &  $0$  &  $54$  &  $0$   \\
            J0605+3757  &  $15$  &  $4$  &  $15$  &  $7$  &  J1357\textminus6429 &  $3$  &  $7$  &  $81$  &  $81$   \\
            J0610\textminus2100  &  $90$  &  $81$  &  $36$  &  $4$  &  J1358\textminus6025 &  $21$  &  $11$  &  $78$  &  $67$   \\
            J0613\textminus0200  &  $12$  &  $14$  &  $12$  &  $14$  &  J1400\textminus1431 &  $78$  &  $78$  &  $78$  &  $78$   \\
            J0614\textminus3329  &  $33$  &  $4$  &  $33$  &  $4$  &  J1410\textminus6132 &  $12$  &  $0$  &  $90$  &  $74$   \\
            J0621+2514  &  $87$  &  $56$  &  $84$  &  $46$  &  J1413\textminus6205 &  $69$  &  $46$  &  $69$  &  $42$   \\
            J0622+3749  &  $6$  &  $4$  &  $15$  &  $7$  &  J1418\textminus6058 &  $87$  &  $56$  &  $24$  &  $4$   \\
            J0631+0646  &  $81$  &  $74$  &  $84$  &  $67$  &  J1420\textminus6048 &  $24$  &  $14$  &  $72$  &  $60$   \\
            J0633+0632  &  $18$  &  $25$  &  $51$  &  $4$  &  J1422\textminus6138 &  $24$  &  $14$  &  $21$  &  $11$   \\
            J0633+1746  &  $90$  &  $71$  &  $54$  &  $0$  &  J1429\textminus5911 &  $33$  &  $4$  &  $27$  &  $4$   \\
            J0659+1414  &  $6$  &  $14$  &  $3$  &  $4$  &  J1446\textminus4701 &  $63$  &  $56$  &  $54$  &  $35$   \\
            J0729\textminus1448  &  $42$  &  $46$  &  $90$  &  $35$  &  J1447\textminus5757 &  $3$  &  $4$  &  $81$  &  $78$   \\
            J0734\textminus1559  &  $6$  &  $7$  &  $81$  &  $81$  &  J1455\textminus3330 &  $60$  &  $42$  &  $57$  &  $35$   \\
            J0740+6620  &  $51$  &  $0$  &  $39$  &  $7$  &  J1459\textminus6053 &  $6$  &  $7$  &  $6$  &  $7$   \\
            J0742\textminus2822  &  $30$  &  $32$  &  $87$  &  $35$  &  J1509\textminus5850 &  $15$  &  $11$  &  $78$  &  $71$   \\
            J0744\textminus2525  &  $90$  &  $46$  &  $63$  &  $4$  &  J1513\textminus5908 &  $3$  &  $0$  &  $57$  &  $53$   \\
            J0751+1807  &  $24$  &  $18$  &  $24$  &  $18$  &  J1513\textminus2550 &  $45$  &  $11$  &  $60$  &  $28$   \\
            J0802\textminus5613  &  $69$  &  $64$  &  $54$  &  $46$  &  J1514\textminus4946 &  $75$  &  $53$  &  $72$  &  $49$   \\
            J0835\textminus4510  &  $75$  &  $32$  &  $75$  &  $32$  &  J1522\textminus5735 &  $18$  &  $4$  &  $15$  &  $4$   \\
            J0908\textminus4913  &  $60$  &  $4$  &  $54$  &  $7$  &  J1526\textminus2744 &  $9$  &  $18$  &  $12$  &  $4$   \\
            J0931\textminus1902  &  $6$  &  $7$  &  $51$  &  $39$  &  J1528\textminus5838 &  $9$  &  $11$  &  $81$  &  $78$   \\
            J0940\textminus5428  &  $30$  &  $25$  &  $63$  &  $56$  &  J1531\textminus5610 &  $6$  &  $11$  &  $90$  &  $78$   \\
            J0952\textminus0607  &  $21$  &  $14$  &  $54$  &  $42$  &  J1536\textminus4948 &  $18$  &  $7$  &  $90$  &  $60$   \\
            J0955\textminus6150  &  $87$  &  $56$  &  $87$  &  $56$  &  J1543\textminus5149 &  $78$  &  $78$  &  $42$  &  $35$   \\
            J1012\textminus4235  &  $57$  &  $0$  &  $63$  &  $4$  &  J1552+5437 &  $90$  &  $81$  &  $90$  &  $78$   \\
            J1016\textminus5857  &  $84$  &  $53$  &  $81$  &  $46$  &  J1555\textminus2908 &  $48$  &  $7$  &  $84$  &  $32$   \\
            J1019\textminus5749  &  $9$  &  $14$  &  $48$  &  $49$  &  J1600\textminus3053 &  $30$  &  $21$  &  $60$  &  $53$   \\
            J1024\textminus0719  &  $15$  &  $11$  &  $24$  &  $14$  &  J1614\textminus2230 &  $90$  &  $67$  &  $63$  &  $4$   \\
            J1028\textminus5819  &  $39$  &  $4$  &  $42$  &  $4$  &  J1615\textminus5137 &  $27$  &  $18$  &  $69$  &  $64$   \\
            J1035\textminus6720  &  $15$  &  $14$  &  $48$  &  $49$  &  J1620\textminus4927 &  $12$  &  $7$  &  $81$  &  $74$   \\
            J1036\textminus8317  &  $3$  &  $4$  &  $12$  &  $0$  &  J1623\textminus5005 &  $75$  &  $56$  &  $75$  &  $56$   \\
            J1044\textminus5737  &  $33$  &  $14$  &  $33$  &  $14$  &  J1624\textminus4041 &  $27$  &  $4$  &  $30$  &  $4$   \\
            J1048\textminus5832  &  $72$  &  $32$  &  $42$  &  $11$  &  J1625\textminus0021 &  $21$  &  $11$  &  $3$  &  $4$   \\
            J1048+2339  &  $48$  &  $53$  &  $45$  &  $0$  &  J1627+3219 &  $54$  &  $46$  &  $51$  &  $42$   \\
            J1055\textminus6028  &  $9$  &  $7$  &  $84$  &  $85$  &  J1628\textminus3205 &  $87$  &  $71$  &  $87$  &  $67$   \\
            J1057\textminus5226  &  $81$  &  $67$  &  $78$  &  $74$  &  J1630+3734 &  $39$  &  $39$  &  $51$  &  $7$   \\
            J1057\textminus5851  &  $90$  &  $67$  &  $15$  &  $4$  &  J1640+2224 &  $54$  &  $39$  &  $51$  &  $32$   \\
            J1105\textminus6107  &  $51$  &  $0$  &  $48$  &  $0$  &  J1641+8049 &  $33$  &  $4$  &  $84$  &  $32$   \\
            J1105\textminus6037  &  $18$  &  $7$  &  $21$  &  $11$  &  J1641\textminus5317 &  $6$  &  $14$  &  $72$  &  $74$   \\
            J1111\textminus6039  &  $87$  &  $81$  &  $87$  &  $81$  &  J1648\textminus4611 &  $21$  &  $14$  &  $75$  &  $67$   \\\hline
        \end{tabular}
        \label{tab:parameters_pulsars_lightcurves_1}
    \end{table*}

    \addtocounter{table}{-1}  
    
    \begin{table*}
        \centering
        \caption{- \emph{continued}}
        \begin{tabular}{crccc|crccc}
            \hline\hline
            \multirow{2}{*}{Pulsar}  & \multicolumn{2}{c}{Time domain}  & \multicolumn{2}{c}{Frequency domain} & \multirow{2}{*}{Pulsar}  & \multicolumn{2}{c}{Time domain}  & \multicolumn{2}{c}{Frequency domain} 
            \\
             &$\psi_{\Omega} [^{\circ}]$ & $|\theta_{obs}| [^{\circ}]$ & $\psi_{\Omega} [^{\circ}]$ & $|\theta_{obs}| [^{\circ}]$ &  & $\psi_{\Omega} [^{\circ}]$ & $|\theta_{obs}| [^{\circ}]$ & $\psi_{\Omega} [^{\circ}]$ & $|\theta_{obs}| [^{\circ}]$   \\\hline
             J1649\textminus3012  &  $21$  &  $18$  &  $42$  &  $35$  &  J1903\textminus7051 &  $6$  &  $14$  &  $51$  &  $49$   \\
            J1650\textminus4601  &  $18$  &  $7$  &  $78$  &  $64$  &  J1906+0722 &  $24$  &  $18$  &  $24$  &  $18$   \\
            J1653\textminus0158  &  $9$  &  $4$  &  $90$  &  $81$  &  J1907+0602 &  $81$  &  $64$  &  $81$  &  $64$   \\
            J1658\textminus5324  &  $3$  &  $4$  &  $45$  &  $35$  &  J1908+2105 &  $30$  &  $35$  &  $39$  &  $7$   \\
            J1702\textminus4128  &  $3$  &  $4$  &  $48$  &  $46$  &  J1913+0904 &  $78$  &  $53$  &  $57$  &  $35$   \\
            J1705\textminus1906  &  $72$  &  $49$  &  $60$  &  $39$  &  J1921+0137 &  $24$  &  $11$  &  $72$  &  $49$   \\
            J1709\textminus4429  &  $24$  &  $18$  &  $24$  &  $18$  &  J1932+1916 &  $6$  &  $11$  &  $78$  &  $78$   \\
            J1713+0747  &  $12$  &  $11$  &  $81$  &  $74$  &  J1939+2134 &  $45$  &  $4$  &  $45$  &  $4$   \\
            J1714\textminus3830  &  $3$  &  $0$  &  $15$  &  $11$  &  J1946\textminus5403 &  $3$  &  $4$  &  $6$  &  $4$   \\
            J1718\textminus3825  &  $6$  &  $11$  &  $6$  &  $11$  &  J1952+3252 &  $42$  &  $4$  &  $39$  &  $4$   \\
            J1730\textminus2304  &  $33$  &  $14$  &  $54$  &  $28$  &  J1954+2836 &  $21$  &  $4$  &  $24$  &  $4$   \\
            J1730\textminus3350  &  $42$  &  $11$  &  $33$  &  $7$  &  J1957+5033 &  $84$  &  $71$  &  $72$  &  $67$   \\
            J1732\textminus3131  &  $78$  &  $56$  &  $78$  &  $35$  &  J1958+2846 &  $84$  &  $56$  &  $87$  &  $64$   \\
            J1732\textminus5049  &  $12$  &  $11$  &  $54$  &  $42$  &  J1959+2048 &  $48$  &  $0$  &  $48$  &  $0$   \\
            J1736\textminus3422  &  $54$  &  $7$  &  $87$  &  $35$  &  J2006+0148 &  $45$  &  $32$  &  $90$  &  $28$   \\
            J1741\textminus2054  &  $75$  &  $64$  &  $72$  &  $60$  &  J2006+3102 &  $15$  &  $18$  &  $48$  &  $49$   \\
            J1741+1351  &  $54$  &  $49$  &  $57$  &  $0$  &  J2017+0603 &  $78$  &  $64$  &  $69$  &  $49$   \\
            J1742\textminus3321  &  $66$  &  $67$  &  $57$  &  $56$  &  J2017+3625 &  $24$  &  $4$  &  $45$  &  $21$   \\
            J1744\textminus1134  &  $54$  &  $42$  &  $48$  &  $49$  &  J2017\textminus1614 &  $90$  &  $71$  &  $84$  &  $49$   \\
            J1744\textminus7619  &  $87$  &  $67$  &  $75$  &  $60$  &  J2021+3651 &  $48$  &  $4$  &  $42$  &  $4$   \\
            J1745+1017  &  $75$  &  $60$  &  $51$  &  $35$  &  J2021+4026 &  $3$  &  $0$  &  $15$  &  $4$   \\
            J1746\textminus3239  &  $9$  &  $11$  &  $33$  &  $25$  &  J2022+3842 &  $33$  &  $7$  &  $87$  &  $64$   \\
            J1747\textminus2958  &  $21$  &  $7$  &  $81$  &  $60$  &  J2028+3332 &  $84$  &  $64$  &  $84$  &  $39$   \\
            J1747\textminus4036  &  $15$  &  $0$  &  $90$  &  $74$  &  J2030+3641 &  $84$  &  $67$  &  $72$  &  $60$   \\
            J1801\textminus2451  &  $54$  &  $0$  &  $54$  &  $0$  &  J2032+4127 &  $18$  &  $25$  &  $54$  &  $4$   \\
            J1803\textminus2149  &  $72$  &  $46$  &  $72$  &  $46$  &  J2034+3632 &  $3$  &  $11$  &  $51$  &  $32$   \\
            J1805+0615  &  $87$  &  $74$  &  $72$  &  $60$  &  J2039\textminus3616 &  $39$  &  $39$  &  $45$  &  $46$   \\
            J1809\textminus2332  &  $36$  &  $18$  &  $27$  &  $11$  &  J2039\textminus5617 &  $12$  &  $11$  &  $72$  &  $64$   \\
            J1810+1744  &  $6$  &  $7$  &  $84$  &  $81$  &  J2042+0246 &  $3$  &  $4$  &  $48$  &  $35$   \\
            J1811\textminus2405  &  $24$  &  $7$  &  $33$  &  $11$  &  J2043+1711 &  $21$  &  $4$  &  $24$  &  $4$   \\
            J1813\textminus1246  &  $15$  &  $0$  &  $15$  &  $0$  &  J2043+2740 &  $81$  &  $53$  &  $75$  &  $35$   \\
            J1816+4510  &  $90$  &  $56$  &  $39$  &  $7$  &  J2047+1053 &  $69$  &  $53$  &  $60$  &  $42$   \\
            J1817\textminus1742  &  $30$  &  $35$  &  $90$  &  $35$  &  J2051\textminus0827 &  $84$  &  $64$  &  $54$  &  $39$   \\
            J1823\textminus3021A  &  $51$  &  $25$  &  $51$  &  $21$  &  J2052+1219 &  $63$  &  $32$  &  $54$  &  $25$   \\
            J1824\textminus2452A  &  $9$  &  $0$  &  $90$  &  $78$  &  J2055+2539 &  $12$  &  $14$  &  $78$  &  $67$   \\
            J1826\textminus1256  &  $39$  &  $4$  &  $36$  &  $4$  &  J2111+4606 &  $78$  &  $64$  &  $72$  &  $53$   \\
            J1827\textminus0849  &  $12$  &  $14$  &  $90$  &  $81$  &  J2115+5448 &  $75$  &  $49$  &  $63$  &  $25$   \\
            J1828\textminus1101  &  $6$  &  $0$  &  $90$  &  $14$  &  J2124\textminus3358 &  $9$  &  $7$  &  $51$  &  $42$   \\
            J1824\textminus0621  &  $69$  &  $60$  &  $48$  &  $39$  &  J2129\textminus0429 &  $90$  &  $56$  &  $60$  &  $0$   \\
            J1827\textminus1446  &  $15$  &  $14$  &  $54$  &  $49$  &  J2139+4716 &  $6$  &  $14$  &  $45$  &  $39$   \\
            J1833\textminus1034  &  $42$  &  $46$  &  $87$  &  $60$  &  J2214+3000 &  $18$  &  $4$  &  $18$  &  $4$   \\
            J1836+5925  &  $9$  &  $4$  &  $9$  &  $4$  &  J2215+5135 &  $84$  &  $46$  &  $81$  &  $35$   \\
            J1837\textminus0604  &  $21$  &  $0$  &  $48$  &  $4$  &  J2229+6114 &  $9$  &  $11$  &  $30$  &  $25$   \\
            J1838\textminus0537  &  $18$  &  $11$  &  $78$  &  $71$  &  J2234+0944 &  $54$  &  $46$  &  $57$  &  $49$   \\
            J1843\textminus1113  &  $9$  &  $4$  &  $81$  &  $64$  &  J2238+5903 &  $54$  &  $0$  &  $57$  &  $0$   \\
            J1844\textminus0346  &  $9$  &  $14$  &  $78$  &  $74$  &  J2240+5832 &  $27$  &  $32$  &  $87$  &  $32$   \\
            J1846+0919  &  $84$  &  $74$  &  $78$  &  $74$  &  J2241\textminus5236 &  $12$  &  $4$  &  $45$  &  $11$   \\
            J1855\textminus1436  &  $12$  &  $14$  &  $90$  &  $56$  &  J2256\textminus1024 &  $39$  &  $0$  &  $45$  &  $4$   \\
            J1858\textminus2216  &  $78$  &  $60$  &  $72$  &  $49$  &  J2302+4442 &  $60$  &  $35$  &  $90$  &  $21$   \\
            J1901\textminus0125  &  $42$  &  $18$  &  $36$  &  $14$  &  J2310\textminus0555 &  $84$  &  $53$  &  $42$  &  $4$   \\
            J1902\textminus5105  &  $24$  &  $4$  &  $21$  &  $4$  &  J2339\textminus0533 &  $75$  &  $46$  &  $78$  &  $53$   \\ \hline
        \end{tabular}
        \label{tab:parameters_pulsars_lightcurves_2}
    \end{table*}

    \label{lastpage}
    \end{document}